\documentclass[twocolumn]{aastex631}

\usepackage{natbib,aas_macros,amsmath}
\usepackage{multirow,color}

\begin{document}

\shorttitle{Morphologies of Galaxies at $\lowercase{z} \gtrsim 9$ 
}
\shortauthors{Ono et al.}
\submitjournal{ApJ in press}

\title{%
Morphologies of Galaxies at $\lowercase{z} \gtrsim 9$ 
Uncovered by JWST/NIRCam Imaging:\\
Cosmic Size Evolution and an Identification of an Extremely Compact Bright Galaxy at $z\sim 12$
}


\author[0000-0001-9011-7605]{Yoshiaki Ono}
\affiliation{Institute for Cosmic Ray Research, The University of Tokyo, 5-1-5 Kashiwanoha, Kashiwa, Chiba 277-8582, Japan}

\author[0000-0002-6047-430X]{Yuichi Harikane}
\affiliation{Institute for Cosmic Ray Research, The University of Tokyo, 5-1-5 Kashiwanoha, Kashiwa, Chiba 277-8582, Japan}

\author[0000-0002-1049-6658]{Masami Ouchi}
\affiliation{National Astronomical Observatory of Japan, 2-21-1 Osawa, Mitaka, Tokyo 181-8588, Japan}
\affiliation{Institute for Cosmic Ray Research, The University of Tokyo, 5-1-5 Kashiwanoha, Kashiwa, Chiba 277-8582, Japan}
\affiliation{Kavli Institute for the Physics and Mathematics of the Universe (WPI), The University of Tokyo, 5-1-5 Kashiwanoha, Kashiwa-shi, Chiba, 277-8583, Japan}

\author[0000-0002-1319-3433]{Hidenobu Yajima}
\affiliation{Center for Computational Sciences, University of Tsukuba, Ten-nodai, 1-1-1 Tsukuba, Ibaraki 305-8577, Japan}

\author{Makito Abe}
\affiliation{Center for Computational Sciences, University of Tsukuba, Ten-nodai, 1-1-1 Tsukuba, Ibaraki 305-8577, Japan}

\author[0000-0001-7730-8634]{Yuki Isobe}
\affiliation{Institute for Cosmic Ray Research, The University of Tokyo, 5-1-5 Kashiwanoha, Kashiwa, Chiba 277-8582, Japan}
\affiliation{Department of Physics, Graduate School of Science, The University of Tokyo, 7-3-1 Hongo, Bunkyo, Tokyo 113-0033, Japan}

\author{Takatoshi Shibuya}
\affiliation{Kitami Institute of Technology, 165 Koen-cho, Kitami, Hokkaido 090-8507, Japan}

\author[0000-0003-1173-8847]{John H. Wise}
\affiliation{Center for Relativistic Astrophysics, School of Physics, Georgia Institute of Technology, Atlanta, GA 30332, USA}

\author[0000-0003-3817-8739]{Yechi Zhang}
\affiliation{Institute for Cosmic Ray Research, The University of Tokyo, 5-1-5 Kashiwanoha, Kashiwa, Chiba 277-8582, Japan}
\affiliation{Department of Astronomy, Graduate School of Science, the University of Tokyo, 7-3-1 Hongo, Bunkyo, Tokyo 113-0033, Japan}

\author[0000-0003-2965-5070]{Kimihiko Nakajima}
\affiliation{National Astronomical Observatory of Japan, 2-21-1 Osawa, Mitaka, Tokyo 181-8588, Japan}

\author{Hiroya Umeda}
\affiliation{Institute for Cosmic Ray Research, The University of Tokyo, 5-1-5 Kashiwanoha, Kashiwa, Chiba 277-8582, Japan}
\affiliation{Department of Physics, Graduate School of Science, The University of Tokyo, 7-3-1 Hongo, Bunkyo, Tokyo 113-0033, Japan}

\begin{abstract}
We present morphologies of galaxies at $z \gtrsim 9$ resolved by JWST/NIRCam $2$--$5\mu$m imaging. 
Our sample consists of $22$ galaxy candidates 
identified by stringent dropout and photo-$z$ criteria in GLASS, CEERS, SMACS J0723, and Stephan's Quintet flanking fields, 
one of which has been spectroscopically identified at $z=11.44$.
We perform surface brightness (SB) profile fitting with GALFIT 
for $6$ bright galaxies with S/N $=10$--$40$ on an individual basis 
and for stacked faint galaxies with secure point-spread functions (PSFs) of the NIRCam real data, 
carefully evaluating systematics by Monte-Carlo simulations. 
We compare our results with those of previous JWST studies, 
and confirm that effective radii $r_{\rm e}$ of our measurements are consistent with those of previous measurements at $z\sim 9$.
We obtain $r_{\rm e}\simeq 200$--$300$ pc with the exponential-like profiles, 
S\'ersic indexes of $n\simeq 1$--$1.5$, for galaxies at $z\sim 12$--$16$,
indicating that the relation of $r_{\rm e}\propto (1+z)^s$ for $s=-1.22^{+0.17}_{-0.16}$
explains cosmic evolution over $z\sim 0$--$16$ for $\sim L^*_{z=3}$ galaxies.
One bright ($M_{\rm UV}=-21$ mag) galaxy at $z\sim 12$, GL-z12-1, 
has an extremely compact profile with $r_{\rm e}=39 \pm 11$ pc 
that is surely extended over the PSF. 
Even in the case that the GL-z12-1 SB is fit by AGN$+$galaxy composite profiles, 
the best-fit galaxy component is again compact, 
$r_{\rm e}=48^{+38}_{-15}$ pc that is significantly ($>5\sigma$) smaller than the typical $r_{\rm e}$ value at $z\sim 12$. 
Comparing with numerical simulations, 
we find that such a compact galaxy naturally forms at $z\gtrsim 10$, 
and that frequent mergers at the early epoch produce more extended galaxies following the $r_{\rm e}\propto (1+z)^s$ relation.
\end{abstract}

\keywords{%
\href{http://astrothesaurus.org/uat/734}{High-redshift galaxies (734)}; 
\href{http://astrothesaurus.org/uat/979}{Lyman-break galaxies (979)}; 
\href{http://astrothesaurus.org/uat/594}{Galaxy evolution (594)}; 
\href{http://astrothesaurus.org/uat/595}{Galaxy formation (595)}; 
\href{http://astrothesaurus.org/uat/622}{Galaxy structure (622)}; 
\href{http://astrothesaurus.org/uat/617}{Galaxy radii (617)}; 
\href{http://astrothesaurus.org/uat/573}{Galaxies (573)}; 
\href{http://astrothesaurus.org/uat/563}{Galactic and extragalactic astronomy (563)} 
}

\section{Introduction} \label{sec:introduction}

The James Webb Space Telescope (JWST; \citealt{2023arXiv230404869G}) has opened up the redshift frontier 
by providing deep infrared images with unprecedented high sensitivity and resolution 
(e.g., \citealt{2022ApJ...936L..14P}). 
Early studies have reported 
dozens of high-$z$ galaxies candidates at $z \sim 9$--$16$ 
(\citealt{2022ApJ...940L..14N}; \citealt{2022ApJ...938L..15C}; \citealt{2023ApJ...942L..26L}; 
\citealt{2023MNRAS.518.4755A}; \citealt{2023ApJ...942L...9Y}; \citealt{2023MNRAS.519.1201A})
and the evolution of the UV luminosity density and the stellar mass density at very early epochs 
(\citealt{2023MNRAS.518.6011D}; \citealt{2022ApJ...940L..55F}; \citealt{2022arXiv220712446L}; 
\citealt{2023ApJS..265....5H}; 
see also, \citealt{2022ApJ...938L..10I}; \citealt{2023MNRAS.518.2511L}) 
as well as their various galaxy properties 
such as star formation histories and UV continuum slopes 
have begun to be discussed 
(\citealt{2023MNRAS.519..157W}; \citealt{2022ApJ...941..153T}; 
\citealt{2023MNRAS.520...14C}; \citealt{2023MNRAS.519.3064F}; 
see also, \citealt{2022arXiv220713860N}).

Characterizing the evolution of galaxy sizes is useful for understanding galaxy formation history 
(\citealt{2014ARA&A..52..291C}).  
Before the arrival of JWST, 
the sizes of galaxies at $z \sim 7$--$10$ have been measured 
based on deep near-infrared images taken with the Hubble Space Telescope (HST) 
and their size-luminosity relation and the size evolution have been intensively investigated 
(e.g., \citealt{2010ApJ...709L..21O}; 
\citealt{2012A&A...547A..51G}; 
\citealt{2013ApJ...777..155O}; 
\citealt{2015ApJ...804..103K}; 
\citealt{2015ApJ...808....6H}; 
\citealt{2015ApJS..219...15S}; 
\citealt{2016MNRAS.457..440C}; 
\citealt{2017ApJ...834L..11A}; 
\citealt{2017MNRAS.466.3612B}; 
\citealt{2017ApJ...843...41B}; 
\citealt{2018ApJ...855....4K}; 
\citealt{2018ApJ...864L..22S}; 
\citealt{2019ApJ...882...42B}; 
\citealt{2020AJ....160..154H}; 
\citealt{2021AJ....162..255B}; 
\citealt{2022ApJ...927...81B}). 
Based on the size measurement results with HST legacy data 
for about 190,000 galaxies at $z\sim0$--$10$, 
\cite{2015ApJS..219...15S} have performed 
a power-law fitting for the size-luminosity relation, 
and found that 
the slope of the power-law is almost constant, $\alpha = -0.27 \pm 0.01$, 
and the characteristic half-light radius at $M_{\rm UV} = -21.0$ mag 
becomes smaller with increasing redshift 
scaling as $\propto (1+z)^{-1.20 \pm 0.04}$. 
Now that JWST has allowed us to find galaxies with similar luminosities at even higher redshifts 
thanks to the better sensitivity and longer wavelength coverage, 
and to obtain more accurate size measurements 
because of the better spatial resolution compared to HST,  
it would be interesting to investigate if this trend continues toward higher redshifts, 
to study star formation activities in galaxies well before the end of the cosmic reionization 
(\citealt{2020ARA&A..58..617O}; \citealt{2020A&A...641A...6P}; \citealt{2022ARA&A..60..121R}). 
Recently, theoretical studies tailored for JWST observations for galaxies at lower redshifts $z=3$--$6$ 
based on cosmological simulations have been reported (\citealt{2022arXiv220800007C}), 
and comparisons with such theoretical study results are expected to  
become increasingly important as JWST data are obtained.

\begin{figure}[h]
\begin{center}
   \includegraphics[width=0.5\textwidth]{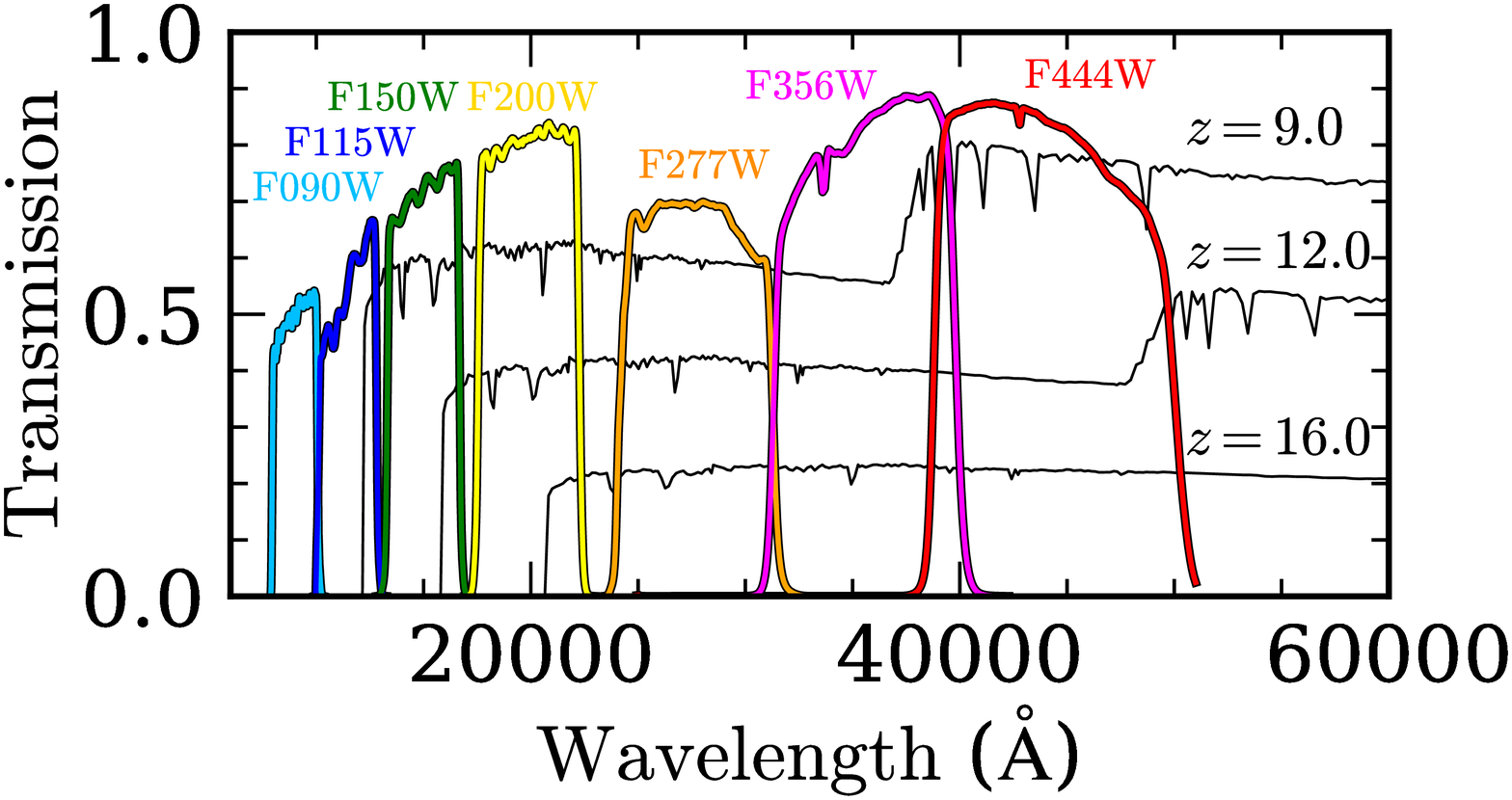}
\caption{
Transmissions of the seven NIRCam broadband filters 
(cyan: F090W, blue: F115W, green: F150W, yellow: F200W, 
orange: F277W, magenta: F356W, red: F444W) 
together with three spectra of star-forming galaxies at 
$z=9.0$, $12.0$, and $16.0$ from the \cite{2003MNRAS.344.1000B} library (black lines). 
}
\label{fig:SED}
\end{center}
\end{figure}

\begin{deluxetable*}{ccccccc} 
\tablecolumns{7} 
\tablewidth{0pt} 
\tablecaption{$z \sim 9$ Galaxy Candidates (F115W-dropouts) Used in Our Size Analysis  
\label{tab:z9cand}}
\tablehead{
    \colhead{ID} 
    &  \colhead{R.A.}
    &  \colhead{Decl.}
    &  \colhead{$z_{\rm photo}$}
    &  \colhead{$m_{\rm UV}^{\rm (ap)}$}
    &  \colhead{$m_{\rm opt}^{\rm (ap)}$}
    &  \colhead{$M_{\rm UV}$}
\\
    \colhead{ } 
    &  \colhead{(h:m:s)}
    &  \colhead{(d:m:s)}
    &  \colhead{ }
    &  \colhead{(mag)}
    &  \colhead{(mag)}
    &  \colhead{(mag)}
\\
    \colhead{(1)} 
    &  \colhead{(2)}
    &  \colhead{(3)}
    &  \colhead{(4)}
    &  \colhead{(5)}
    &  \colhead{(6)}
    &  \colhead{(7)}
}
\startdata 
\multicolumn{7}{c}{$L/L^\ast_{z=3} = 0.3$--$1$} \\  
GL-z9-1 & 00:14:02.85 & $-$30:22:18.6 & $10.49_{-0.72}^{+0.53}$ & $27.9$ & $27.6$ & $-20.9$ \\ \hline 
\multicolumn{7}{c}{$L/L^\ast_{z=3} = 0.12$--$0.3$} \\  
GL-z9-2 & 00:14:03.28 & $-$30:21:05.6 & $10.46_{-0.99}^{+0.45}$ & $29.6$ & $29.6$ & $-19.7$ \\ 
GL-z9-4 & 00:14:00.27 & $-$30:21:25.9 & $10.19_{-0.55}^{+0.63}$ & $29.2$ & $29.9$ & $-19.4$ \\ 
GL-z9-6 & 00:14:04.37 & $-$30:20:39.6 & $8.97_{-0.36}^{+0.36}$ & $29.6$ & $29.7$ & $-18.9$ \\ 
GL-z9-3 & 00:14:00.09 & $-$30:19:06.9 & $8.93_{-0.38}^{+0.39}$ & $29.5$ & $29.0$ & $-18.8$ \\ 
GL-z9-5 & 00:14:03.10 & $-$30:22:26.3 & $8.69_{-0.15}^{+0.42}$ & $29.4$ & $29.7$ & $-18.8$ \\ \hline 
\multicolumn{7}{c}{$L/L^\ast_{z=3} = 0.048$--$0.12$} \\  
GL-z9-11 & 00:14:02.49 & $-$30:22:00.9 & $9.89_{-0.74}^{+0.21}$ & $29.2$ & $30.3$ & $-18.6$ \\ 
GL-z9-7 & 00:14:02.52 & $-$30:21:57.0 & $10.32_{-0.82}^{+0.74}$ & $29.8$ & $29.3$ & $-18.2$ \\ 
GL-z9-10 & 00:14:03.47 & $-$30:19:00.9 & $8.73_{-0.41}^{+0.68}$ & $29.7$ & $29.8$ & $-18.2$ \\ 
GL-z9-12 & 00:14:06.85 & $-$30:22:02.0 & $9.07_{-0.23}^{+1.02}$ & $29.9$ & $30.3$ & $-18.2$ \\ 
GL-z9-8 & 00:14:00.83 & $-$30:21:29.8 & $9.08_{-0.32}^{+0.94}$ & $29.3$ & $30.4$ & $-18.1$ \\ 
GL-z9-9 & 00:14:03.71 & $-$30:21:03.6 & $9.27_{-0.61}^{+1.28}$ & $29.9$ & $29.9$ & $-18.1$ \\ 
GL-z9-13 & 00:13:57.45 & $-$30:18:00.0 & $8.74_{-0.28}^{+0.57}$ & $29.2$ & $30.1$ & $-18.1$ \\ 
\enddata 
\tablecomments{The values presented in this table have been obtained in \cite{2023ApJS..265....5H}. 
(1) Object ID. 
(2) Right ascension. (3) Declination. 
(4) Photometric Redshift. 
(5) Aperture magnitude in F150W measured in $0\farcs2$ diameter circular aperture. 
(6) Aperture magnitude in F444W measured in $0\farcs2$ diameter circular aperture. 
(7) Total absolute UV magnitude. 
}
\end{deluxetable*} 

\begin{deluxetable*}{ccccccc} 
\tablecolumns{7} 
\tablewidth{0pt} 
\tablecaption{$z \sim 12$ Galaxy Candidates (F150W-dropouts) Used in Our Size Analysis 
\label{tab:z12cand}}
\tablehead{
    \colhead{ID} 
    &  \colhead{R.A.}
    &  \colhead{Decl.}
    &  \colhead{$z_{\rm photo}$}
    &  \colhead{$z_{\rm spec}$}
    &  \colhead{$m_{\rm UV}^{\rm (ap)}$}
    &  \colhead{$M_{\rm UV}$}
\\
    \colhead{ } 
    &  \colhead{(h:m:s)}
    &  \colhead{(d:m:s)}
    &  \colhead{ }
    &  \colhead{ }
    &  \colhead{(mag)}
    &  \colhead{(mag)}
\\
    \colhead{(1)} 
    &  \colhead{(2)}
    &  \colhead{(3)}
    &  \colhead{(4)}
    &  \colhead{(5)}
    &  \colhead{(6)}
    &  \colhead{(7)}
}
\startdata 
\multicolumn{6}{c}{$L/L^\ast_{z=3} = 0.3$--$1$} \\  
GL-z12-1 & 00:13:59.74 & $-$30:19:29.1 & $12.28_{-0.07}^{+0.08}$ & --- & $27.4$ & $-21.0$ \\ 
S5-z12-1 & 22:36:06.72 & $+$34:00:09.7 & $12.58_{-0.46}^{+1.23}$ & --- & $28.3$ & $-20.2$ \\ 
CR2-z12-1 & 14:19:46.36 & $+$52:56:32.8 & $11.63_{-0.53}^{+0.51}$ & $11.44^{+0.09}_{-0.08}$$^{\dagger 1}$ & $28.5$ & $-19.9$ \\ \hline 
\multicolumn{6}{c}{$L/L^\ast_{z=3} = 0.12$--$0.3$} \\  
CR2-z12-3 & 14:19:41.61 & $+$52:55:07.6 & $11.66_{-0.71}^{+0.69}$ & --- & $28.6$ & $-19.2$ \\ 
CR2-z12-2 & 14:19:42.57 & $+$52:54:42.0 & $11.96_{-0.87}^{+1.44}$ & --- & $28.9$ & $-19.0$ \\ 
CR2-z12-4 & 14:19:24.86 & $+$52:53:13.9 & $12.08_{-1.25}^{+2.11}$ & --- & $29.4$ & $-19.0$ \\ \hline 
\multicolumn{6}{c}{$L/L^\ast_{z=3} = 0.048$--$0.12$} \\  
SM-z12-1 & 07:22:32.59 & $-$73:28:33.3 & $12.47_{+0.72}^{+1.19}$ & --- & $29.0$ & $-18.5$ \\ 
CR3-z12-1 & 14:19:11.11 & $+$52:49:33.6 & $11.05_{-0.47}^{+2.24}$ & --- & $29.5$ & $-18.4$ \\ 
\enddata 
\tablecomments{The values presented in this table have been obtained in \cite{2023ApJS..265....5H}. 
(1) Object ID. 
(2) Right ascension. (3) Declination. 
(4) Photometric Redshift. 
(5) Spectroscopic Redshift. 
(6) Aperture magnitude in F200W measured in $0\farcs2$ diameter circular aperture. 
(7) Total absolute UV magnitude. 
}
\tablenotetext{$^{\dagger 1}$}{
Obtained by \cite{2023arXiv230315431A}.}
\end{deluxetable*} 

\begin{deluxetable*}{cccccc} 
\tablecolumns{6} 
\tablewidth{0pt} 
\tablecaption{$z \sim 16$ Galaxy Candidate (F200W-dropout) Used in Our Size Analysis 
\label{tab:z17cand}}
\tablehead{
    \colhead{ID} 
    &  \colhead{R.A.}
    &  \colhead{Decl.}
    &  \colhead{$z_{\rm photo}$}
    &  \colhead{$m_{\rm UV}^{\rm (ap)}$}
    &  \colhead{$M_{\rm UV}$}
\\
    \colhead{ } 
    &  \colhead{(h:m:s)}
    &  \colhead{(d:m:s)}
    &  \colhead{ }
    &  \colhead{(mag)}
    &  \colhead{(mag)}
\\
    \colhead{(1)} 
    &  \colhead{(2)}
    &  \colhead{(3)}
    &  \colhead{(4)}
    &  \colhead{(5)}
    &  \colhead{(6)}
}
\startdata 
\multicolumn{6}{c}{Stephan's Quintet} \\ 
S5-z16-1 & 22:36:03.81 & $+$33:54:16.7 & $16.41_{-0.55}^{+0.66}$ & $27.6$ & $-21.6$
\enddata 
\tablecomments{The values presented in this table have been obtained in \cite{2023ApJS..265....5H}. 
(1) Object ID. 
(2) Right ascension. (3) Declination. 
(4) Photometric Redshift. 
(5) Aperture magnitude in F277W measured in $0\farcs2$ diameter circular aperture. 
(6) Total absolute UV magnitude. 
}
\end{deluxetable*} 

In this paper, we investigate sizes of galaxies at $z\sim9$--$16$ 
with deep JWST NIRCam (\citealt{2005SPIE.5904....1R}) images 
based on the high-$z$ galaxy candidate samples 
constructed by \cite{2023ApJS..265....5H}, 
who have selected F115W-, F150W-, and F200W-dropouts 
from the JWST deep imaging data publicly available so far 
taken by four 
Early Release Science (ERS) 
and 
Early Release Observation (ERO; \citealt{2022ApJ...936L..14P}) 
programs: 
ERS Grism Lens-Amplified Survey from Space (GLASS; \citealt{2022ApJ...935..110T}), 
ERS The Cosmic Evolution Early Release Science (CEERS; \citealt{2022ApJ...940L..55F}), 
ERO SMACS J0723, 
and 
ERO Stephan's Quintet.

This paper is outlined as follows. 
In Section \ref{sec:data}, we introduce the high-$z$ galaxy samples
and summarize the NIRCam imaging data used in this study. 
In Section \ref{sec:measurements},  
the surface brightness profile fitting is described 
and Monte Carlo simulations are conducted to take into account 
the systematic and statistical uncertainties in the profile fitting.  
We present size and total magnitude measurement results  
in the rest-frame UV and optical 
with object-by-object comparisons with previous results, 
and then investigate the size--luminosity relation and the size evolution 
in Section \ref{sec:results}.   
We discuss the physical origin of very compact galaxy candidates 
and compare our results with cosmological simulation results 
in Section \ref{sec:discussion}. 
We summarize our work in Section \ref{sec:summary}. 
Throughout this paper, 
we use magnitudes in the AB system (\citealt{1983ApJ...266..713O})
and assume a flat universe with 
$\Omega_{\rm m} = 0.3$, $\Omega_\Lambda = 0.7$, and $H_0 = 70$ km s$^{-1}$ Mpc$^{-1}$. 
In this cosmological model, 
an angular dimension of $1.0$ arcsec corresponds to a physical dimension of 
$4.463$ kpc at $z=9.0$, 
$3.659$ kpc at $z=12.0$, and  
$2.953$ kpc at $z=16.0$ 
(e.g., Equation 18 of \citealt{1999astro.ph..5116H}).
Following the previous work, 
we express galaxy UV luminosities in units of the characteristic luminosity 
of $z\sim 3$ galaxies, $L^\ast_{z=3}$, 
which corresponds to $M_{\rm UV} = -21.0$ mag 
(\citealt{1999ApJ...519....1S}).\footnote{In this case, 
$0.048 L^\ast_{z=3}$, $0.12 L^\ast_{z=3}$ and $0.3 L^\ast_{z=3}$ 
correspond to 
$M_{\rm UV} = -17.7$, $-18.7$ and $-19.7$ mag, respectively. 
}


\begin{figure*}[ht]
\begin{center}
   \includegraphics[height=24em]{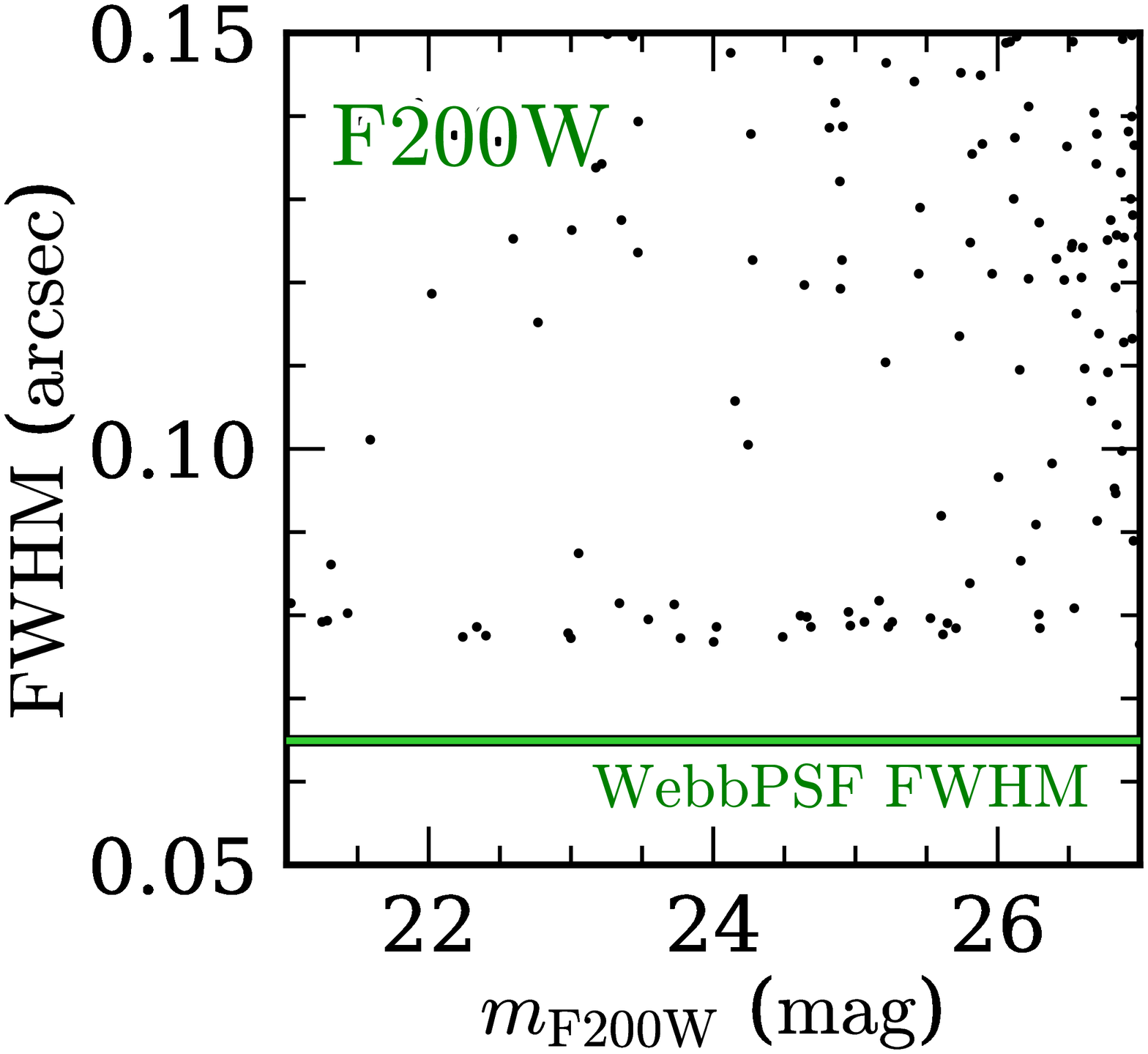}
   \includegraphics[height=24em]{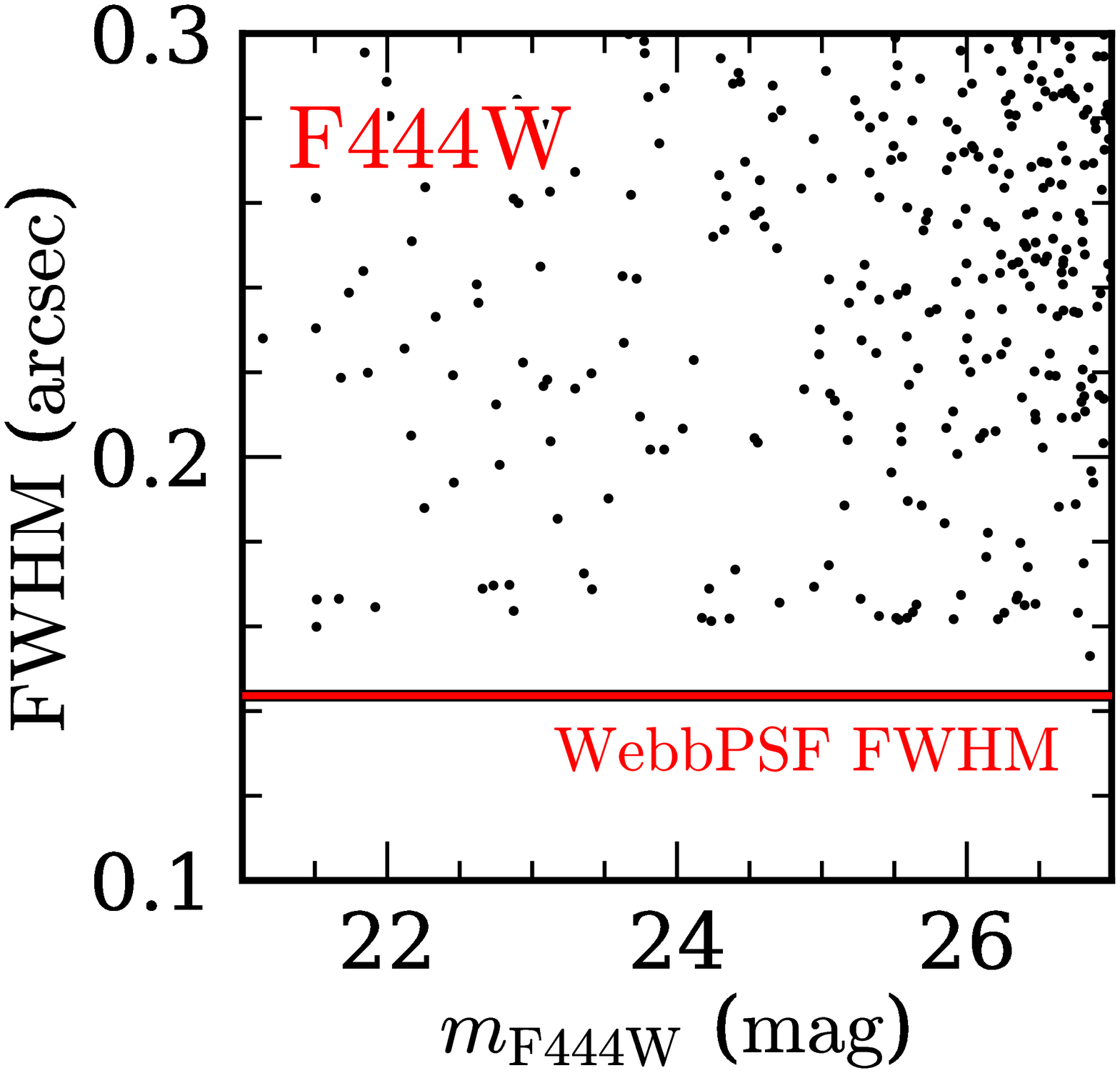}
\caption{
\textbf{Left}:
FWHM vs. $0\farcs2$ diameter aperture magnitude 
for bright objects with relatively small FWHMs 
detected in the F200W image for the GLASS field 
(black dots). 
The solid horizontal green line represents  
the FWHM of PSF created with WebbPSF. 
Note that the PSF created with WebbPSF is not drizzled. 
\textbf{Right}:
Same as the left panel but for F444W.
}
\label{fig:WebbPSF_FWHM}
\end{center}
\end{figure*}

\begin{figure*}[ht]
\begin{center}
   \includegraphics[height=17em]{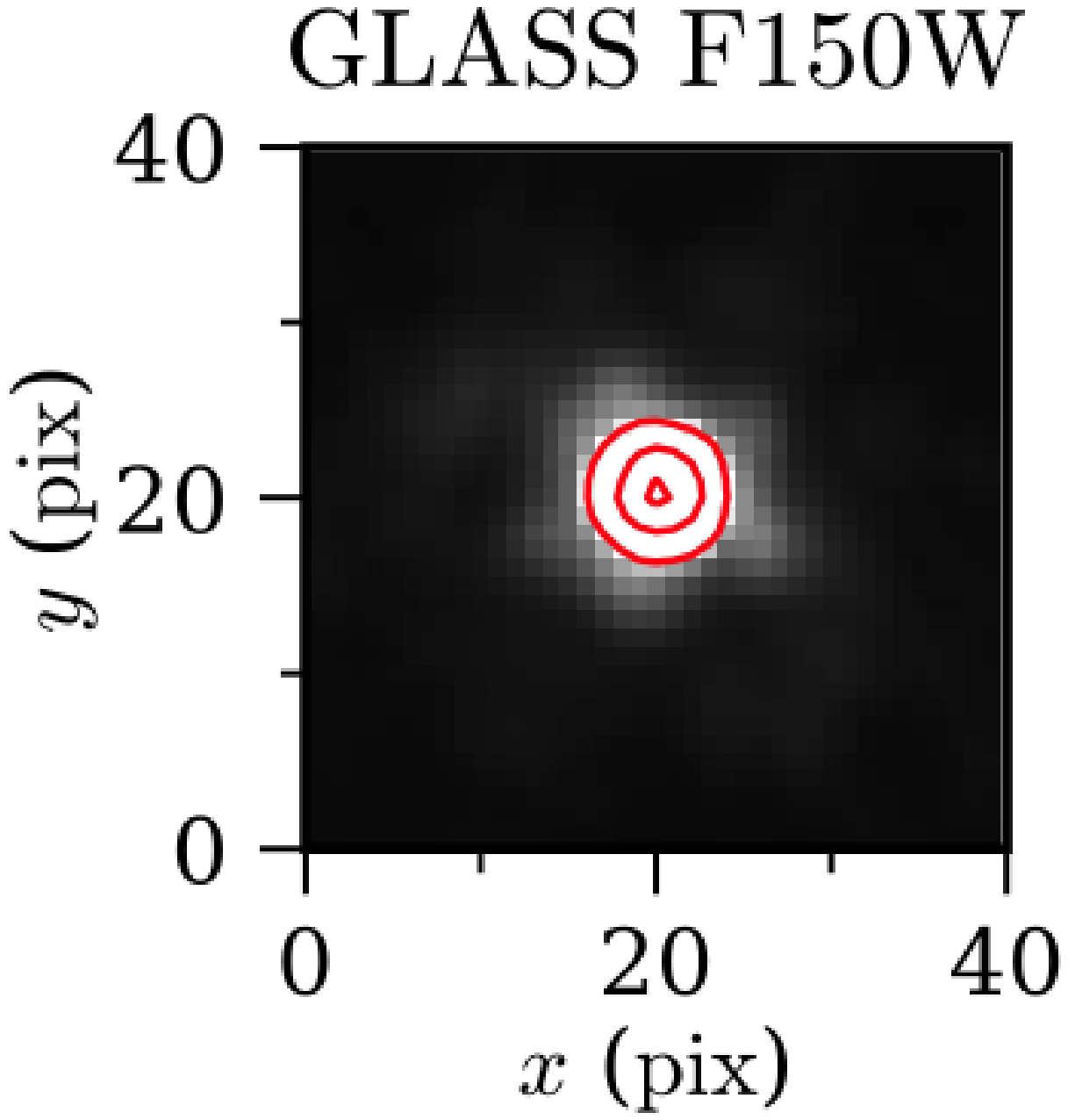}
   \includegraphics[height=17em]{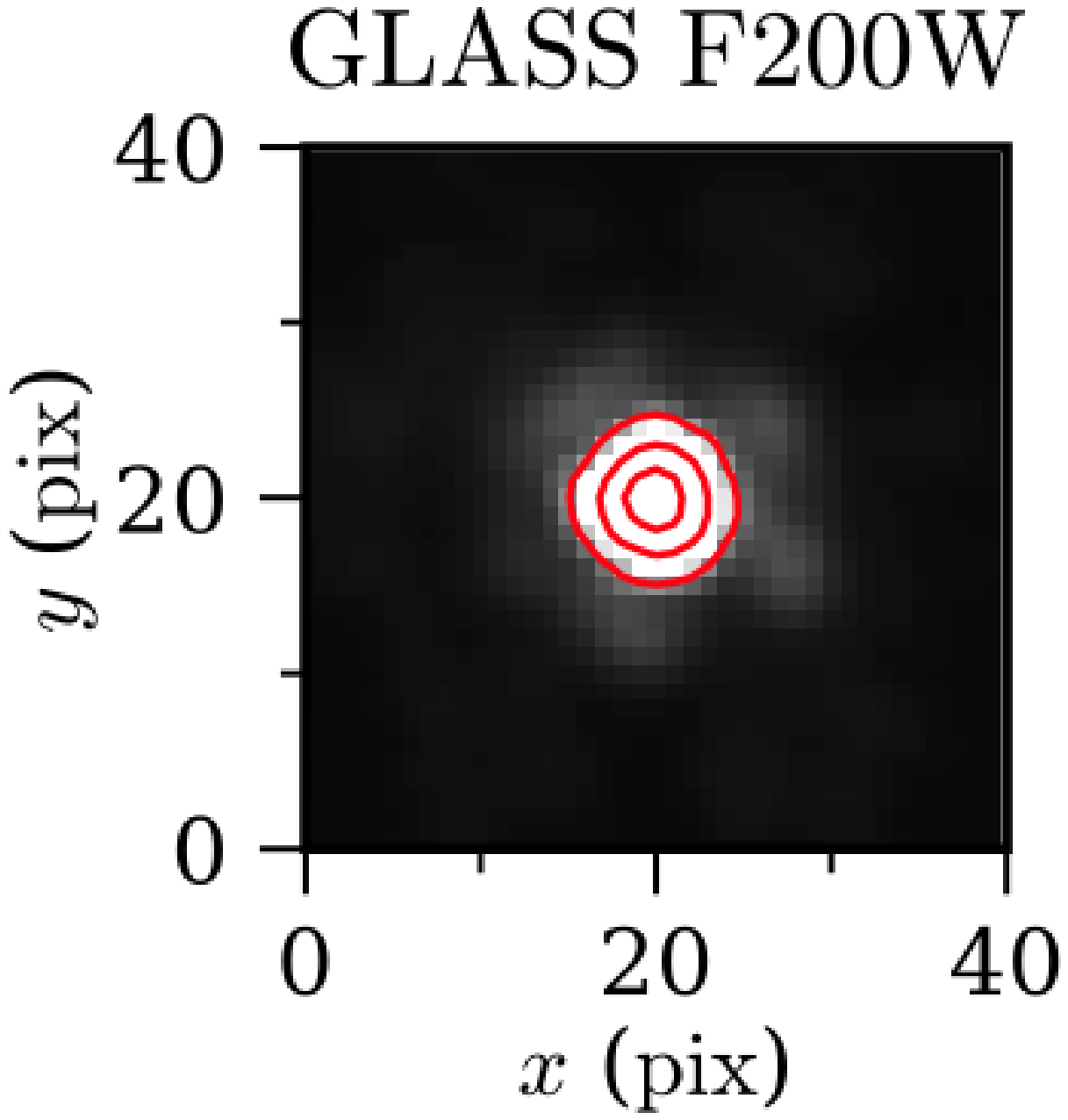}
   \includegraphics[height=17em]{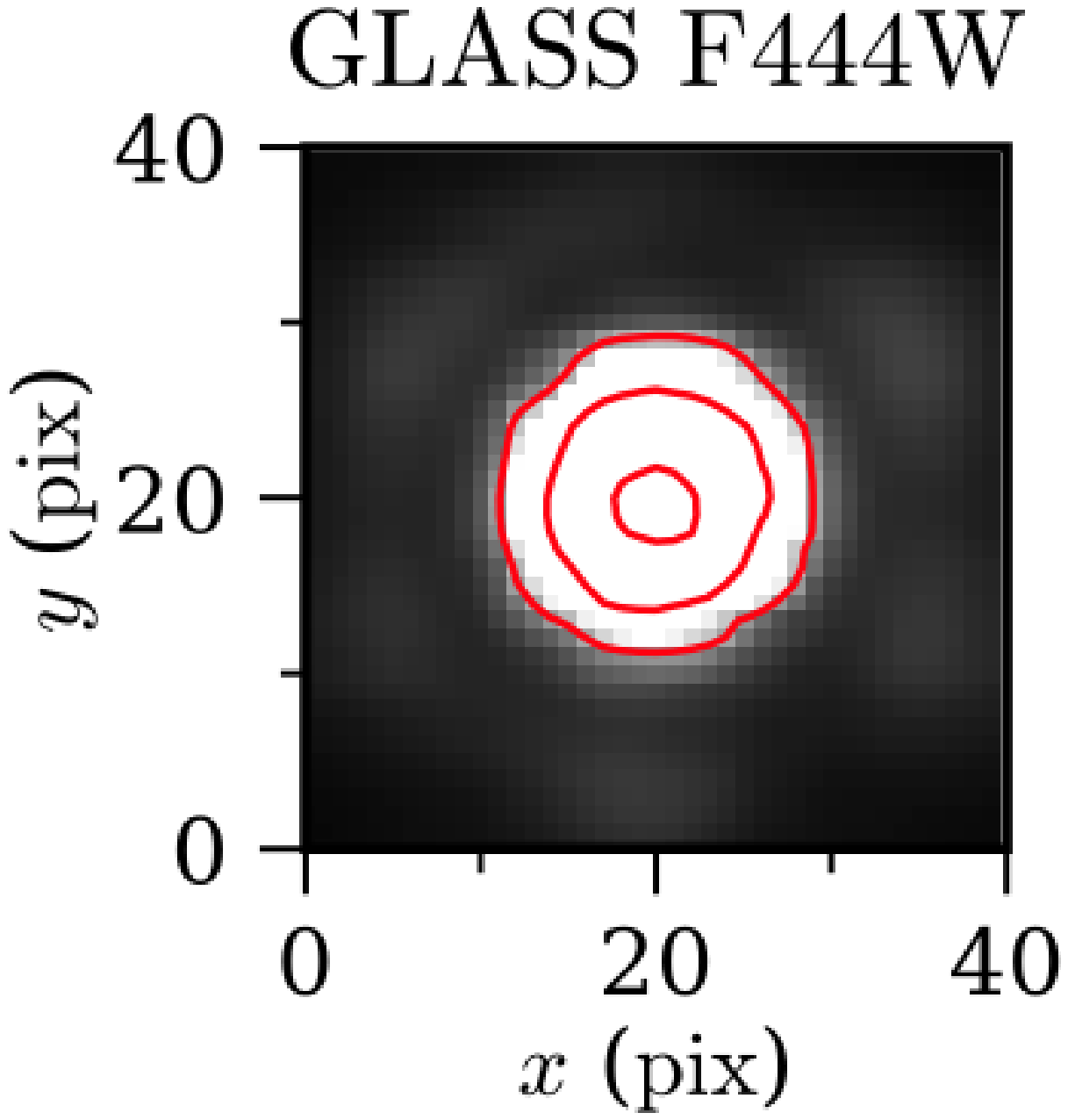}
\caption{
Example PSF images in F150W, F200W, and F444W for the GLASS field. 
The red contours correspond to $300$, $1000$, and $2000 \sigma$ for F150W and F200W, 
and  $100$, $300$, and $700 \sigma$ for F444W. 
The FWHMs of these PSFs are summarized in Table \ref{tab:limitmag}. 
}
\label{fig:psf}
\end{center}
\end{figure*}

\begin{deluxetable*}{ccccc} 
\tablecolumns{5} 
\tablewidth{0pt} 
\tablecaption{Limiting Magnitudes and PSF FWHMs of the JWST NIRCam Images for Size Analysis 
\label{tab:limitmag}}
\tablehead{
\multicolumn{5}{c}{$5\sigma$ Depth / $10 \sigma$ Depth / PSF FWHM} \\ 
    \colhead{Field} 
    &  \colhead{F150W}
    &  \colhead{F200W}
    &  \colhead{F277W}
    &  \colhead{F444W}
}
\startdata 
GLASS 			& $29.4$ / $28.6$ / $0\farcs0704$ 	& $29.6$ / $28.8$ / $0\farcs0776$ 	& --- 							& $29.6$ / $28.8$ / $0\farcs1605$ \\
CEERS 			& --- 							& $29.7$ / $28.9$ / $0\farcs0795$ 	& --- 	& --- \\
SMACS J0723 		& ---  						& $29.6$ / $28.8$ / $0\farcs0765$ 	& --- 							& ---  \\
Stephan's Quintet 	& --- 							& $28.1$ / $27.3$ / $0\farcs0771$	& $28.8$ / $28.0$ / $0\farcs1197$ 	& ---
\enddata 
\tablecomments{
Limiting magnitudes are measured with randomly distributed $0\farcs2$ diameter circular apertures 
(\citealt{2023ApJS..265....5H}).  
}
\end{deluxetable*} 




\section{Data and Samples} \label{sec:data}

We measure sizes of $z \sim 9$--$16$ galaxy candidates 
selected in \cite{2023ApJS..265....5H} 
from four JWST NIRCam early deep public datasets by the ERS and ERO programs: 
ERS GLASS, 
ERS CEERS, 
ERO SMACS J0723, 
and 
ERO Stephan's Quintet. 
See \cite{2023ApJS..265....5H} 
for details of the imaging datasets and the sample construction. 
Briefly, $z \sim 9$ galaxy candidates (F115W-dropouts) are selected by 
\begin{gather}
(F115W-F150W > 1.0) \land, \\
(F150W-F277W < 1.0) \land, \\
(F115W-F150W > (F150W-F277W) +1.0), 
\end{gather}
$z \sim 12$ galaxy candidates (F150W-dropouts) are selected by 
\begin{gather}
(F150W-F200W > 1.0) \land, \\
(F200W-F356W < 1.0) \land, \\
(F150W-F200W > (F200W-F356W) +1.0), 
\end{gather}
and $z \sim 16$ galaxy candidates (F200W-dropouts) are selected by 
\begin{gather}
(F200W-F277W > 1.0), \\
(F277W-F444W < 1.0), \\
(F200W-F277W > 1.5(F277W-F444W) +1.0). 
\end{gather}
These colors are measured with fixed apertures in 
point spread function (PSF) matched 
multi-band images. 
Null detection criteria in bluer bands than the Lyman break, 
and a SExtractor (\citealt{1996A&AS..117..393B}) stellarity parameter criterion, 
CLASS\_STAR $<0.9$, are also applied. 
In addition, based on the results with the CEERS simulated images 
(\citealt{2021MNRAS.502.4858S}; \citealt{2022MNRAS.515.5416Y}), 
it is required that 
the difference between $\chi^2$ values of high-$z$ and low-$z$ solutions for our candidates 
calculated by a photometric redshift code exceed 9, i.e., $\Delta \chi^2 > 9$, 
which is more strict than a frequently adopted criterion of $\Delta \chi^2 > 4$, 
to better exclude low-$z$ interlopers 
(For details, see Section 3.3 of \citealt{2023ApJS..265....5H}). 
As a result, reliable samples of high-$z$ galaxy candidates are constructed; 
the numbers of F115W-dropouts, F150W-dropouts, and F200W-dropouts are 
$13$, $8$, and $1$, 
respectively.\footnote{CR2-z16-1 
(CEERS-93316, \citealt{2023MNRAS.518.6011D}; see also, \citealt{2022arXiv220802794N})
is excluded from our analyses 
because it has been spectroscopically identified as a strong line emitter at $z=4.912$
(\citealt{2023arXiv230315431A}).} 
Among these dropouts, 
CR2-z12-1 (Maisie's Galaxy, \citealt{2022ApJ...940L..55F}) has been spectroscopically identified at $z=11.44$  
with JWST/NIRSpec (\citealt{2023arXiv230315431A}), 
whose spectroscopic redshift is consistent with the previous photometric redshift estimates.
Table \ref{tab:z9cand}, Table \ref{tab:z12cand}, and Table \ref{tab:z17cand} 
summarize $z\sim 9$, $z\sim12$, and $z\sim16$ galaxy candidates used in our size analyses, respectively.

To minimize the effects of morphological $K$-correction 
in comparison with the previous work, 
we measure galaxy sizes with NIRCam images 
that are closest to the rest-frame $1600$--$1700${\AA}. 
Specifically, we use F150W images for $z\sim9$ galaxy candidates, 
F200W images for $z\sim12$ galaxy candidates, 
and F277W images for $z\sim16$ galaxy candidates 
(Figure \ref{fig:SED}).  
In addition, we use F444W images to measure the rest-frame optical ($4000$--$5000${\AA}) sizes 
for $z\sim9$ galaxy candidates. 
The pixel scale of the NIRCam images is $0.015$ arcsec pix$^{-1}$, 
except for the ERO Stephan's Quintet field, 
where the pixel scale is $0.030$ arcsec pix$^{-1}$ to reduce the image size. 
Their $5\sigma$ and $10\sigma$ limiting magnitudes are summarized in Table \ref{tab:limitmag}.

We use empirical PSFs created by stacking bright point sources in the NIRCam real images.  
For this purpose, we select $7$--$15$ unsaturated bright point sources 
with $\simeq 22$--$24.5$ mag in each field. 
PSFs can also be generated by using WebbPSF (\citealt{2012SPIE.8442E..3DP}; \citealt{2014SPIE.9143E..3XP}). 
We create PSFs with WebbPSF by using a jitter value of $4.1$ milli-arcsec (mas)
that corresponds to the square root of the sum of the squares 
of the pointing stability ($1$ mas) and dithering accuracy ($4$ mas at most), 
as described in Sections 3.3 and 3.4 of \cite{2022arXiv220705632R}.
However, as shown in Figure \ref{fig:WebbPSF_FWHM},
we find that the full widths at half maximum (FWHMs) of PSFs generated with WebbPSF are smaller than 
the smallest FWHMs of bright point objects detected in the NIRCam real images, 
indicating that the model PSFs are sharper than the actual PSFs.
This is probably just because 
the PSFs created with WebbPSF here are not drizzled, 
although they should be drizzled in the same way as the science data 
(see also, \citealt{2023arXiv230207234T}). 
We thus use the empirically created PSFs in this study. 
Table \ref{tab:limitmag} presents the PSF FWHMs 
and Figure \ref{fig:psf} shows the PSFs in F150W, F200W, and F444W for the GLASS field as examples. 
The variation of point source FWHMs between different fields is not large; 
in F200W, the difference is at most $\simeq 4${\%} 
($=[0.0795-0.0765]/0.0765 \times 100$).

To obtain reliable galaxy size estimates, 
\cite{2012ApJ...756L..12M} have reported that
a signal-to-noise ratio (S/N) of $10$ is needed, 
because surface brightness profile fittings require a significant detection 
in not only the central region of sources, but also the outer structures. 
We thus analyze our dropouts individually down to S/N of $10$ 
for their apparent magnitudes measured in $0\farcs2$ diameter circular apertures. 
Namely, the object IDs that show S/N $>10$ are 
GL-z9-1 in the F115W-dropout sample, 
GL-z12-1, CR2-z12-1, CR2-z12-2, and CR2-z12-3 in the F150W-dropout sample, 
and 
S5-z16-1 in the F200W-dropout sample. 
For the rest-frame optical analyses, 
we individually investigate GL-z9-1, which is the only object in our F115W-dropout sample 
showing an aperture magnitude brighter than $10\sigma$ in F444W.

To extend our size measurements to fainter objects, 
we divide the samples into three luminosity bins, 
$L/L^\ast_{z=3} = 0.3$--$1$, 
$L/L^\ast_{z=3} = 0.12$--$0.3$, 
and 
$L/L^\ast_{z=3} = 0.048$--$0.12$ 
based on their $M_{\rm UV}$ magnitudes 
as listed in Table \ref{tab:z9cand} and Table \ref{tab:z12cand},   
and make median-stacked images separately for the second and third brightest luminosity bins. 
The number of F115W-dropouts (F150W-dropouts) with 
$L/L^\ast_{z=3} = 0.12$--$0.3$ is 5 (3),
and the number with $L/L^\ast_{z=3} = 0.048$--$0.12$ is 7 (2).
We confirm that the S/Ns of the aperture magnitudes of the stacked objects are $\gtrsim 10$. 
The stacked F150W-dropout with $L/L^\ast_{z=3} = 0.048$--$0.12$ is not used in this study, 
because the number of objects in this luminosity bin is only two, 
and one of them (SM-z12-1) is located behind a galaxy cluster region 
where the effect of gravitational lensing needs to be considered.

\section{Surface Brightness Profile Fitting} \label{sec:measurements}

\begin{figure*}[ht]
\begin{center}
   \includegraphics[width=0.24\textwidth]{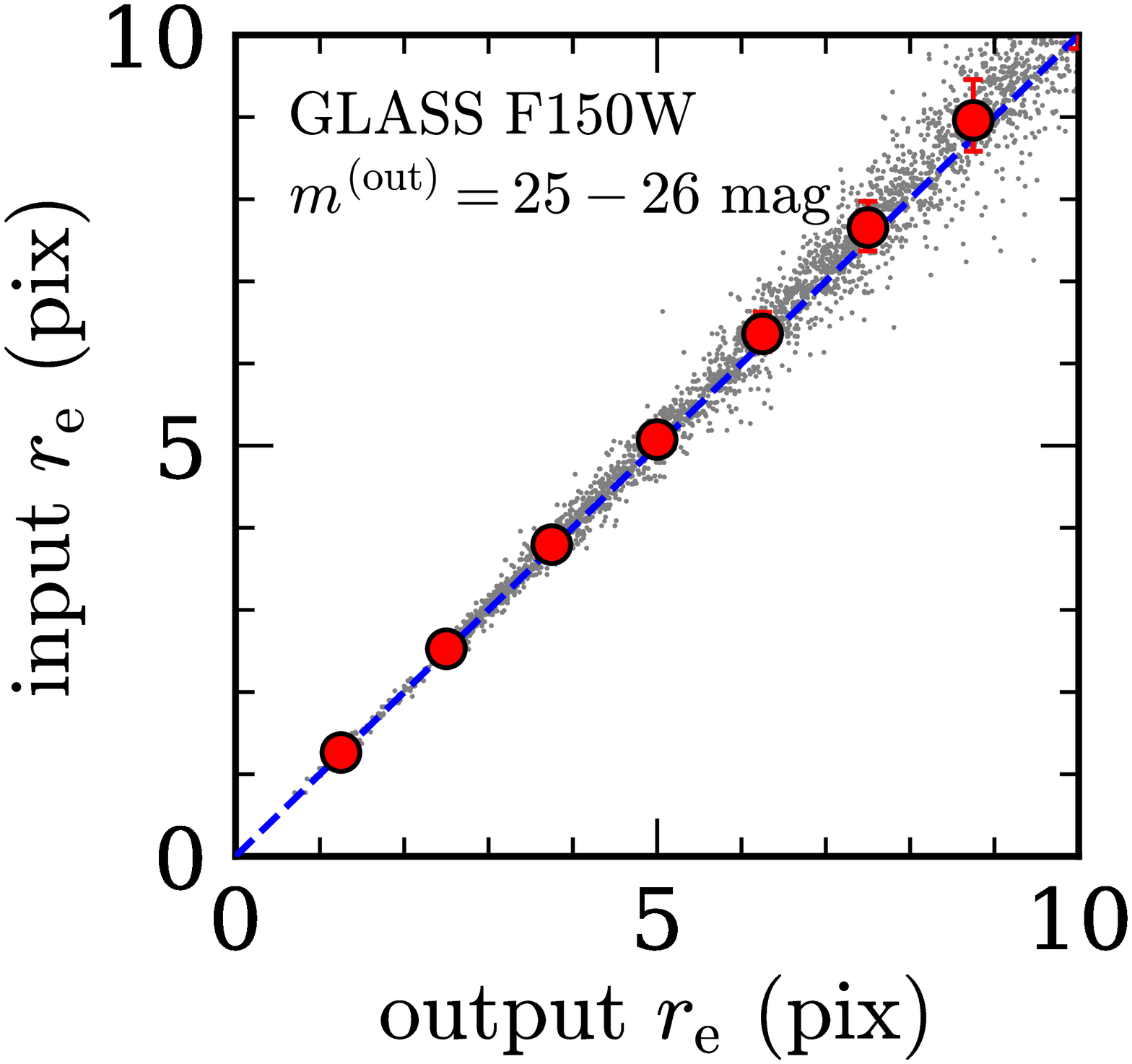}
   \includegraphics[width=0.24\textwidth]{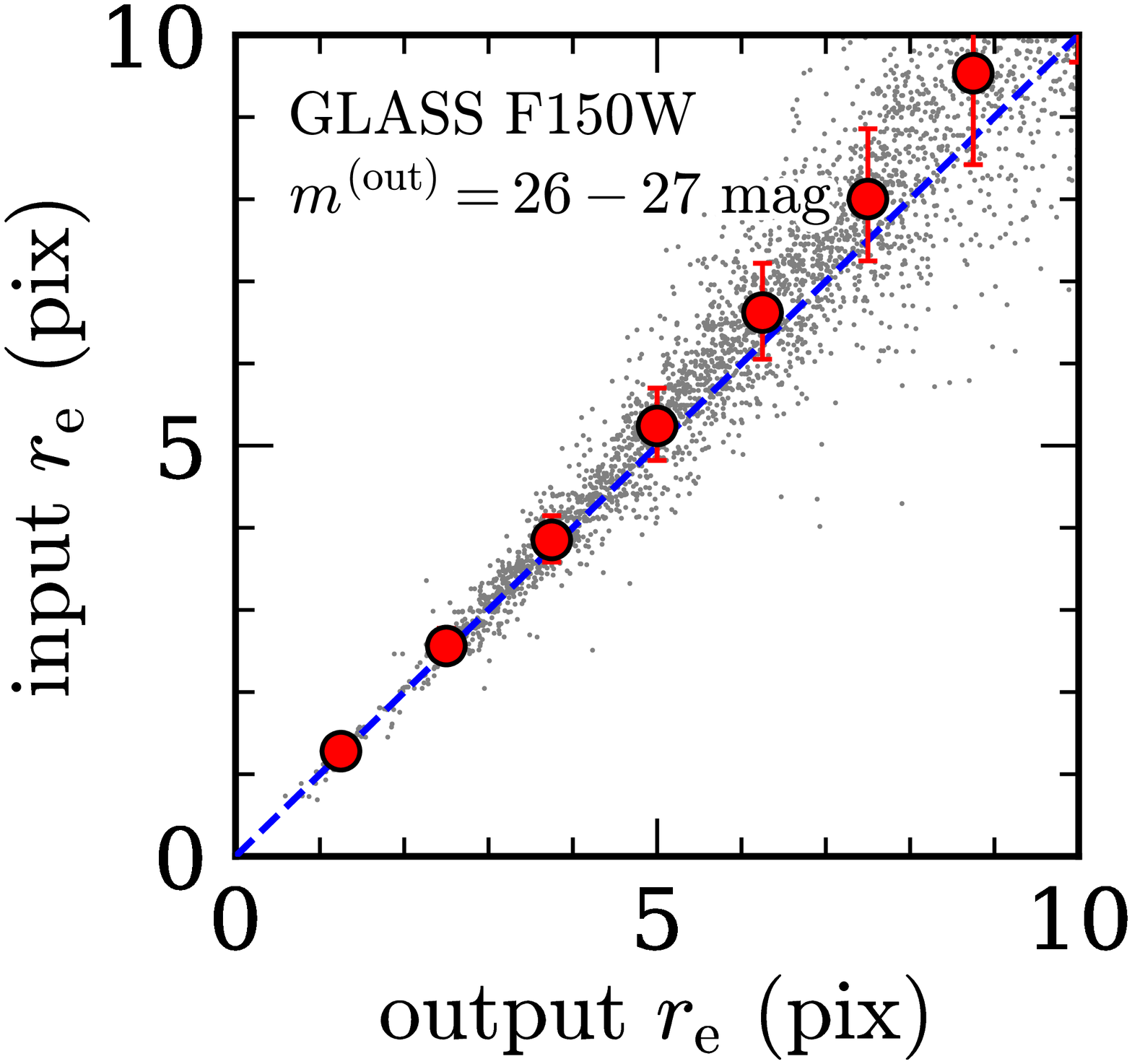}
   \includegraphics[width=0.24\textwidth]{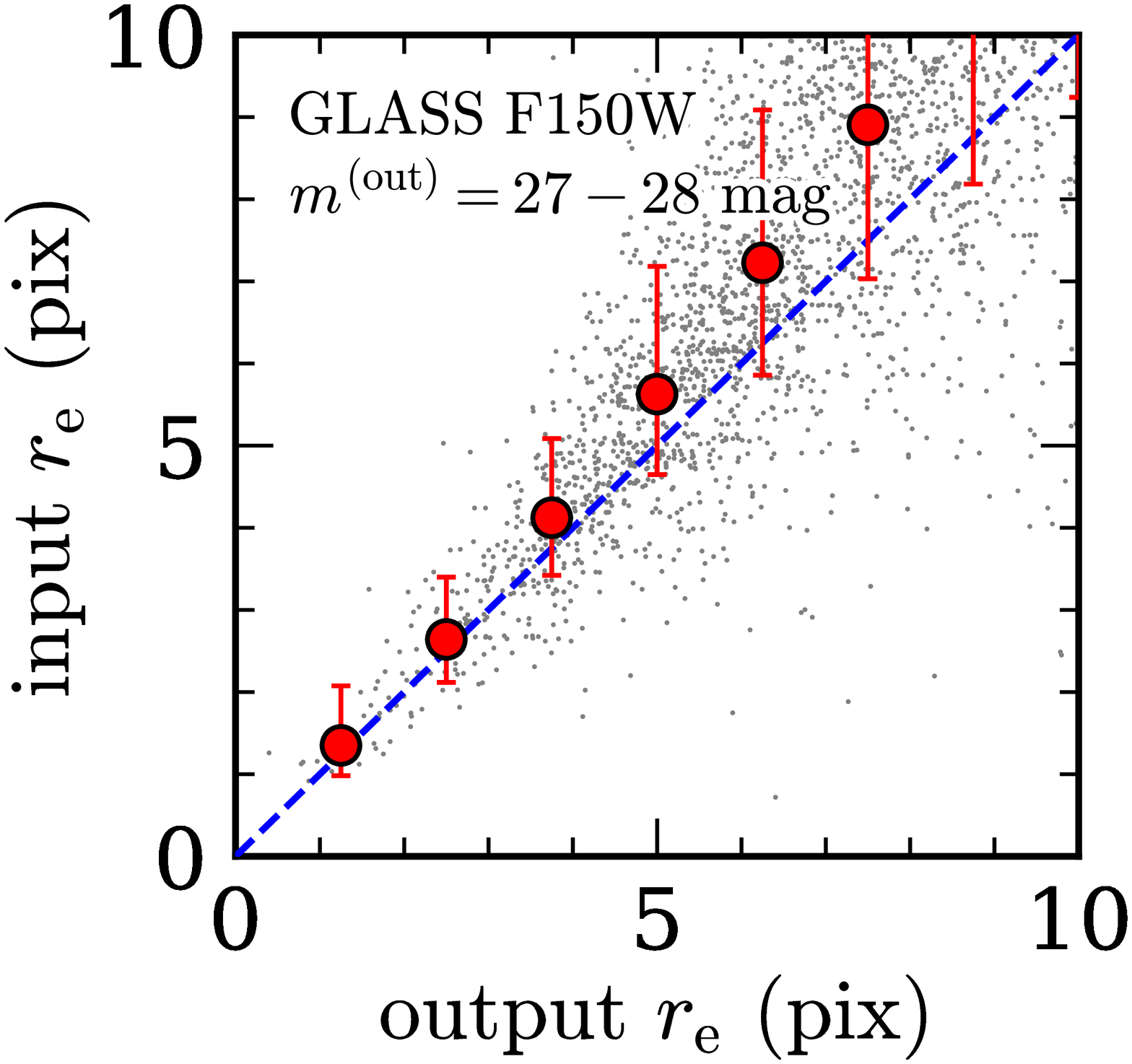}
   \includegraphics[width=0.24\textwidth]{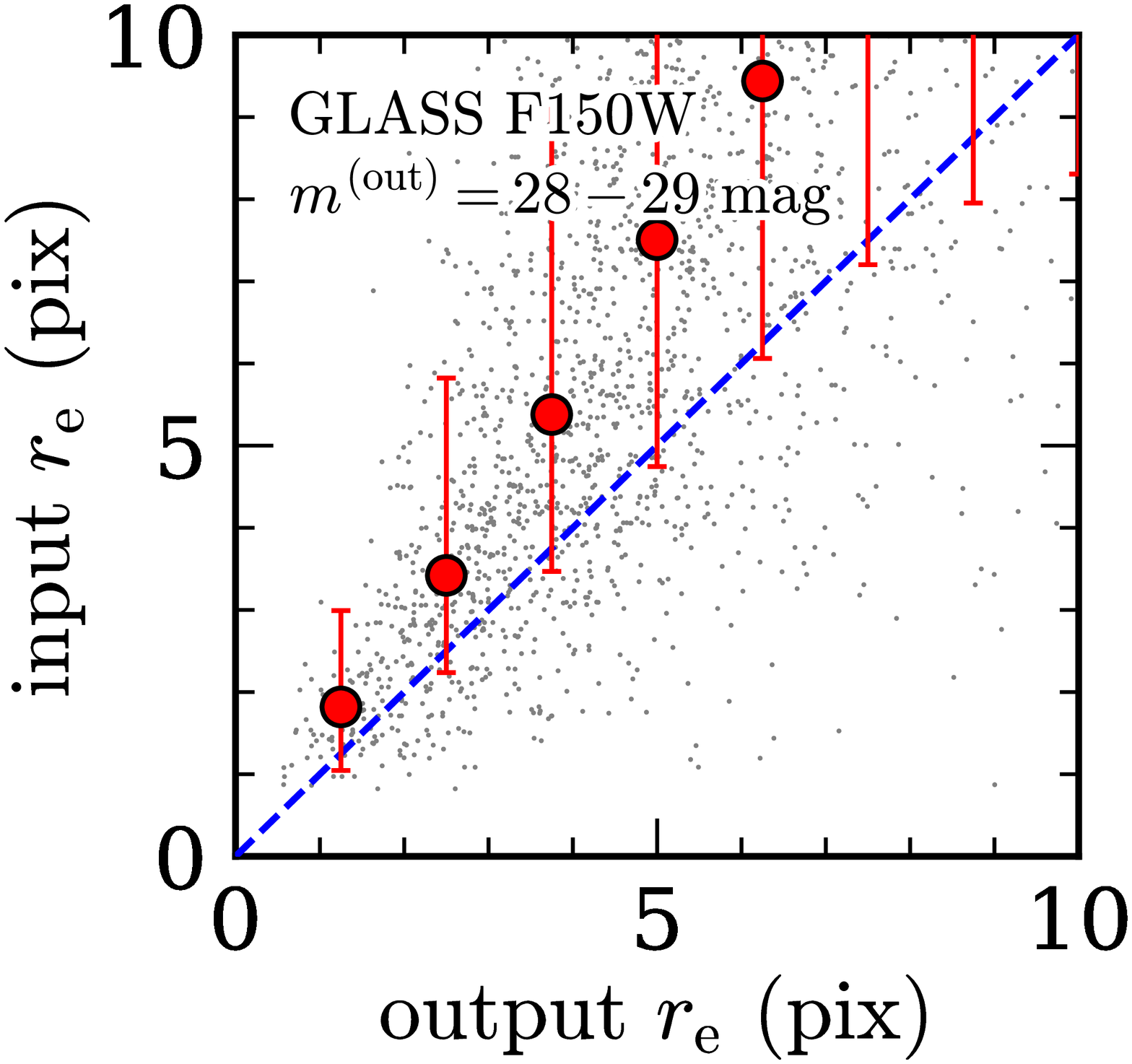}
   \includegraphics[width=0.24\textwidth]{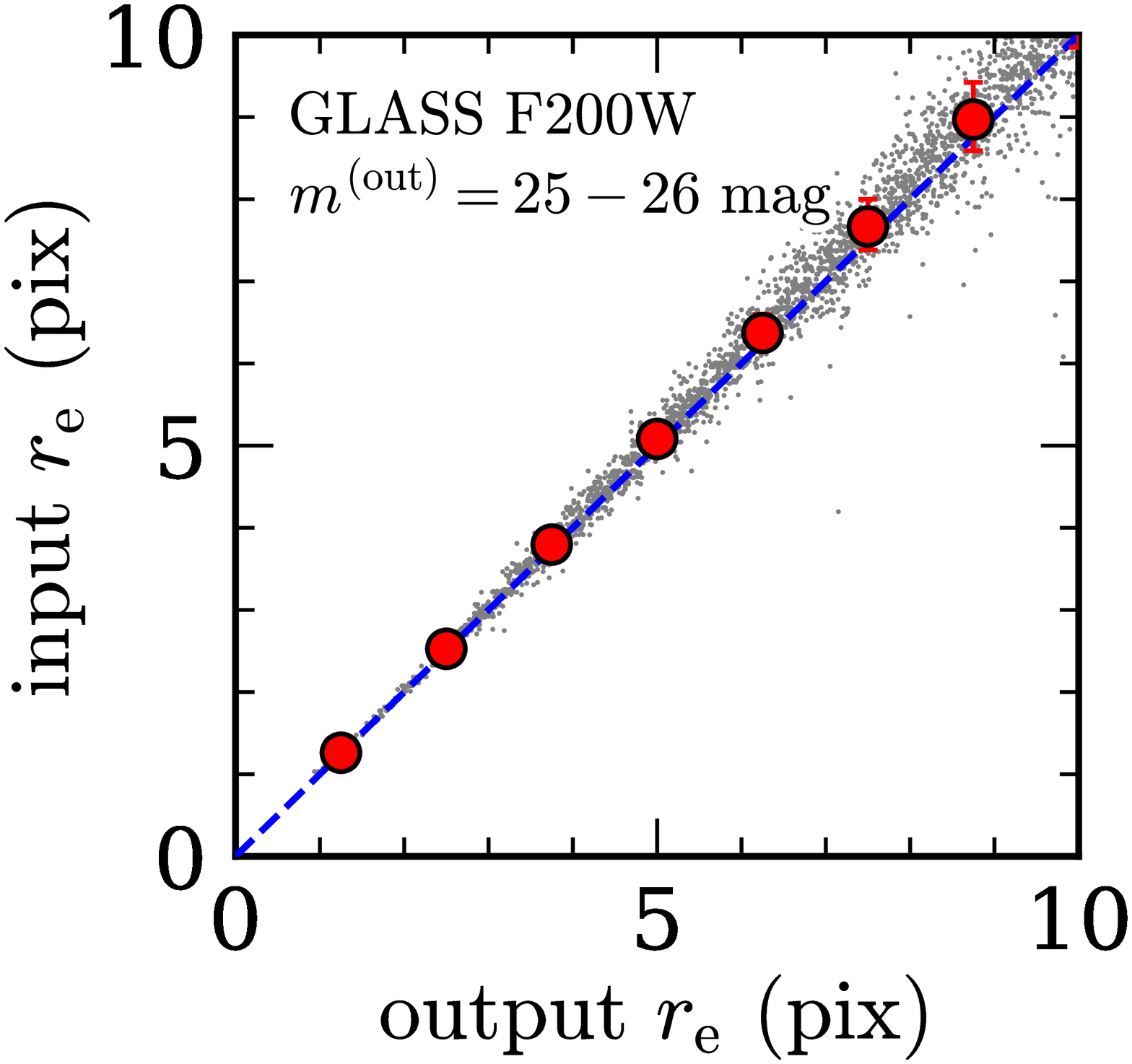}
   \includegraphics[width=0.24\textwidth]{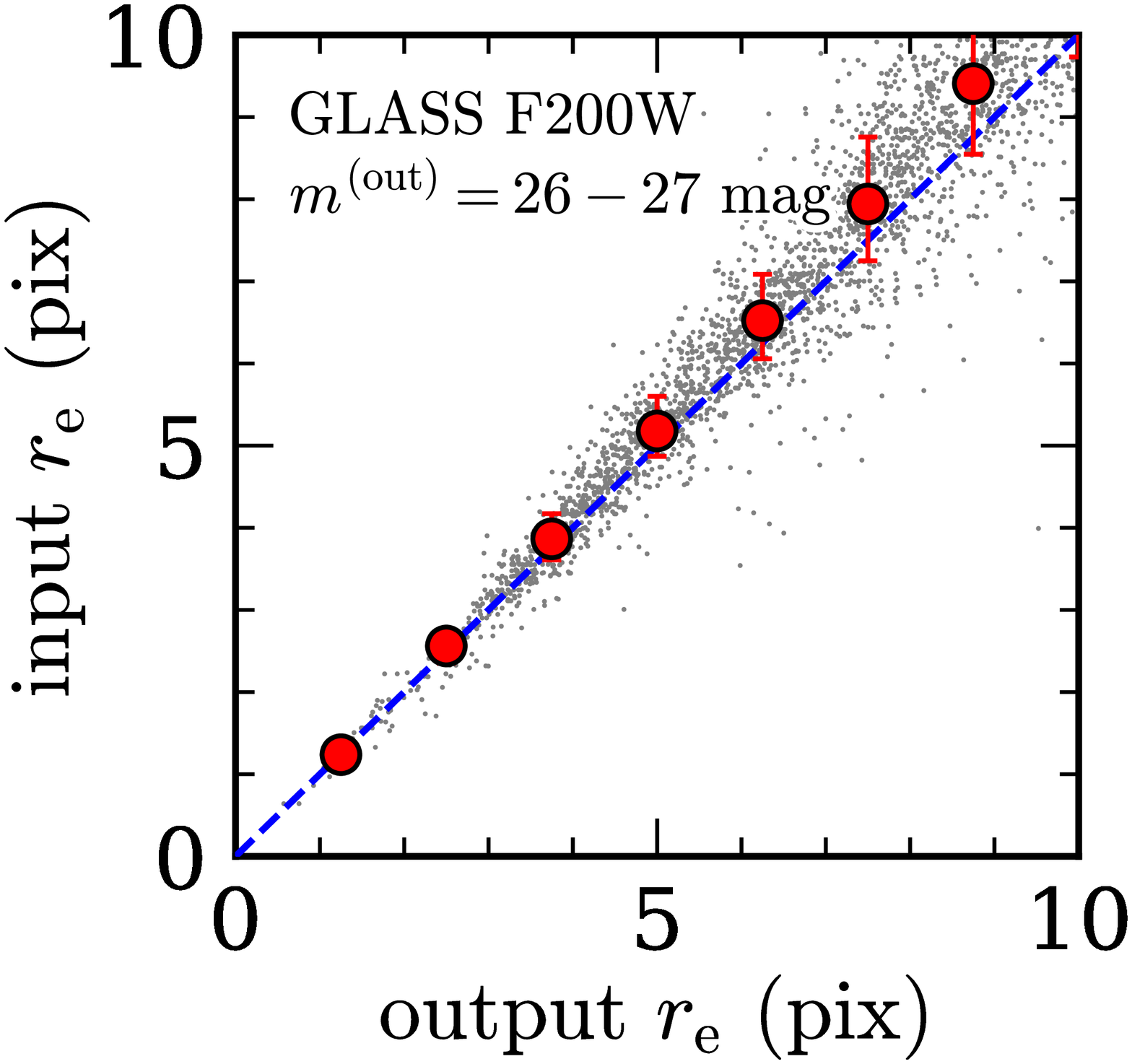}
   \includegraphics[width=0.24\textwidth]{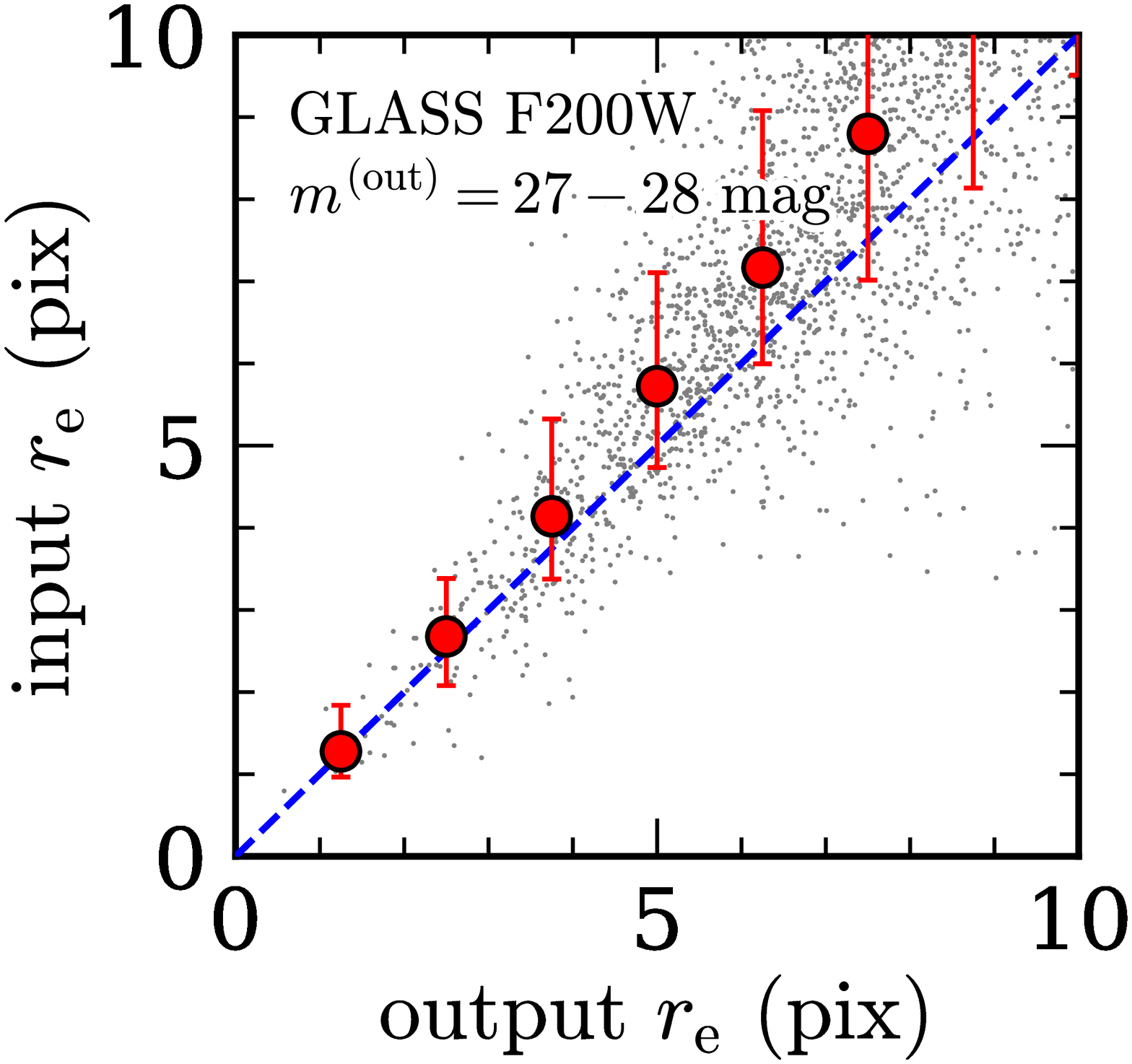}
   \includegraphics[width=0.24\textwidth]{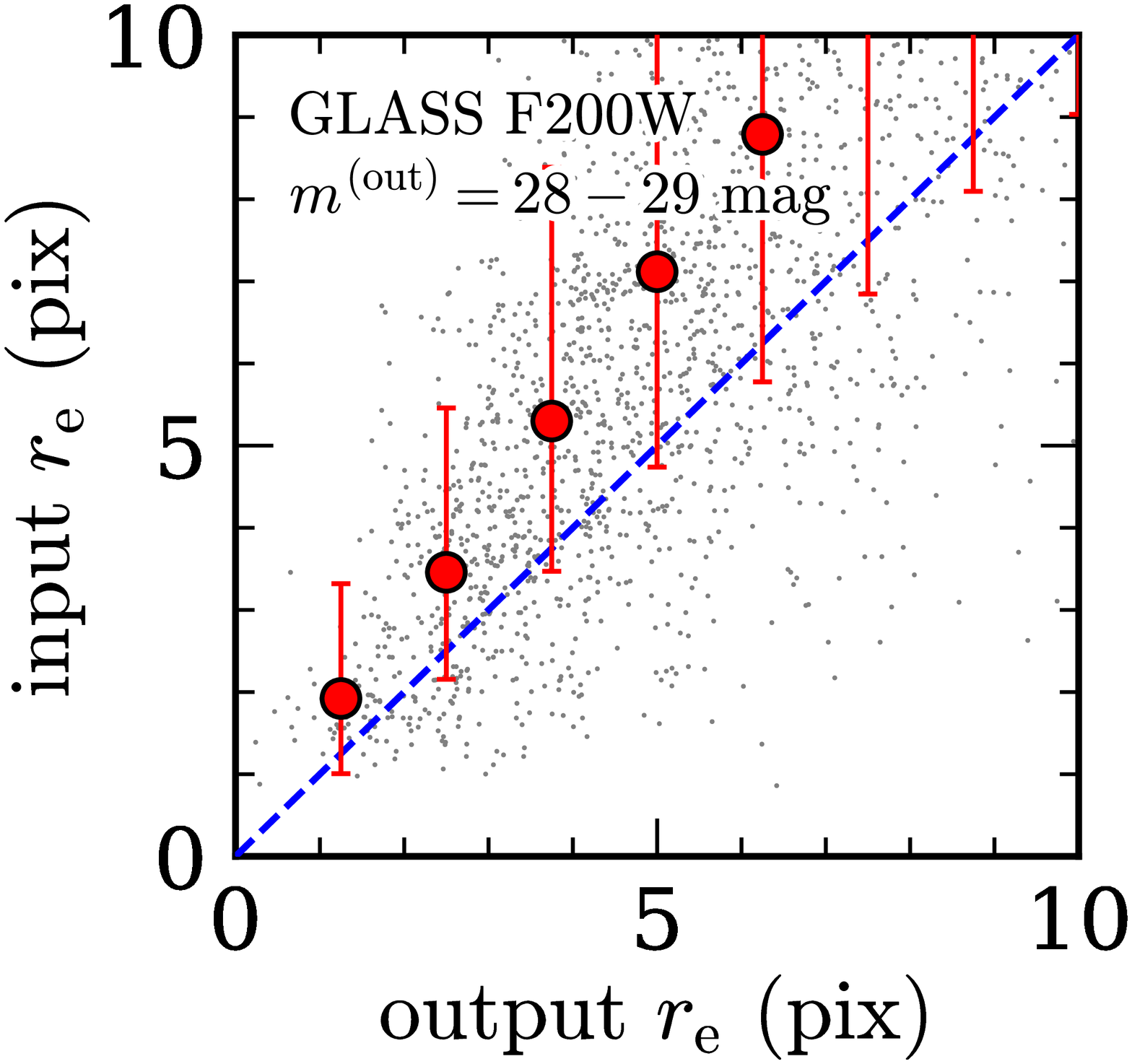}
   \includegraphics[width=0.24\textwidth]{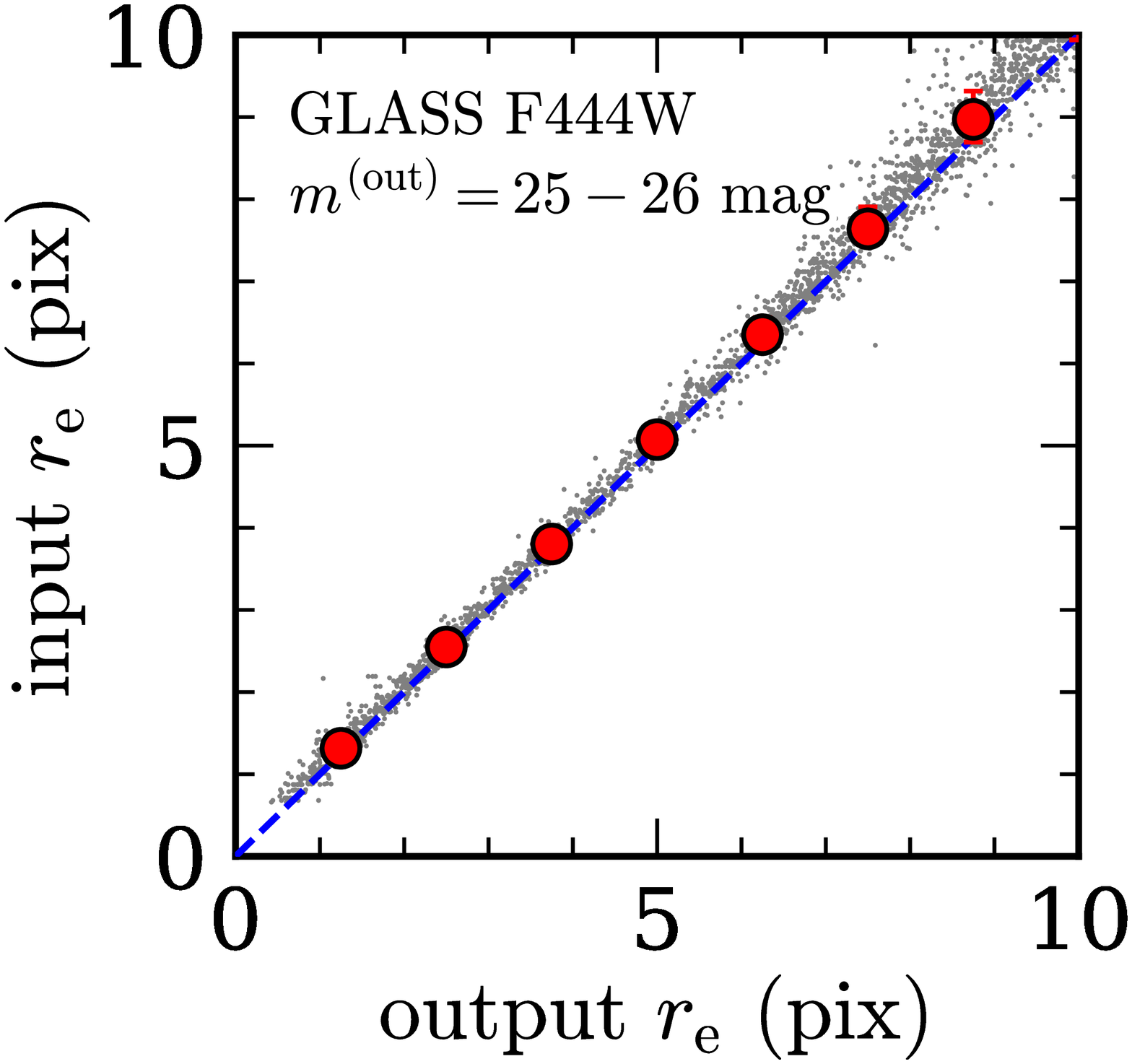}
   \includegraphics[width=0.24\textwidth]{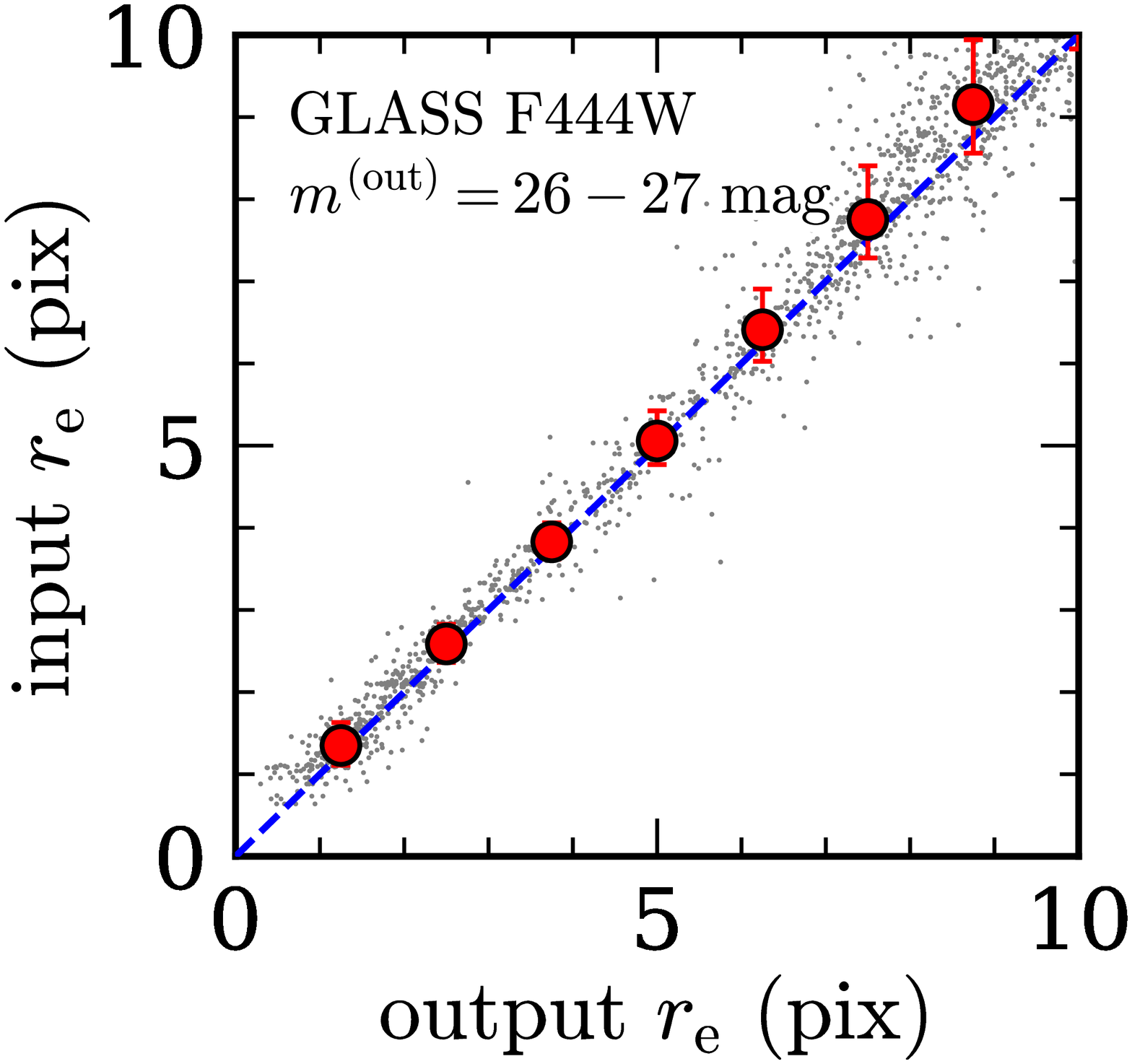}
   \includegraphics[width=0.24\textwidth]{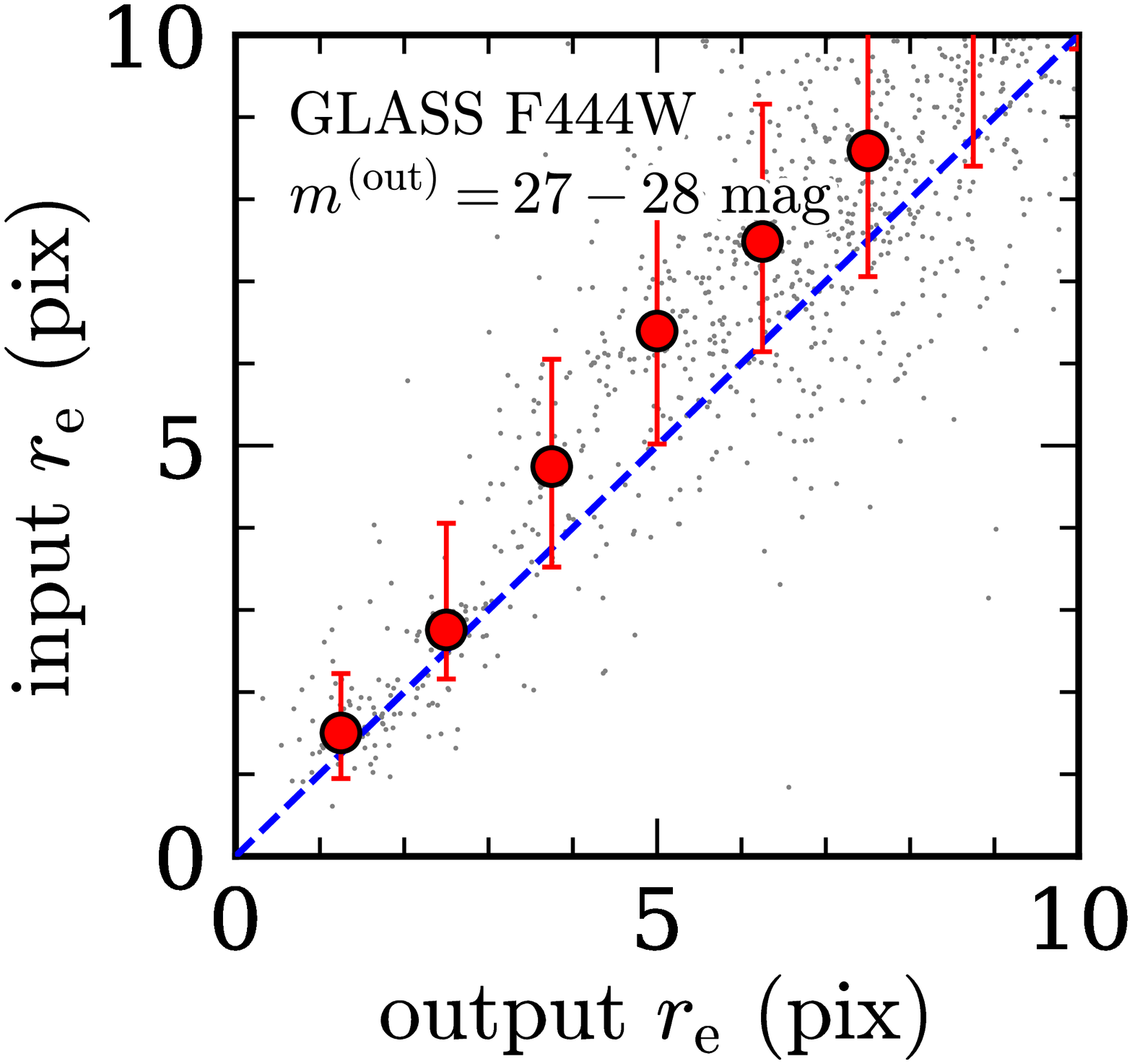}
   \includegraphics[width=0.24\textwidth]{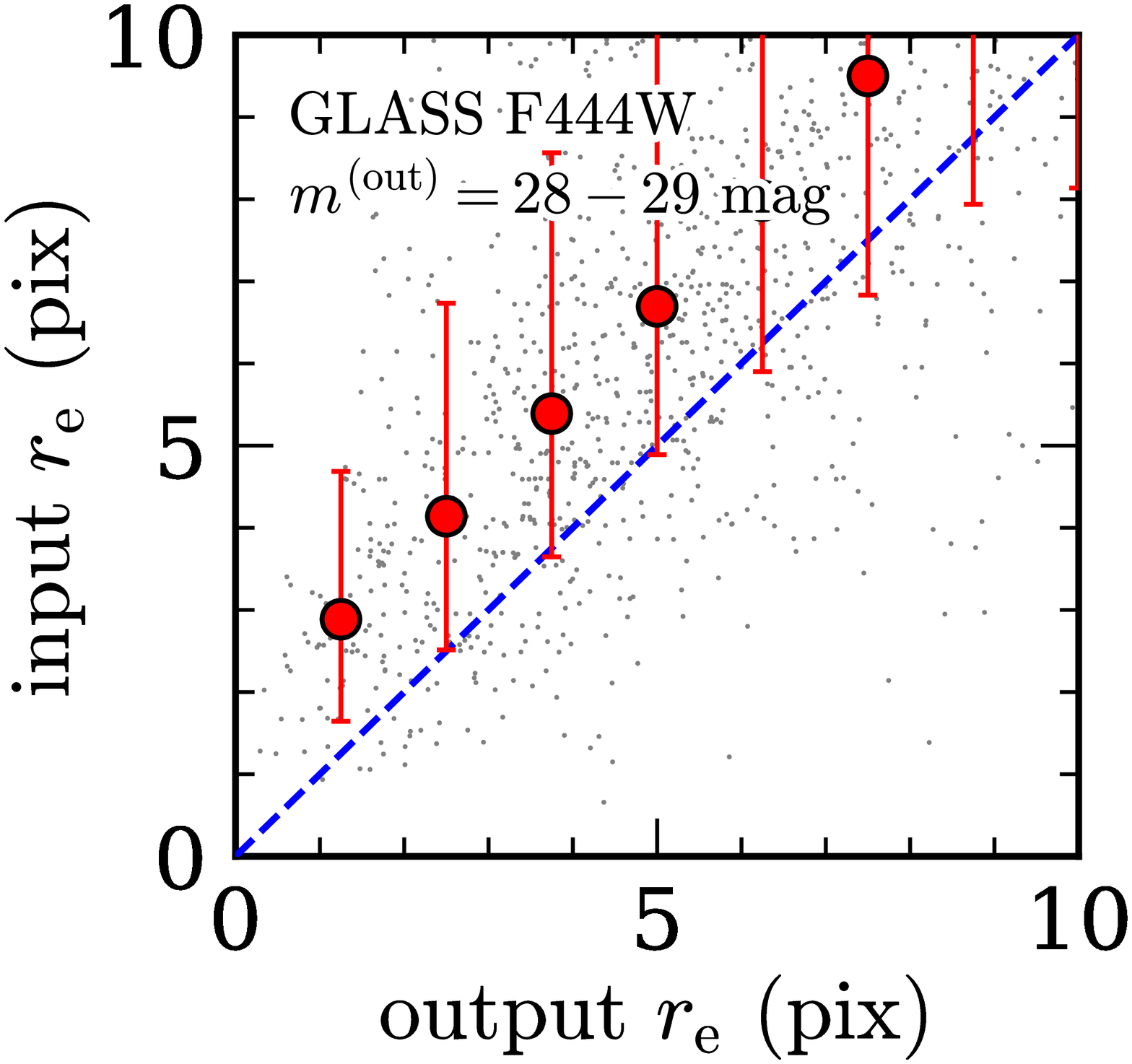}
\caption{
Input circularized radius  vs. output circularized radius 
for a range of output total magnitude 
$m^{\rm (out)} = 25$--$26$ mag, $26$--$27$ mag, $27$--$28$ mag, and $28$--$29$ mag from left to right, 
based on our GALFIT Monte Carlo simulations. 
From top to bottom, the results for 
the GLASS field in F150W,  F200W, and  F444W 
are presented. 
The red filled circles and the red error bars correspond to 
the median values of the difference between the input and output magnitudes 
and the 68 percentile ranges, respectively. 
The gray dots are the results for individual simulated objects. 
The blue dashed line represents the relation that 
the input and output circularized radius are equal.  
}
\label{fig:input_output_re}
\end{center}
\end{figure*}

\begin{figure*}[ht]
\begin{center}
   \includegraphics[width=0.24\textwidth]{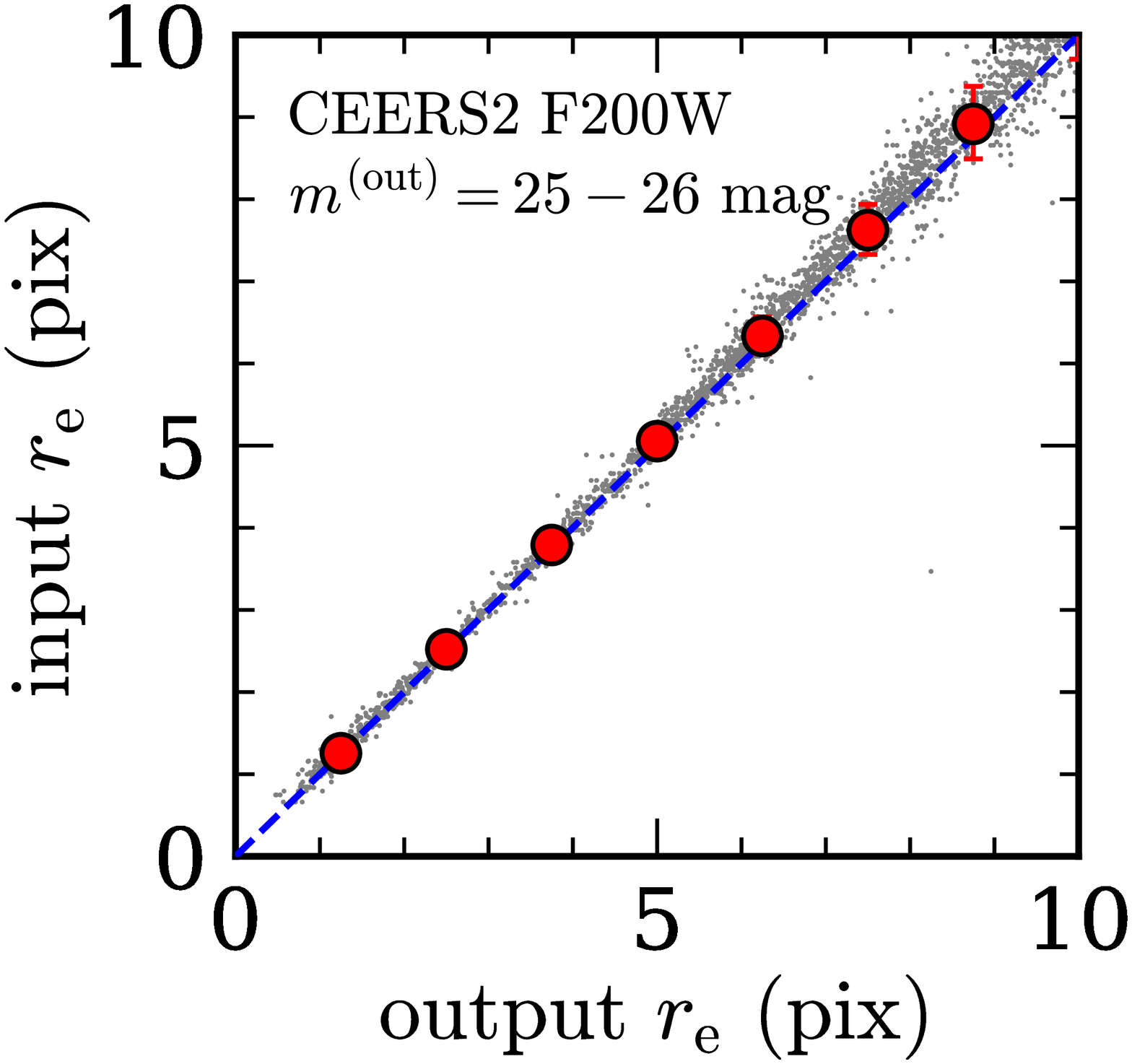}
   \includegraphics[width=0.24\textwidth]{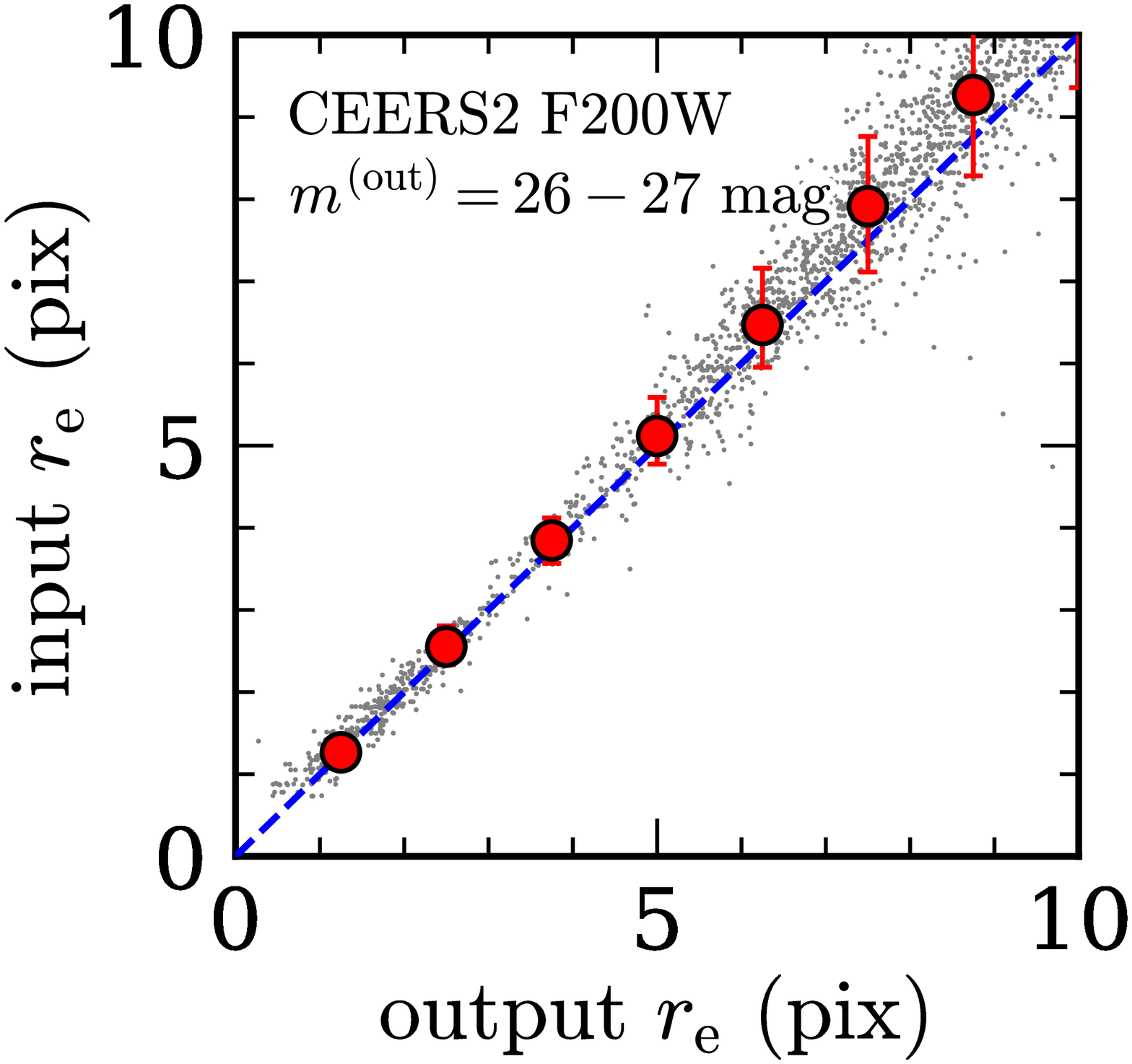}
   \includegraphics[width=0.24\textwidth]{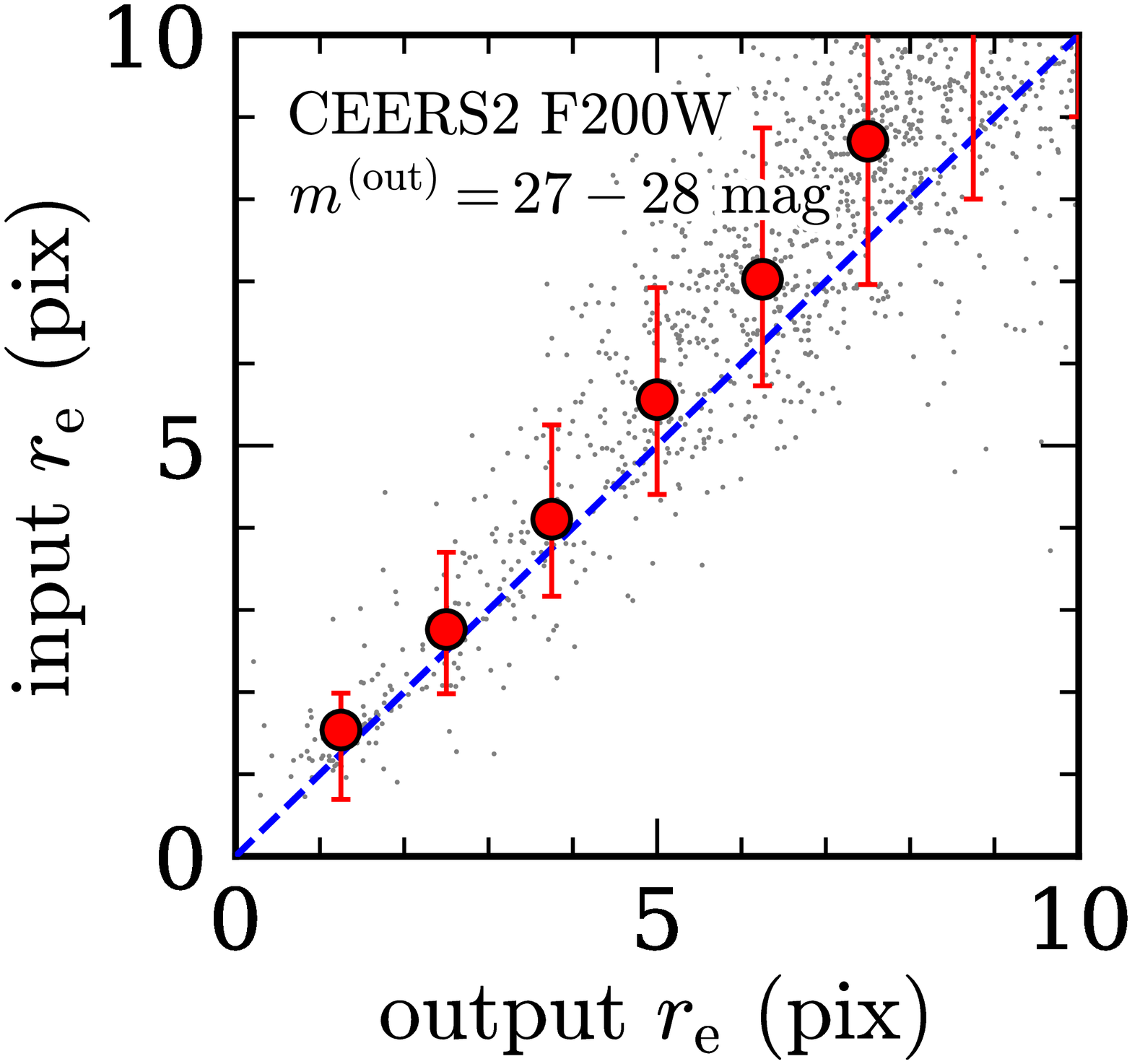}
   \includegraphics[width=0.24\textwidth]{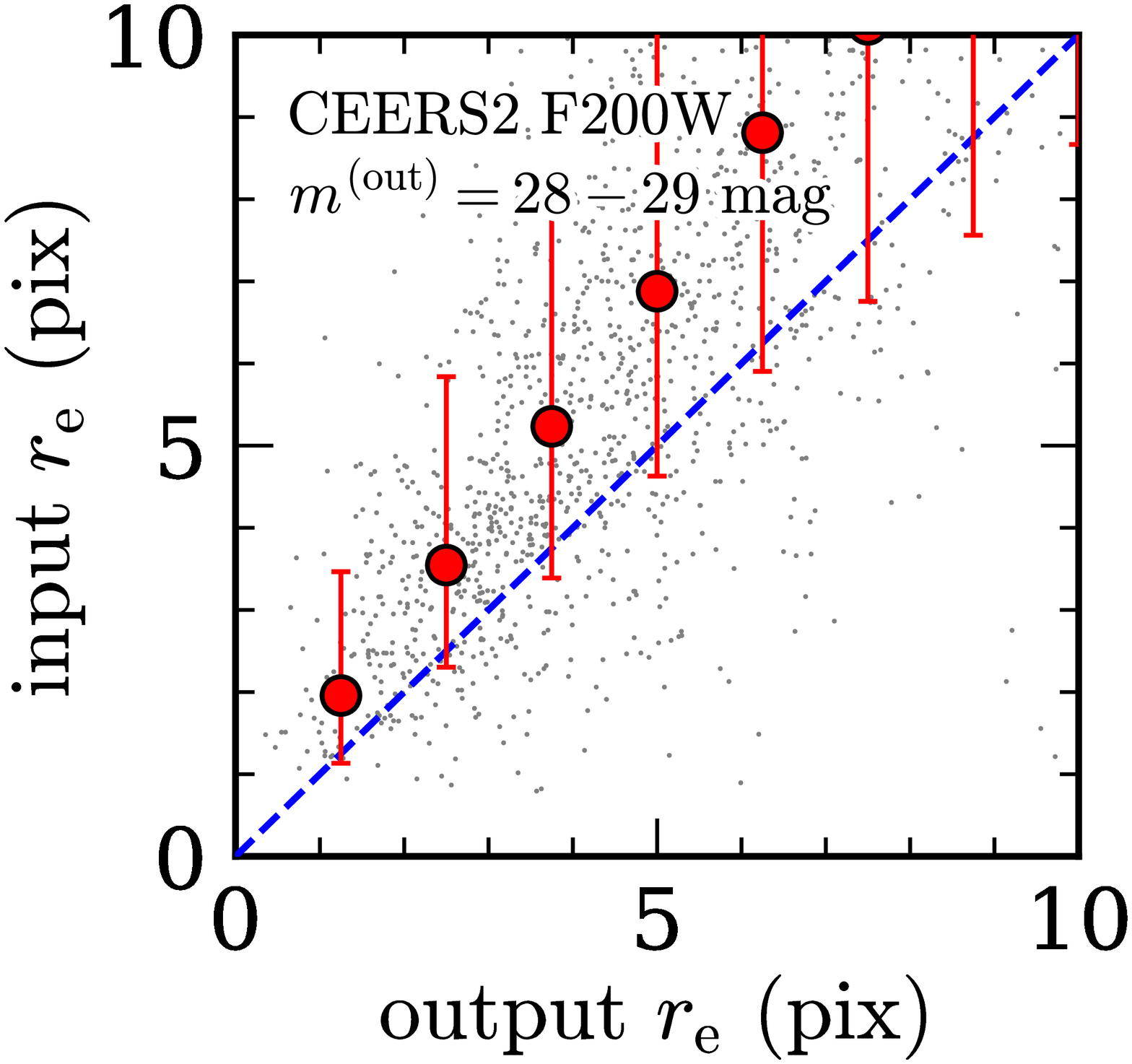}
   \includegraphics[width=0.24\textwidth]{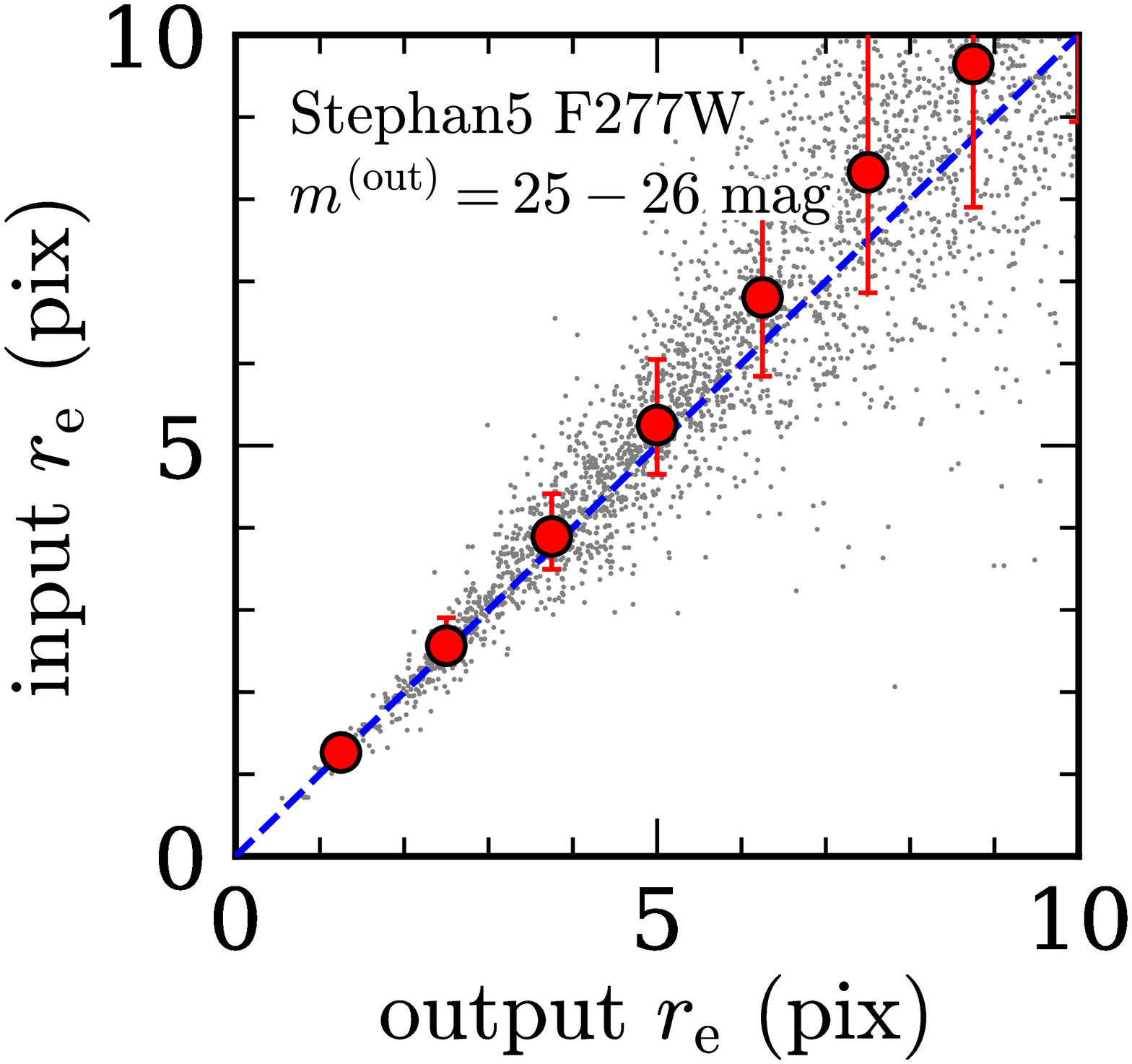}
   \includegraphics[width=0.24\textwidth]{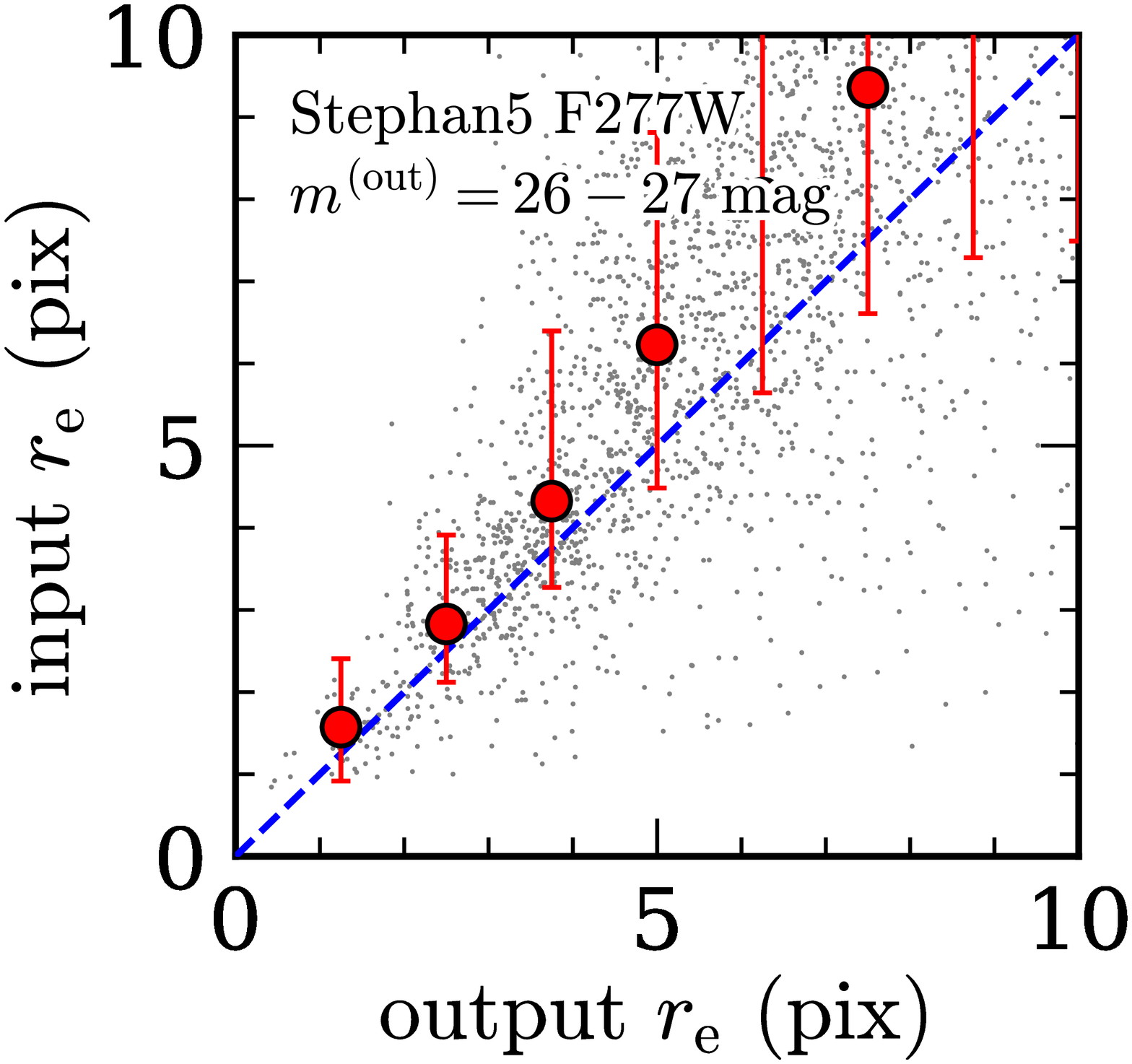}
   \includegraphics[width=0.24\textwidth]{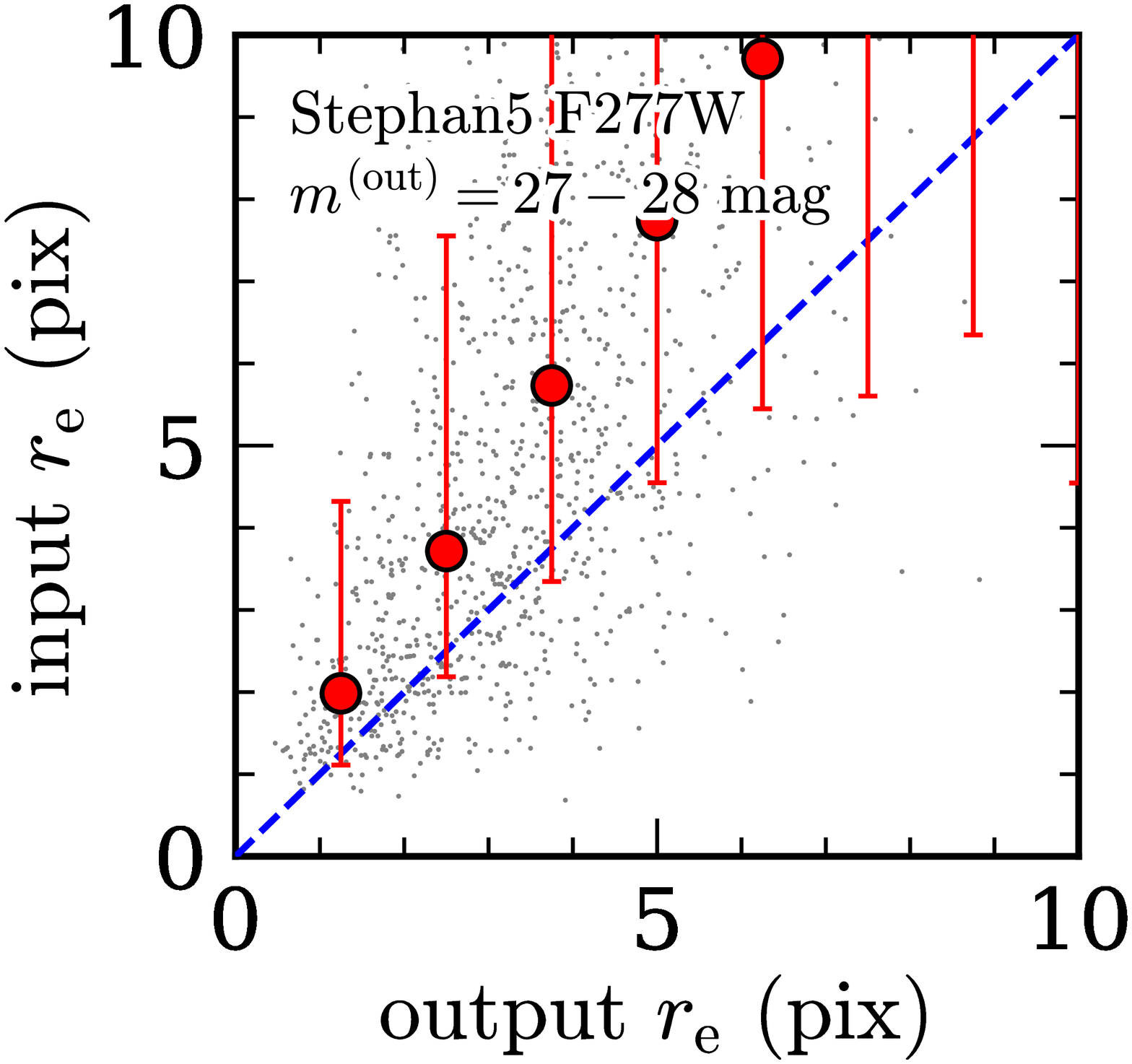}
\caption{
Continuation of Figure \ref{fig:input_output_re}. 
From top to bottom, the results for 
the CEERS2 field in F200W, 
and the Stephan's Quintet field in F277W 
are presented. 
}
\label{fig:input_output_re2}
\end{center}
\end{figure*}

\begin{deluxetable*}{ccccc} 
\tablecolumns{5} 
\tablewidth{0pt} 
\tablecaption{Output and Median Input Circularized Radii with 68 Percentile Ranges Based on Our GALFIT Monte Carlo Simulations 
\label{tab:input_output_re}}
\tablehead{
    \colhead{output $r_{\rm e}$} 
    &  \colhead{input $r_{\rm e}$} 
    &  \colhead{input $r_{\rm e}$} 
    &  \colhead{input $r_{\rm e}$} 
    &  \colhead{input $r_{\rm e}$} \\
    \colhead{ } 
    &  \colhead{at $25$--$26$ mag} 
    &  \colhead{at $26$--$27$ mag} 
    &  \colhead{at $27$--$28$ mag} 
    &  \colhead{at $28$--$29$ mag} \\
    \colhead{(pix)} 
    &  \colhead{(pix)} 
    &  \colhead{(pix)} 
    &  \colhead{(pix)} 
    &  \colhead{(pix)} \\
    \colhead{(1)} 
    &  \colhead{(2)} 
    &  \colhead{(3)} 
    &  \colhead{(4)} 
    &  \colhead{(5)} 
}
\startdata 
\multicolumn{5}{c}{GLASS F150W} \\ 
$1.25$ & $1.26^{+0.05}_{-0.05}$ 	       & $1.28^{+0.14}_{-0.15}$ 	       & $1.35^{+0.72}_{-0.37}$ 	       & $1.82^{+1.17}_{-0.77}$ \\ 
$2.50$ & $2.53^{+0.08}_{-0.07}$ 	       & $2.56^{+0.21}_{-0.18}$ 	       & $2.64^{+0.76}_{-0.52}$ 	       & $3.42^{+2.40}_{-1.18}$ \\ 
$3.75$ & $3.79^{+0.12}_{-0.10}$ 	       & $3.85^{+0.30}_{-0.27}$ 	       & $4.12^{+0.96}_{-0.70}$ 	       & $5.38^{+3.84}_{-1.91}$ \\ 
$5.00$ & $5.07^{+0.18}_{-0.17}$ 	       & $5.24^{+0.46}_{-0.42}$ 	       & $5.63^{+1.55}_{-0.98}$ 	       & $7.51^{+4.11}_{-2.76}$ \\ 
$6.25$ & $6.36^{+0.26}_{-0.22}$ 	       & $6.62^{+0.60}_{-0.57}$ 	       & $7.23^{+1.86}_{-1.37}$ 	       & $9.44^{+4.13}_{-3.38}$ \\ 
$7.50$ & $7.65^{+0.32}_{-0.29}$ 	       & $8.00^{+0.86}_{-0.75}$ 	       & $8.91^{+2.45}_{-1.88}$ 	       & $11.12^{+4.41}_{-3.92}$ \\ 
$8.75$ & $8.96^{+0.49}_{-0.38}$ 	       & $9.53^{+1.02}_{-1.11}$ 	       & $10.51^{+3.06}_{-2.32}$ 	       & $12.15^{+4.93}_{-4.19}$ \\ 
\multicolumn{5}{c}{GLASS F200W} \\ 
$1.25$ & $1.26^{+0.04}_{-0.03}$ 	       & $1.24^{+0.11}_{-0.13}$ 	       & $1.28^{+0.56}_{-0.31}$ 	       & $1.92^{+1.40}_{-0.92}$ \\ 
$2.50$ & $2.53^{+0.07}_{-0.06}$ 	       & $2.56^{+0.20}_{-0.21}$ 	       & $2.68^{+0.71}_{-0.60}$ 	       & $3.46^{+2.00}_{-1.30}$ \\ 
$3.75$ & $3.80^{+0.12}_{-0.10}$ 	       & $3.87^{+0.31}_{-0.26}$ 	       & $4.14^{+1.19}_{-0.76}$ 	       & $5.30^{+3.09}_{-1.83}$ \\ 
$5.00$ & $5.08^{+0.17}_{-0.15}$ 	       & $5.18^{+0.42}_{-0.31}$ 	       & $5.72^{+1.38}_{-0.99}$ 	       & $7.12^{+3.98}_{-2.38}$ \\ 
$6.25$ & $6.37^{+0.23}_{-0.20}$ 	       & $6.52^{+0.56}_{-0.46}$ 	       & $7.17^{+1.91}_{-1.17}$ 	       & $8.79^{+4.68}_{-3.01}$ \\ 
$7.50$ & $7.67^{+0.33}_{-0.28}$ 	       & $7.94^{+0.81}_{-0.69}$ 	       & $8.80^{+2.32}_{-1.78}$ 	       & $10.53^{+4.89}_{-3.68}$ \\ 
$8.75$ & $8.97^{+0.45}_{-0.38}$ 	       & $9.41^{+0.97}_{-0.86}$ 	       & $10.55^{+2.59}_{-2.41}$ 	       & $12.02^{+4.72}_{-3.92}$ \\ 
\multicolumn{5}{c}{GLASS F444W} \\ 
$1.25$ & $1.32^{+0.13}_{-0.10}$ 	       & $1.35^{+0.28}_{-0.24}$ 	       & $1.50^{+0.72}_{-0.56}$ 	       & $2.89^{+1.80}_{-1.24}$ \\ 
$2.50$ & $2.55^{+0.12}_{-0.09}$ 	       & $2.59^{+0.24}_{-0.22}$ 	       & $2.76^{+1.30}_{-0.60}$ 	       & $4.14^{+2.59}_{-1.63}$ \\ 
$3.75$ & $3.80^{+0.12}_{-0.09}$ 	       & $3.83^{+0.23}_{-0.21}$ 	       & $4.75^{+1.31}_{-1.22}$ 	       & $5.39^{+3.17}_{-1.74}$ \\ 
$5.00$ & $5.07^{+0.15}_{-0.11}$ 	       & $5.06^{+0.37}_{-0.29}$ 	       & $6.39^{+1.53}_{-1.38}$ 	       & $6.70^{+3.83}_{-1.80}$ \\ 
$6.25$ & $6.35^{+0.21}_{-0.14}$ 	       & $6.41^{+0.50}_{-0.38}$ 	       & $7.48^{+1.67}_{-1.34}$ 	       & $7.98^{+3.98}_{-2.08}$ \\ 
$7.50$ & $7.64^{+0.27}_{-0.19}$ 	       & $7.75^{+0.65}_{-0.47}$ 	       & $8.59^{+1.85}_{-1.53}$ 	       & $9.50^{+3.48}_{-2.67}$ \\ 
$8.75$ & $8.97^{+0.35}_{-0.28}$ 	       & $9.15^{+0.79}_{-0.59}$ 	       & $10.32^{+2.31}_{-1.92}$ 	       & $11.26^{+4.36}_{-3.32}$ \\ 
\multicolumn{5}{c}{CEERS2 F200W} \\ 
$1.25$ & $1.26^{+0.05}_{-0.07}$ 	       & $1.27^{+0.09}_{-0.17}$ 	       & $1.54^{+0.45}_{-0.84}$ 	       & $1.96^{+1.51}_{-0.82}$ \\ 
$2.50$ & $2.52^{+0.09}_{-0.08}$ 	       & $2.55^{+0.25}_{-0.22}$ 	       & $2.76^{+0.94}_{-0.78}$ 	       & $3.55^{+2.29}_{-1.24}$ \\ 
$3.75$ & $3.79^{+0.13}_{-0.11}$ 	       & $3.85^{+0.27}_{-0.28}$ 	       & $4.10^{+1.15}_{-0.94}$ 	       & $5.23^{+2.99}_{-1.85}$ \\ 
$5.00$ & $5.05^{+0.18}_{-0.18}$ 	       & $5.12^{+0.47}_{-0.34}$ 	       & $5.56^{+1.36}_{-1.15}$ 	       & $6.88^{+4.05}_{-2.25}$ \\ 
$6.25$ & $6.33^{+0.23}_{-0.22}$ 	       & $6.47^{+0.69}_{-0.51}$ 	       & $7.03^{+1.84}_{-1.30}$ 	       & $8.81^{+3.88}_{-2.91}$ \\ 
$7.50$ & $7.62^{+0.31}_{-0.30}$ 	       & $7.91^{+0.85}_{-0.80}$ 	       & $8.70^{+2.41}_{-1.74}$ 	       & $10.13^{+3.76}_{-3.37}$ \\ 
$8.75$ & $8.92^{+0.46}_{-0.43}$ 	       & $9.27^{+1.09}_{-0.99}$ 	       & $10.28^{+2.71}_{-2.28}$ 	       & $11.07^{+4.70}_{-3.51}$ \\ 
\multicolumn{5}{c}{Stephan's Quintet F277W} \\ 
$1.25$ & $1.26^{+0.15}_{-0.17}$ 	       & $1.57^{+0.83}_{-0.66}$ 	       & $1.98^{+2.34}_{-0.87}$ 	       & --- \\ 
$2.50$ & $2.56^{+0.34}_{-0.22}$ 	       & $2.83^{+1.08}_{-0.71}$ 	       & $3.72^{+3.84}_{-1.53}$ 	       & --- \\ 
$3.75$ & $3.89^{+0.52}_{-0.40}$ 	       & $4.32^{+2.07}_{-1.05}$ 	       & $5.73^{+4.40}_{-2.39}$ 	       & --- \\ 
$5.00$ & $5.25^{+0.80}_{-0.60}$ 	       & $6.23^{+2.59}_{-1.74}$ 	       & $7.74^{+5.11}_{-3.20}$ 	       & --- \\ 
$6.25$ & $6.80^{+1.35}_{-0.96}$ 	       & $8.01^{+3.20}_{-2.37}$ 	       & $9.71^{+5.58}_{-4.26}$ 	       & --- \\ 
$7.50$ & $8.33^{+1.77}_{-1.47}$ 	       & $9.36^{+4.22}_{-2.75}$ 	       & $11.21^{+6.52}_{-5.61}$ 	       & --- \\ 
$8.75$ & $9.65^{+2.08}_{-1.74}$ 	       & $10.93^{+4.39}_{-3.64}$ 	       & $11.07^{+9.52}_{-4.72}$ 	       & --- \\ 
\enddata 
\tablecomments{
(1) Output circularized radii. 
(2)--(5) Median input circularized radii 
for a range of output total magnitudes 
$m^{\rm (out)} = 25$--$26$ mag, $26$--$27$ mag, $27$--$28$ mag, and $28$--$29$ mag, 
respectively. 
}
\end{deluxetable*} 

We measure the half-light radii of our high-$z$ galaxy candidates 
by fitting the S\'ersic profile (\citealt{1968adga.book.....S}) 
to the observed two-dimensional (2D) surface brightness profiles. 
The S\'ersic profile has the following functional form, 
\begin{equation}
\Sigma (r)
	= \Sigma_e \exp \left( - b_n \left[ \left( \frac{r}{r_e} \right)^{1/n} -1 \right] \right), 
\end{equation}
where 
$\Sigma_e$ is the surface brightness at the half-light radius $r_e$, 
and $n$ is the S\'ersic index.  
The variable $b_n$ is determined 
to make $r_e$ hold half of the total flux inside. 
For the profile fitting, 
we use GALFIT version 3 
(\citealt{2002AJ....124..266P,2010AJ....139.2097P}), 
which convolves a galaxy surface brightness profile with a PSF profile 
and optimizes the fits using the Levenberg-Marquardt algorithm for $\chi^2$ minimization. 
The output parameters of GALFIT 
include 
the centroid coordinates of a fitted object, 
its total magnitude, 
radius along the semi-major axis ($a$),
S\'ersic index ($n$), 
axis ratio ($b/a$), 
and position angle. 
From the radius along the semi-major axis and the axis ratio, 
we calculate the circularized half-light radius, $r_e = a \sqrt{b/a}$, for each object, 
because it is frequently used for galaxy size measurements 
in previous studies (e.g., \citealt{2012ApJ...746..162N}; \citealt{2012ApJ...756L..12M}; 
\citealt{2013ApJ...777..155O}; \citealt{2015ApJS..219...15S}; \citealt{2018ApJ...855....4K}). 
We provide initial parameters used for the GALFIT profile fitting 
by running SExtractor. 
All of the parameters, except for the S\'ersic index, 
are allowed to vary during the profile fitting. 
The S\'ersic index is fixed at $n=1.5$, 
which corresponds to the median value 
obtained in previous work with HST 
for star-forming galaxies with similar UV luminosities to those of our high-$z$ galaxy candidates 
(\citealt{2015ApJS..219...15S}).\footnote{We confirm that 
our results are almost the same when we fix the S\'ersic index at $n=1.0$.} 
To weight individual pixels during the profile fitting, 
we use noise images that are obtained from the inverse square root of the weight maps. 
We also use segmentation images that are produced by SExtractor,
for masking objects other than the object we are interested in.

As demonstrated in previous work (e.g., \citealt{2013ApJ...777..155O}; \citealt{2015ApJS..219...15S}), 
GALFIT cannot completely trace the outskirts of a galaxy, 
providing systematically low half-light radii and faint total magnitudes, particularly for faint objects.  
To quantify and correct for such systematic effects, we conduct the following Monte Carlo (MC) simulations. 
We use GALFIT to produce galaxy images 
whose S\'ersic index $n$ is fixed at $1.5$,\footnote{
Although this fixed value corresponds to the median value 
obtained for star-forming galaxies at lower redshifts using the HST data as described above, 
the S\'ersic index is not constrained well for star-forming galaxies 
at high redshifts comparable to those investigated in this study. 
In Appendix \ref{sec:MCsimulation_for_varying_Sersic_Index}, 
we present the results of MC simulations with input S\'ersic index values 
ranging from $n=0.5$ to $5.0$, 
to demonstrate an extreme case where the S\'ersic index has a large scatter, 
resulting in a large systematic uncertainty.
The S\'ersic index of high-$z$ galaxies should be well constrained in future studies.}
half-light radius $r_e$ is randomly chosen between $0.5$ and $27.0$ pixels, 
and total magnitude is randomly chosen between $24.5$ and $30.0$ mag. 
We convolve them with a PSF image that 
is a composite of bright and unsaturated stellar objects in each observed field and each band. 
The PSF-convolved galaxy images are then inserted into blank regions of the real NIRCam images 
and being analyzed in the same manner as that for our high-$z$ galaxy candidates.

In Figure \ref{fig:input_output_re} 
and Figure \ref{fig:input_output_re2}, 
we present the results of size measurements of our MC simulated galaxies.
The panels show the input circularized half-light radius, $r_e^{\rm (in)}$, vs. 
the output circularized half-light radius, $r_e^{\rm (out)}$, 
for each image at four different output total magnitude ranges of 
$m^{\rm (out)} = 25$--$26$, $26$--$27$, $27$--$28$, and $28$--$29$ mag.  
As expected, we find that 
measured sizes for all the images show small systematic offsets for objects with small sizes, 
although the fitting progressively underestimate the sizes at large sizes. 
The systematic offsets and statistical uncertainties are larger for fainter objects. 
The median values of input circularized half-light radii 
as a function of output circularized half-light radius are presented 
in Table \ref{tab:input_output_re}.

Figure \ref{fig:input_output_mag} 
and Figure \ref{fig:input_output_mag2} 
show the results of total magnitude measurements 
at two different output size ranges of $r_e^{\rm (out)} = 0$--$5$ and $5$--$10$ pixels 
for each image.  
As expected, it is found that measured total magnitudes show small systematic offsets 
for objects with relatively bright magnitudes, 
but they are systematically fainter than the input values for faint objects.
The systematic offsets and statistical uncertainties are larger for objects with larger sizes. 
The median values of input total magnitudes 
as a function of output total magnitudes are presented 
in Table \ref{tab:input_output_mag}.

In summary, 
our MC simulations indicate that 
GALFIT measurements of half-light radii and total magnitudes 
are systematically underestimated for faint objects.
We correct for these systematic effects 
and also estimate statistical uncertainties in size and total magnitude measurements 
based on our MC simulation results. 
More specifically, for size measurements, 
we use the MC simulation results in the output magnitude bins 
corresponding to the objects we study 
to correct for the output sizes 
by the differences between the input and output sizes. 
Similarly, for total magnitude measurements, 
we refer to the MC simulation results in the output size bins 
corresponding to the objects we investigate. 
We then correct for the output magnitudes 
by the differences between the input and output magnitudes.

\begin{figure}[h]
\begin{center}
   \includegraphics[width=0.23\textwidth]{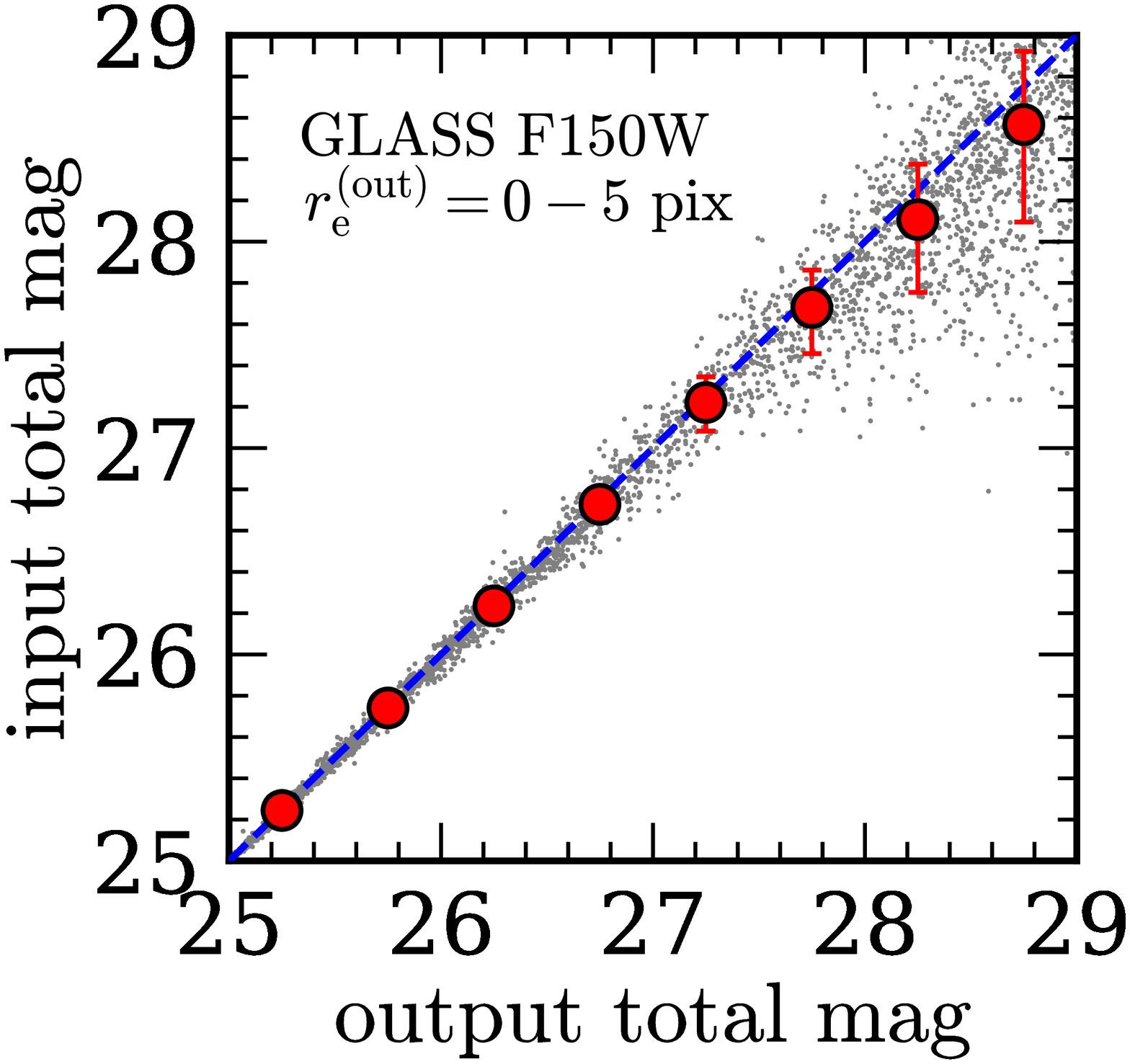}
   \includegraphics[width=0.23\textwidth]{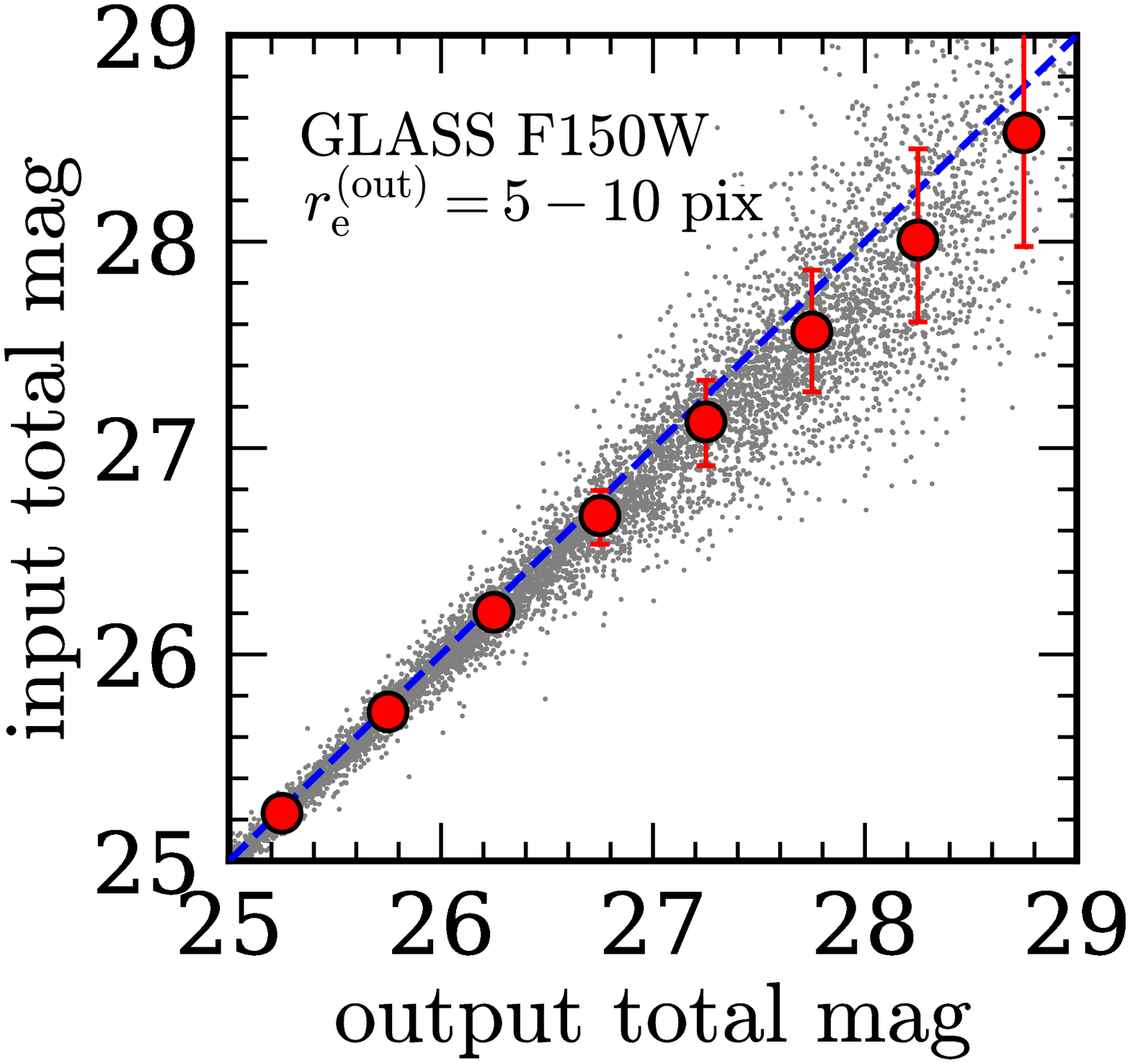}
   \includegraphics[width=0.23\textwidth]{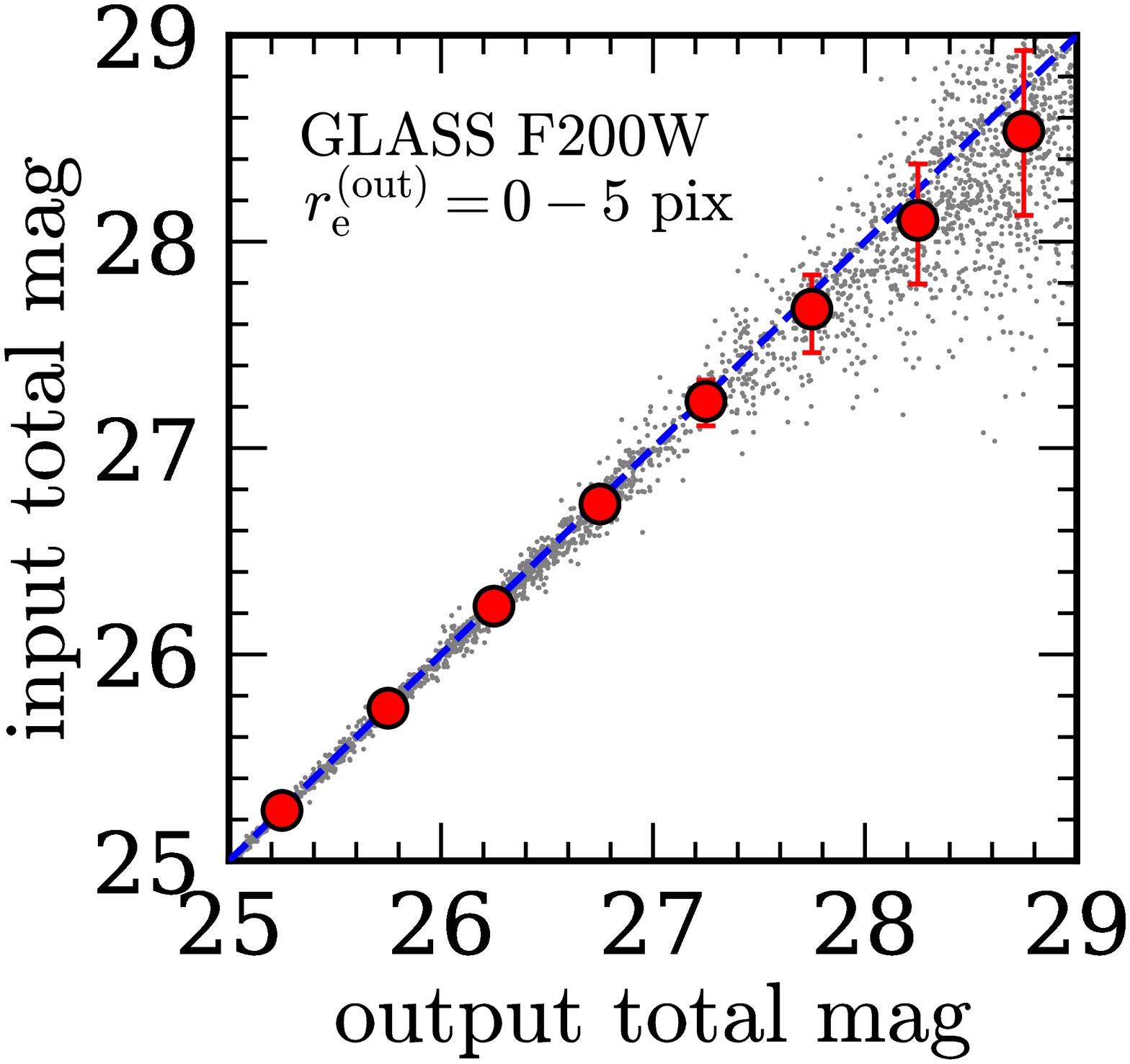}
   \includegraphics[width=0.23\textwidth]{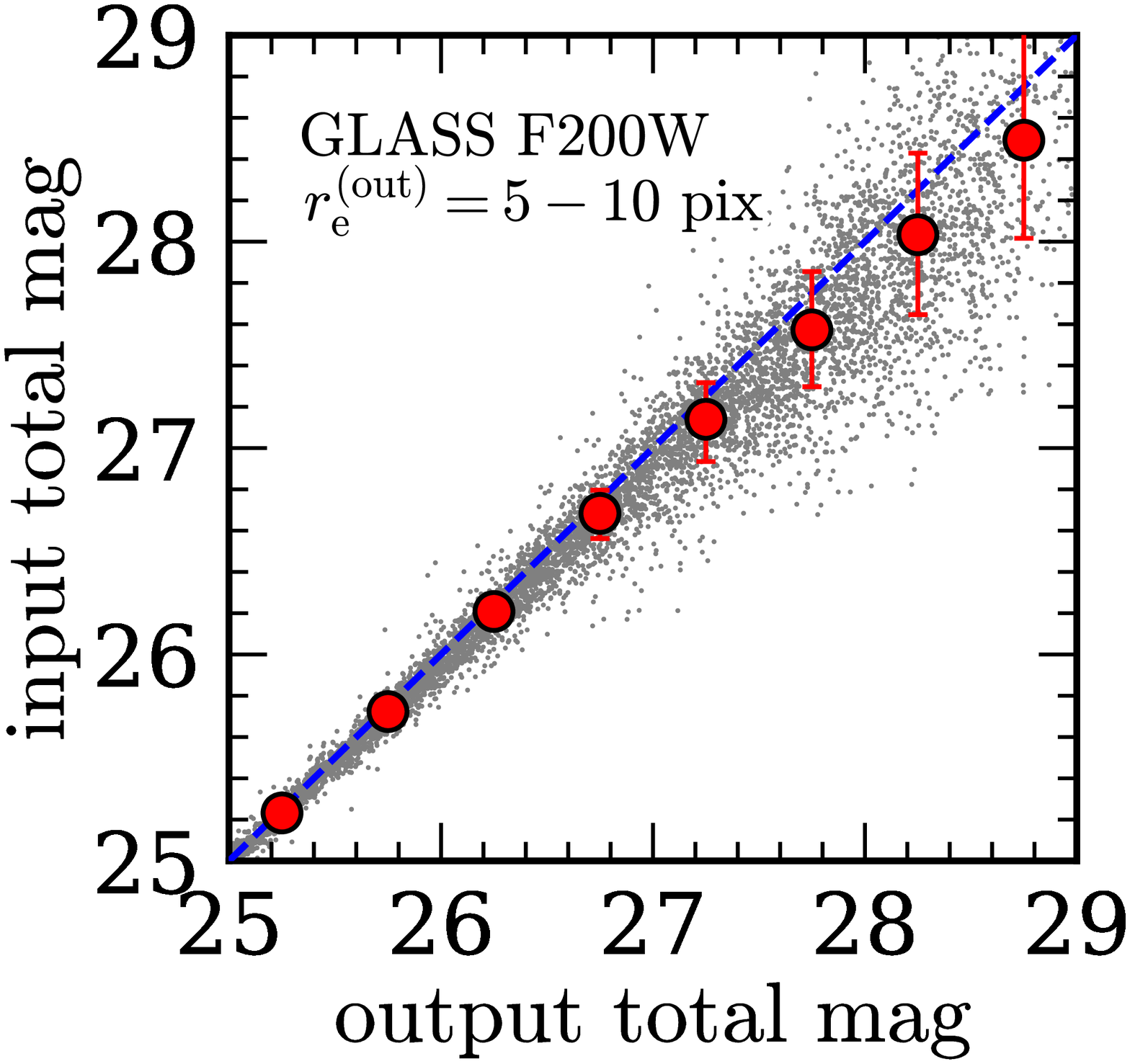}
   \includegraphics[width=0.23\textwidth]{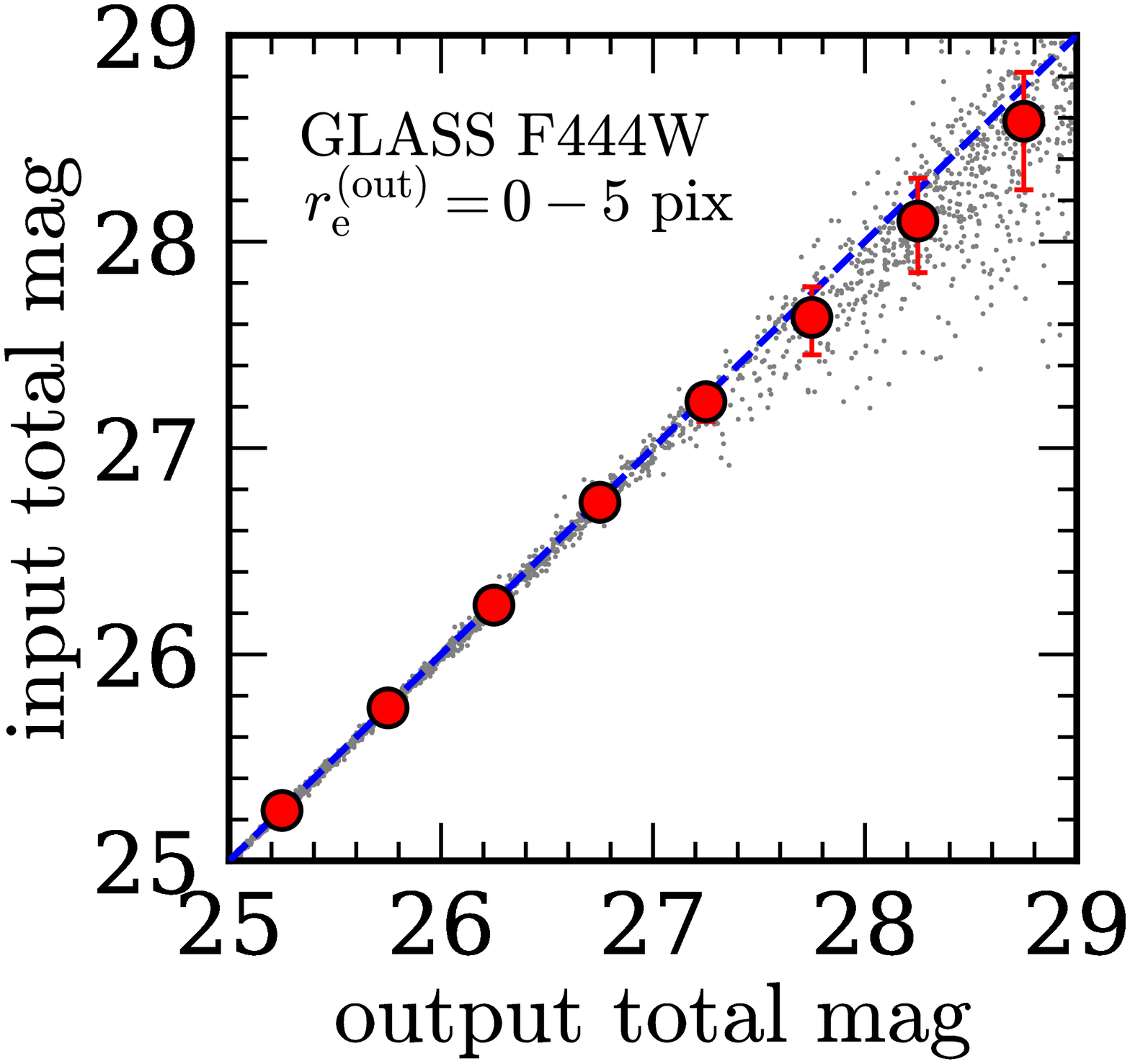}
   \includegraphics[width=0.23\textwidth]{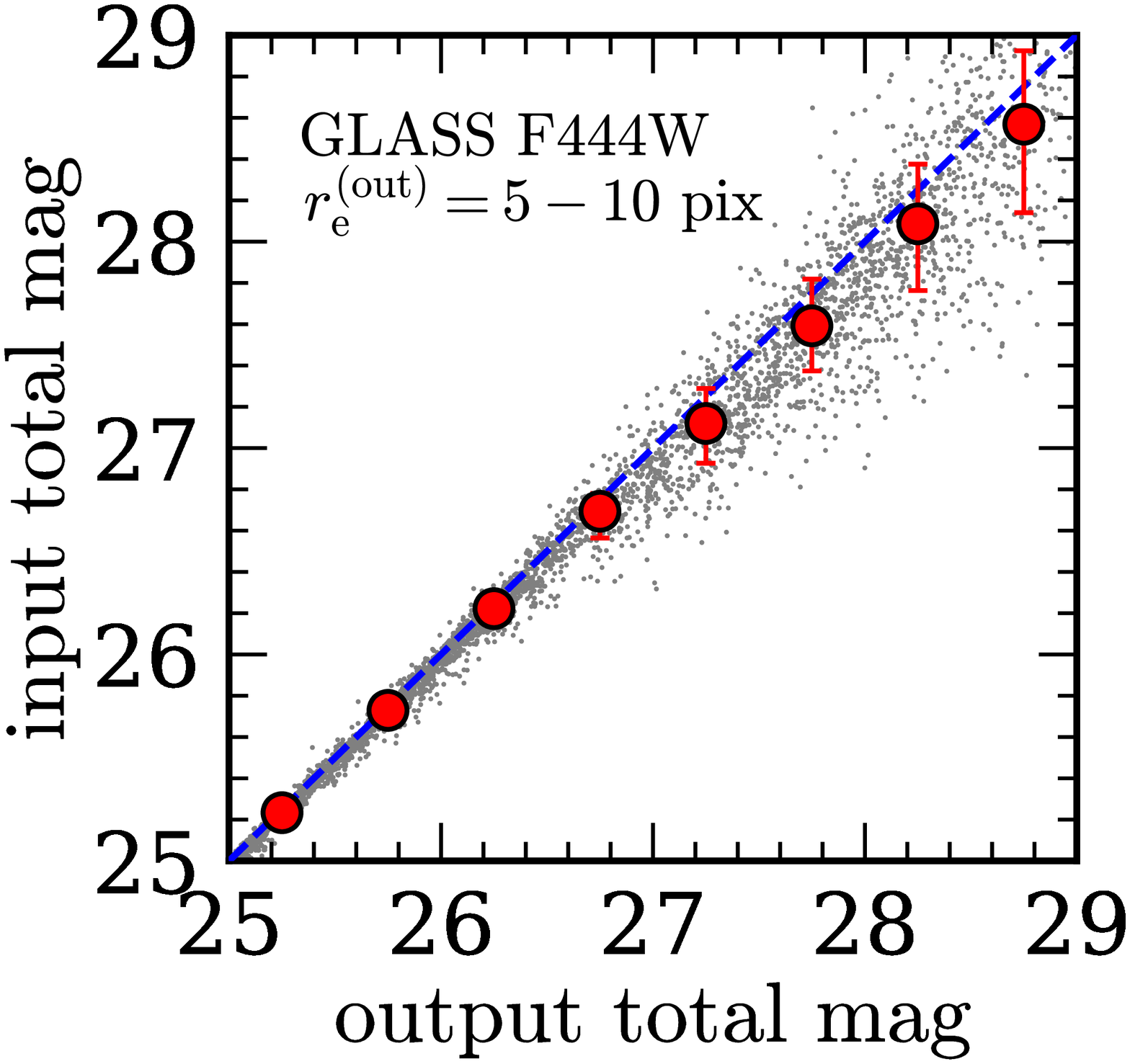}
\caption{
Input total magnitude vs. output total magnitude 
for a range of output half-light radii 
$r_{\rm e}^{\rm (out)} = 0$--$5$ pixels (left) and $5$--$10$ pixels (right) 
based on our  GALFIT Monte Carlo simulations. 
From top to bottom, the results for 
the GLASS field in F150W, F200W, and F444W 
are presented. 
The red filled circles and the red error bars correspond to 
the median values of the difference between the input and output magnitudes 
and the 68 percentile ranges, respectively. 
The gray dots are the results for individual simulated objects. 
The blue dashed line represents the relation that 
the input and output magnitudes are equal.  
}
\label{fig:input_output_mag}
\end{center}
\end{figure}

\begin{figure}[h]
\begin{center}
   \includegraphics[width=0.23\textwidth]{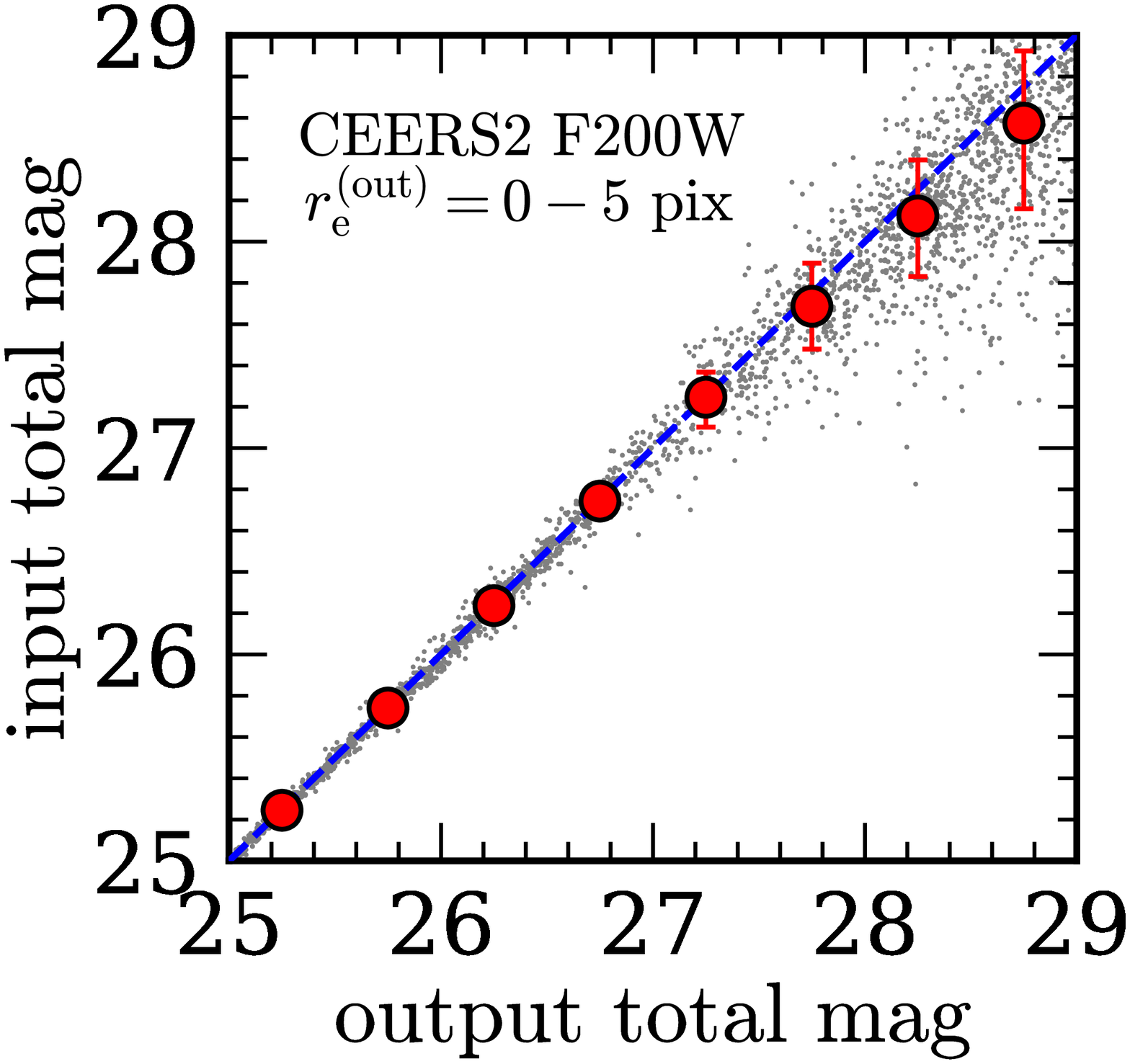}
   \includegraphics[width=0.23\textwidth]{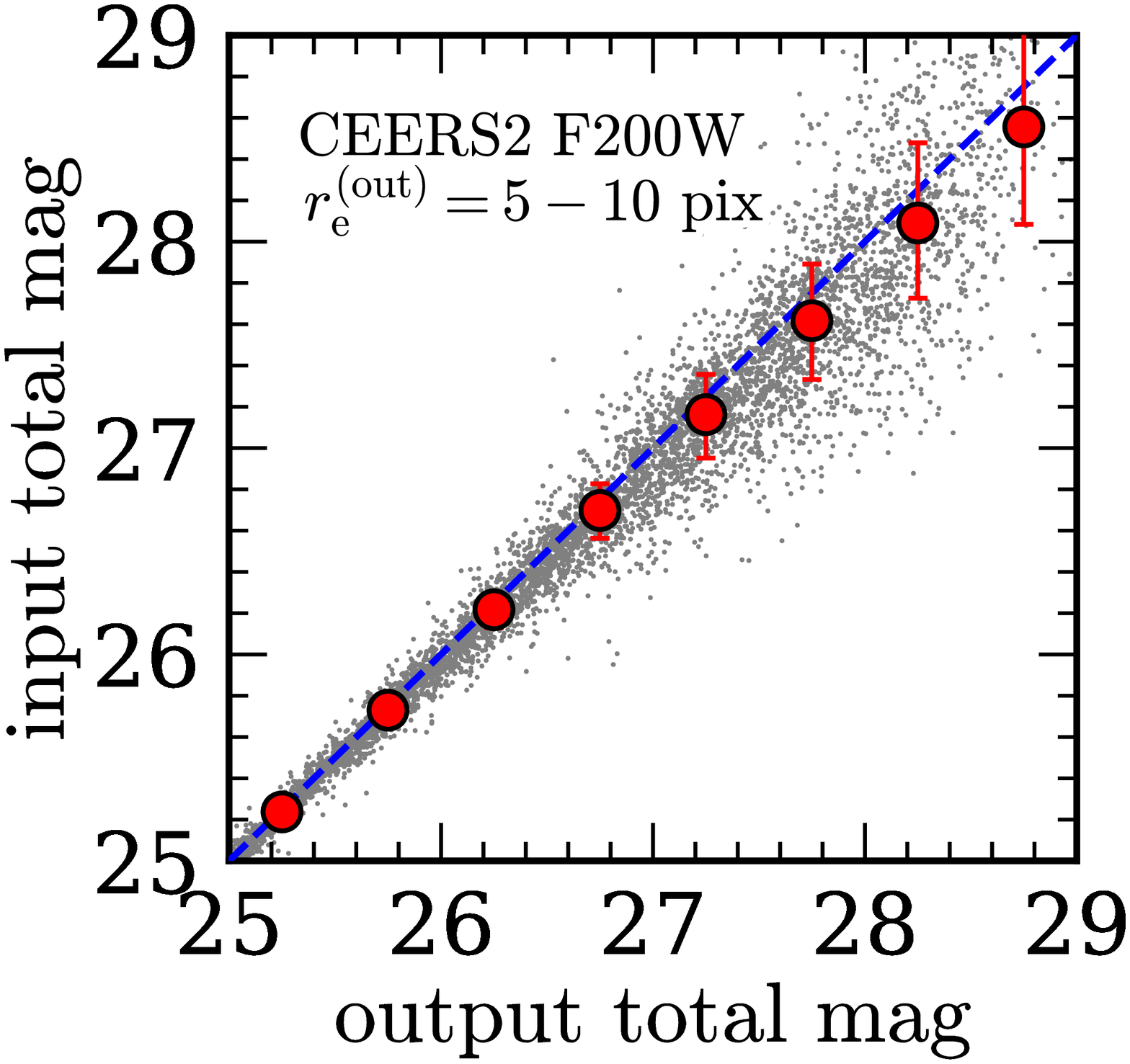}
   \includegraphics[width=0.23\textwidth]{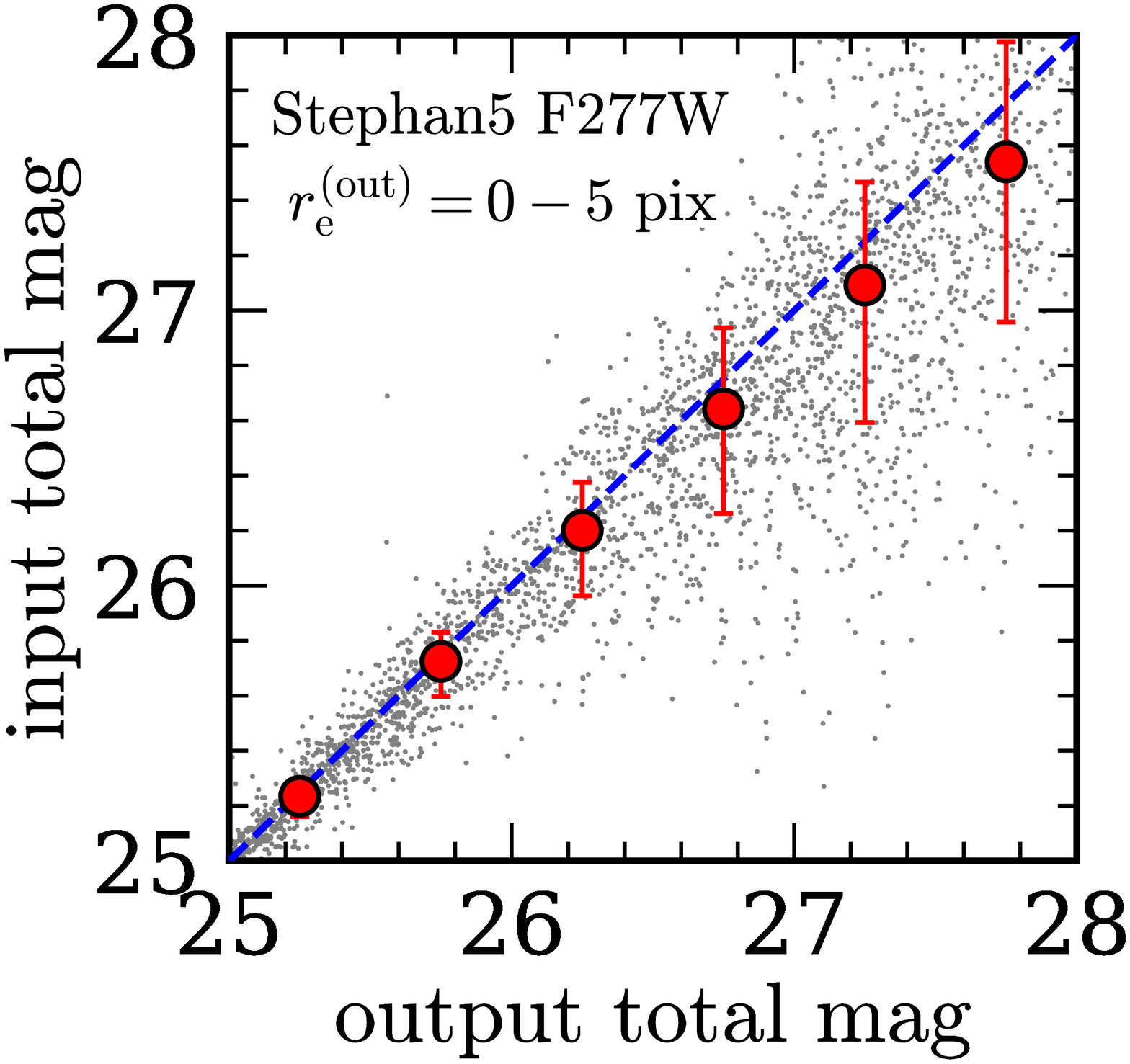}
   \includegraphics[width=0.23\textwidth]{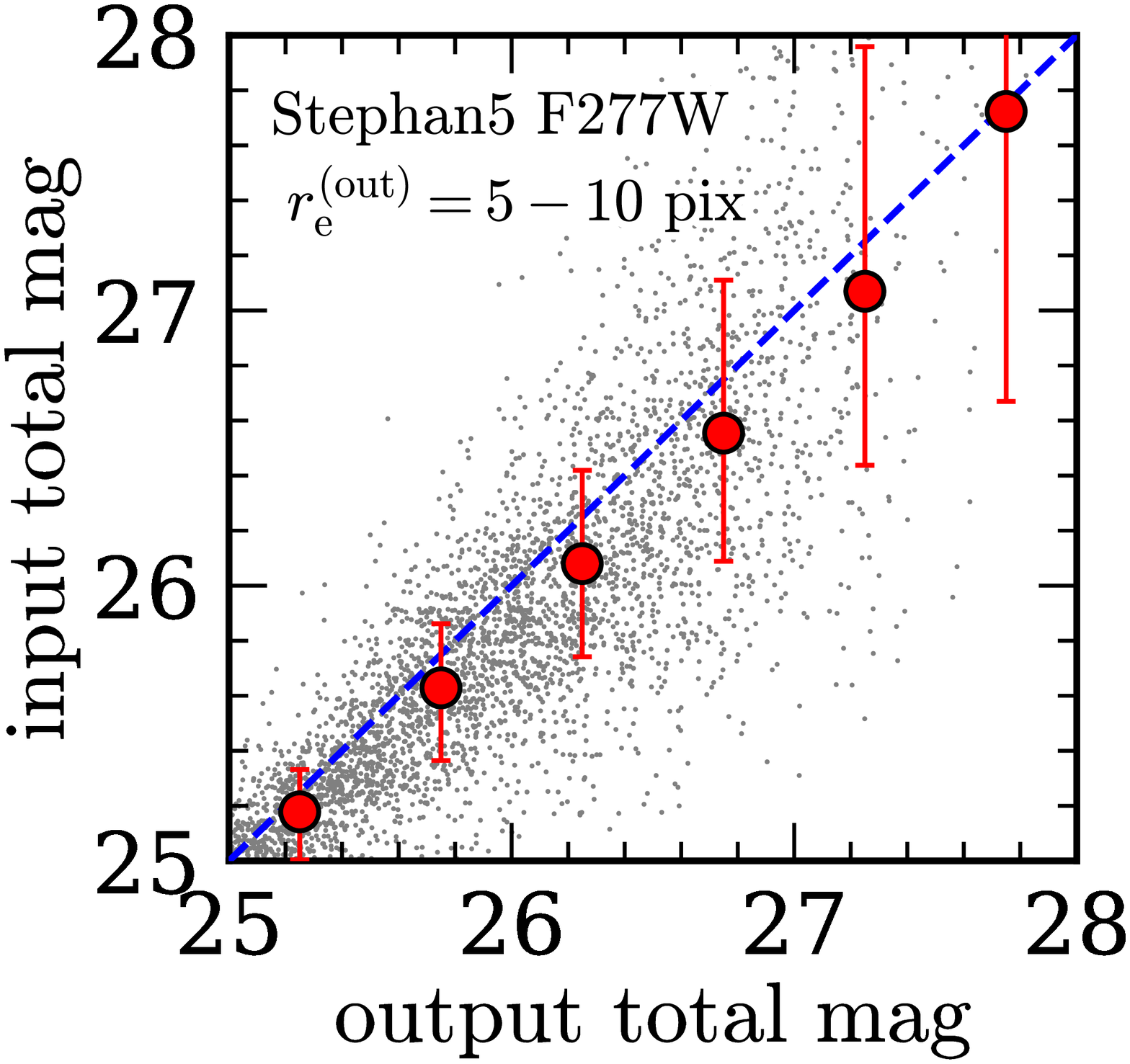}
\caption{
Continuation of Figure \ref{fig:input_output_mag}. 
From top to bottom, the results for 
the CEERS2 field in F200W, 
and the Stephan's Quintet field in F277W 
are presented. 
}
\label{fig:input_output_mag2}
\end{center}
\end{figure}

\begin{deluxetable}{ccc} 
\tablecolumns{3} 
\tablewidth{0pt} 
\tablecaption{Output and Median Input Total Magnitudes with 68 Percentile Ranges Based on Our GALFIT Monte Carlo Simulations 
\label{tab:input_output_mag}}
\tablehead{
    \colhead{output total mag}
    &  \colhead{input total mag}
    &  \colhead{input total mag} \\
    \colhead{ } 
    &  \colhead{at $0$--$5$ pix} 
    &  \colhead{at $5$--$10$ pix} \\
    \colhead{(mag)} 
    &  \colhead{(mag)} 
    &  \colhead{(mag)} \\
    \colhead{(1)} 
    &  \colhead{(2)} 
    &  \colhead{(3)} 
}
\startdata 
\multicolumn{3}{c}{GLASS F150W} \\ 
$25.25$ & $25.24^{+0.02}_{-0.02}$ 	       & $25.23^{+0.03}_{-0.04}$ \\ 
$25.75$ & $25.74^{+0.03}_{-0.03}$ 	       & $25.72^{+0.05}_{-0.06}$ \\ 
$26.25$ & $26.24^{+0.04}_{-0.05}$ 	       & $26.20^{+0.08}_{-0.09}$ \\ 
$26.75$ & $26.73^{+0.07}_{-0.07}$ 	       & $26.67^{+0.12}_{-0.14}$ \\ 
$27.25$ & $27.22^{+0.12}_{-0.14}$ 	       & $27.13^{+0.20}_{-0.21}$ \\ 
$27.75$ & $27.68^{+0.18}_{-0.22}$ 	       & $27.56^{+0.30}_{-0.29}$ \\ 
$28.25$ & $28.11^{+0.27}_{-0.35}$ 	       & $28.01^{+0.44}_{-0.40}$ \\ 
$28.75$ & $28.57^{+0.36}_{-0.47}$ 	       & $28.53^{+0.57}_{-0.55}$ \\ 
\multicolumn{3}{c}{GLASS F200W} \\ 
$25.25$ & $25.24^{+0.01}_{-0.02}$ 	       & $25.23^{+0.03}_{-0.03}$ \\ 
$25.75$ & $25.74^{+0.02}_{-0.03}$ 	       & $25.72^{+0.04}_{-0.05}$ \\ 
$26.25$ & $26.23^{+0.04}_{-0.04}$ 	       & $26.21^{+0.06}_{-0.08}$ \\ 
$26.75$ & $26.73^{+0.07}_{-0.06}$ 	       & $26.68^{+0.11}_{-0.12}$ \\ 
$27.25$ & $27.23^{+0.10}_{-0.12}$ 	       & $27.14^{+0.18}_{-0.20}$ \\ 
$27.75$ & $27.67^{+0.17}_{-0.21}$ 	       & $27.57^{+0.28}_{-0.28}$ \\ 
$28.25$ & $28.10^{+0.27}_{-0.31}$ 	       & $28.03^{+0.40}_{-0.39}$ \\ 
$28.75$ & $28.53^{+0.39}_{-0.41}$ 	       & $28.49^{+0.59}_{-0.48}$ \\ 
\multicolumn{3}{c}{GLASS F444W} \\ 
$25.25$ & $25.24^{+0.01}_{-0.01}$ 	       & $25.23^{+0.02}_{-0.03}$ \\ 
$25.75$ & $25.74^{+0.01}_{-0.02}$ 	       & $25.73^{+0.03}_{-0.04}$ \\ 
$26.25$ & $26.24^{+0.02}_{-0.02}$ 	       & $26.22^{+0.05}_{-0.06}$ \\ 
$26.75$ & $26.74^{+0.04}_{-0.04}$ 	       & $26.69^{+0.09}_{-0.13}$ \\ 
$27.25$ & $27.22^{+0.06}_{-0.10}$ 	       & $27.12^{+0.17}_{-0.19}$ \\ 
$27.75$ & $27.63^{+0.15}_{-0.18}$ 	       & $27.59^{+0.23}_{-0.22}$ \\ 
$28.25$ & $28.10^{+0.21}_{-0.25}$ 	       & $28.09^{+0.29}_{-0.32}$ \\ 
$28.75$ & $28.58^{+0.24}_{-0.33}$ 	       & $28.57^{+0.36}_{-0.43}$ \\ 
\multicolumn{3}{c}{CEERS2 F200W} \\ 
$25.25$ & $25.25^{+0.02}_{-0.02}$ 	       & $25.24^{+0.03}_{-0.03}$ \\ 
$25.75$ & $25.74^{+0.03}_{-0.03}$ 	       & $25.73^{+0.05}_{-0.06}$ \\ 
$26.25$ & $26.24^{+0.04}_{-0.04}$ 	       & $26.22^{+0.08}_{-0.09}$ \\ 
$26.75$ & $26.74^{+0.06}_{-0.06}$ 	       & $26.70^{+0.13}_{-0.13}$ \\ 
$27.25$ & $27.25^{+0.12}_{-0.14}$ 	       & $27.16^{+0.20}_{-0.21}$ \\ 
$27.75$ & $27.69^{+0.21}_{-0.21}$ 	       & $27.62^{+0.27}_{-0.28}$ \\ 
$28.25$ & $28.12^{+0.27}_{-0.29}$ 	       & $28.09^{+0.39}_{-0.36}$ \\ 
$28.75$ & $28.57^{+0.35}_{-0.42}$ 	       & $28.56^{+0.53}_{-0.47}$ \\ 
\multicolumn{3}{c}{Stephan's Quintet F277W} \\ 
$25.25$ & $25.23^{+0.06}_{-0.07}$ 	       & $25.18^{+0.15}_{-0.17}$ \\ 
$25.75$ & $25.72^{+0.11}_{-0.13}$ 	       & $25.63^{+0.23}_{-0.26}$ \\ 
$26.25$ & $26.20^{+0.18}_{-0.24}$ 	       & $26.08^{+0.34}_{-0.34}$ \\ 
$26.75$ & $26.64^{+0.30}_{-0.38}$ 	       & $26.55^{+0.56}_{-0.47}$ \\ 
$27.25$ & $27.09^{+0.37}_{-0.50}$ 	       & $27.07^{+0.89}_{-0.63}$ \\ 
$27.75$ & $27.54^{+0.44}_{-0.58}$ 	       & $27.72^{+1.23}_{-1.05}$ \\ 
\enddata 
\tablecomments{
(1) Output total magnitude. 
(2)--(3) Median input total magnitude 
for a range of output half-light radii 
$r_{\rm e}^{\rm (out)} = 0$--$5$ pixels and $5$--$10$ pixels, 
respectively. 
}
\end{deluxetable} 

\section{Results} \label{sec:results}

\subsection{Surface Brightness Profile Fitting Results for the Rest-frame UV Continuum} 

We perform surface brightness profile fitting 
for our $z \sim 9$--$16$ galaxy candidates with GALFIT. 
We individually measure the sizes of the bright objects, 
for which the S/Ns of the aperture magnitudes are greater than $10$, 
and extend the measurements to fainter ones by stacking their images. 
We correct for the systematic effects 
by using the MC simulation results presented in Section \ref{sec:measurements}.

\begin{figure}[h]
\begin{center}
   \includegraphics[width=0.5\textwidth]{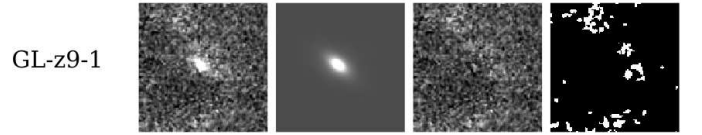}
\caption{
S\'ersic profile fitting results for bright F150W dropouts ($z\sim9$ galaxy candidates). 
From left to right, the $1\farcs5 \times 1\farcs5$ cutouts of the original image, 
the best-fit S\'ersic model profile images, 
the residual images that are made by subtracting the best-fit images from the original ones, 
and the segmentation maps used for masking all the neighboring objects during the profile fitting are presented. 
}
\label{fig:fit_results_f150w_z9}
\end{center}
\end{figure}

Figure \ref{fig:fit_results_f150w_z9} shows 
the results of the surface brightness profile fitting 
for the one F115W-dropout with S/N $>10$, GL-z9-1.
The $1\farcs5 \times 1\farcs5$ cutouts of the original image, 
the best-fit model image, 
the residual image (i.e., the original image cutout $-$ the best-fit model image), 
and the segmentation map are presented from left to right. 
The best-fit parameters of the total magnitudes and 
the circularized half-light radii are summarized in Table \ref{tab:z9result}, 
where the systematic effects and statistical uncertainties are taken into account 
based on our MC simulation results.  
Although the total magnitude of GL-z9-1, whose photo-$z$ is relatively high, 
may be underestimated in F150W due to the Lyman break,  
we confirm that its total magnitude in F200W is almost the same 
by performing the profile fitting with the F200W image.

\begin{figure}[h]
\begin{center}
   \includegraphics[width=0.5\textwidth]{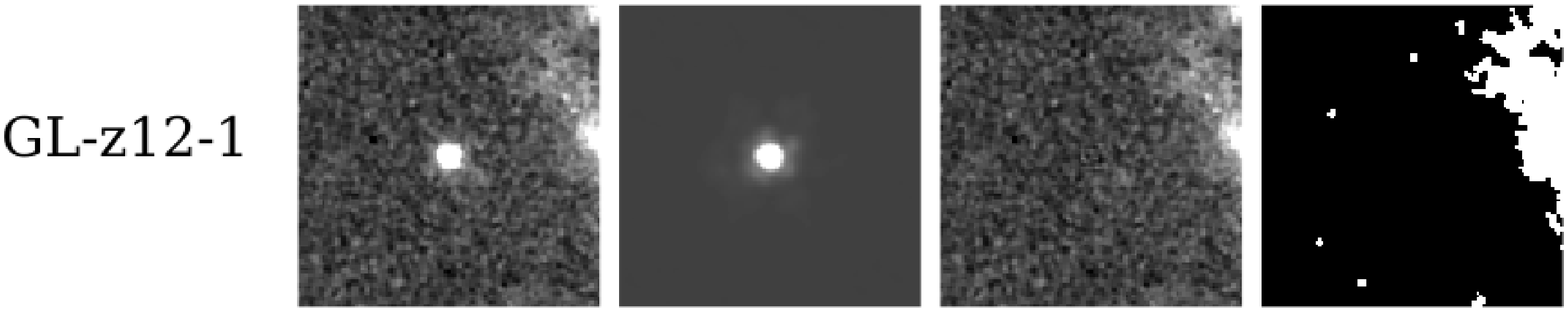}
   \includegraphics[width=0.5\textwidth]{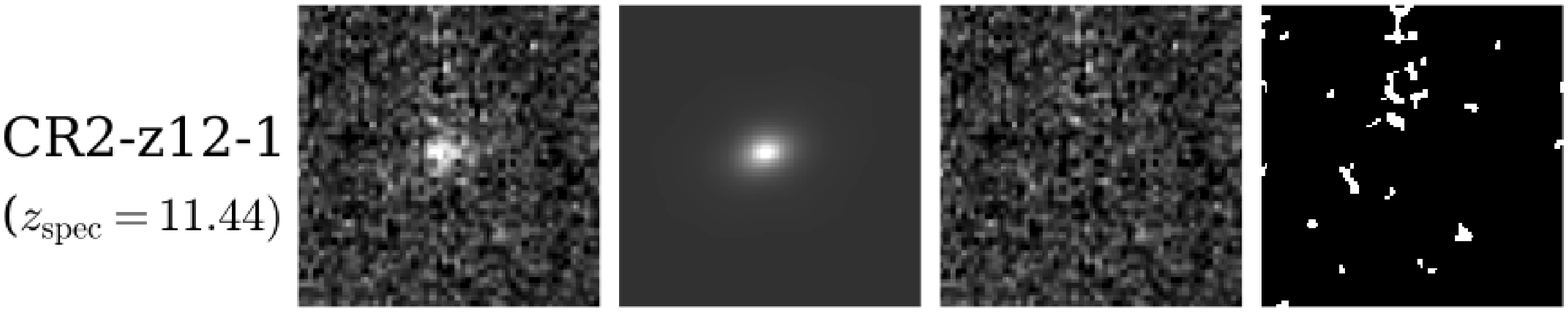}
   \includegraphics[width=0.5\textwidth]{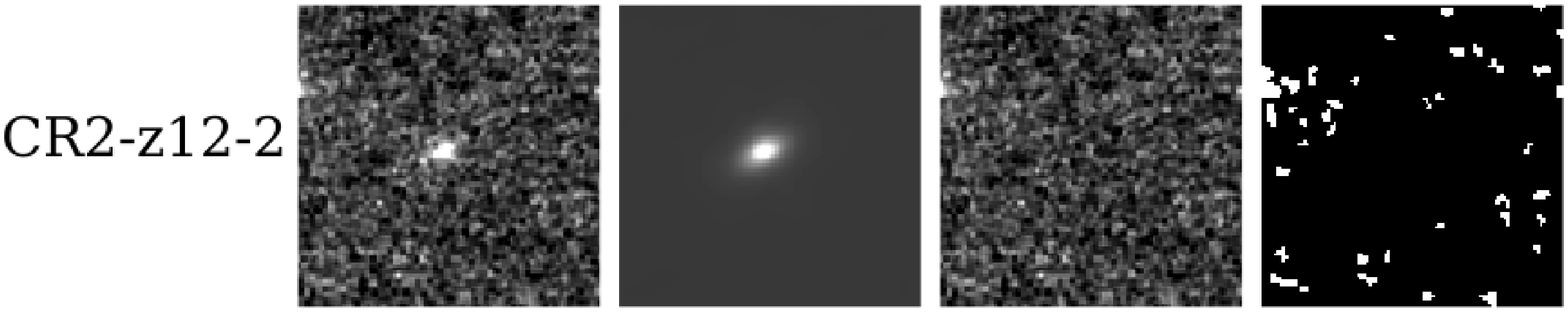}
   \includegraphics[width=0.5\textwidth]{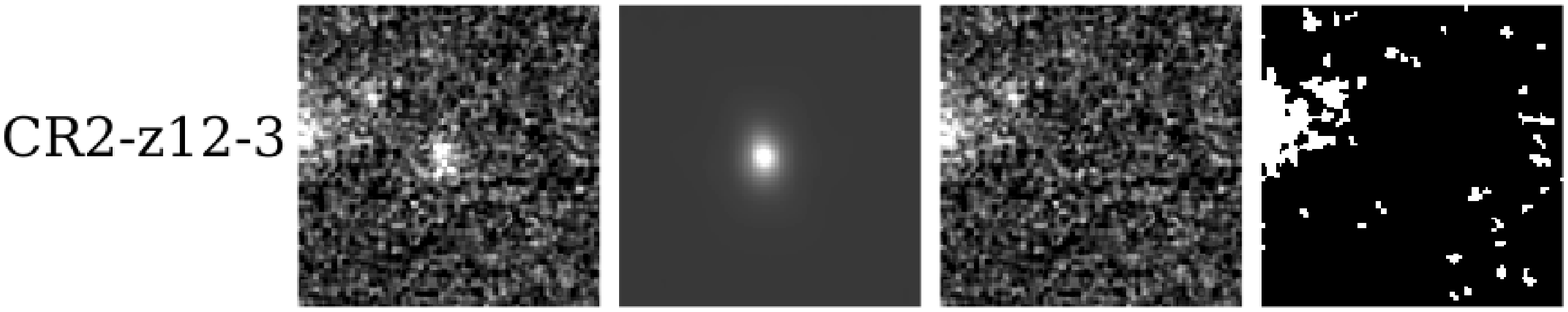}
\caption{
Same as Figure \ref{fig:fit_results_f150w_z9}, 
but for bright F200W dropouts ($z\sim12$ galaxy candidates). 
}
\label{fig:fit_results_f200w_z12}
\end{center}
\end{figure}

\begin{figure}[h]
\begin{center}
   \includegraphics[width=0.5\textwidth]{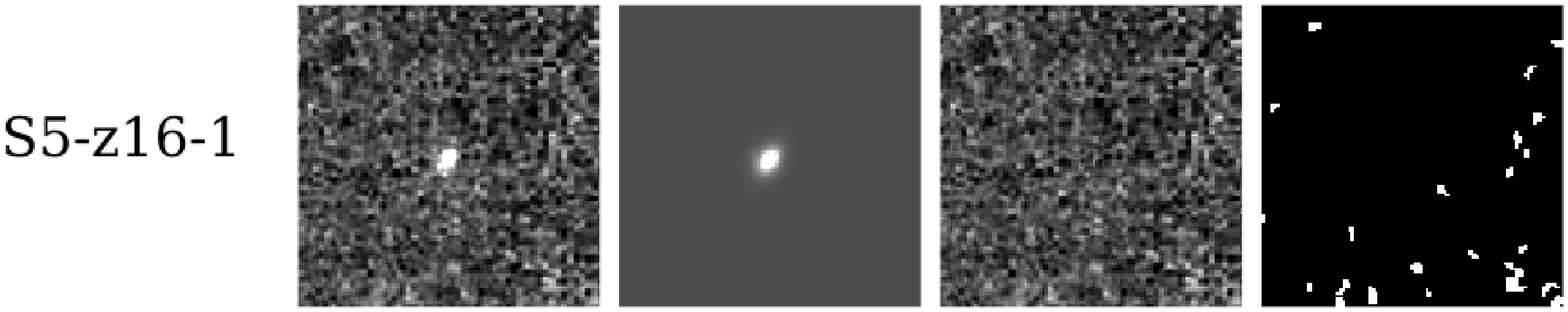}
\caption{
Same as Figure \ref{fig:fit_results_f150w_z9}, 
but for bright F277W dropouts ($z\sim16$ galaxy candidates). 
}
\label{fig:fit_results_f277w_z17}
\end{center}
\end{figure}

Similarly, 
Figure \ref{fig:fit_results_f200w_z12} and Figure \ref{fig:fit_results_f277w_z17} 
present the results of the surface brightness profile fitting 
for the four F150W-dropouts with S/N $>10$, GL-z12-1, CR2-z12-1, CR2-z12-2, and CR2-z12-3, 
and the one F200W-dropouts with S/N $>10$, S5-z16-1, respectively.
Their best-fit total magnitudes and half-light radii are summarized 
in Table \ref{tab:z12result} and Table \ref{tab:z17result}.

\begin{figure}[h]
\begin{center}
   \includegraphics[width=0.45\textwidth]{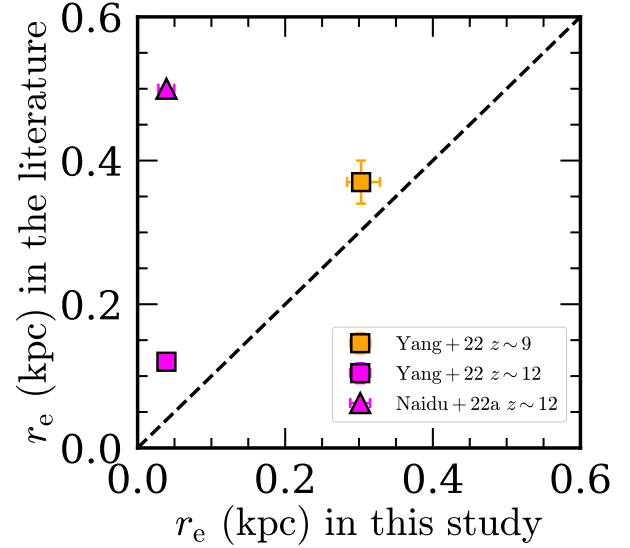}
   \includegraphics[width=0.45\textwidth]{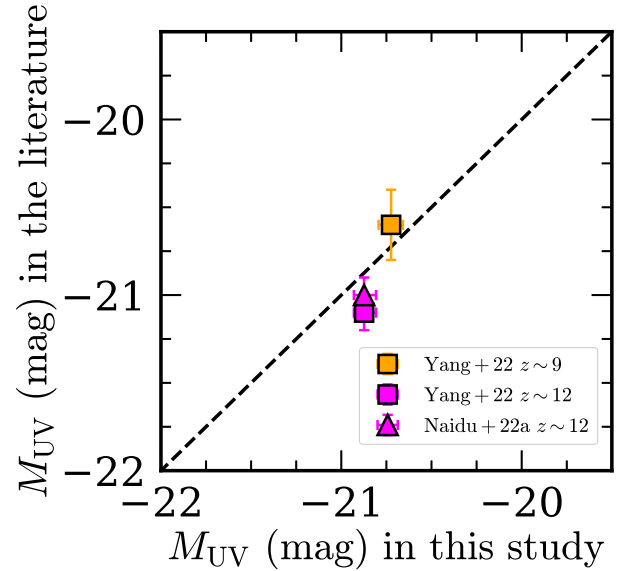}
\caption{
\textbf{Top}: 
Comparison of size measurement results for high-$z$ galaxy candidates on an individual basis in the rest-frame UV.  
The horizontal axis denotes our size measurements and the vertical axis represents those in the literature. 
The orange square compares our results for the bright $z\sim9$ galaxy candidate 
with those in \cite{2022ApJ...938L..17Y}.  
The magenta square and triangle 
show comparisons of our results for the bright galaxy candidate at $z\sim12$ 
with those in \cite{2022ApJ...938L..17Y} and \cite{2022ApJ...940L..14N}, respectively.  
The black dashed line corresponds to the cases that 
our estimates and those in the literature are equal. 
\textbf{Bottom}:
Same as the top panel, 
but for total absolute magnitudes in the rest-frame UV. 
}
\label{fig:comparison_UV}
\end{center}
\end{figure}

The surface brightness profile fittings for two bright candidates in our samples (GL-z9-1 and GL-z12-1) 
have also been conducted in other studies. 
Their sizes and total magnitudes are 
summarized in Table \ref{tab:result_comparison}  
and compared with our results in Figure \ref{fig:comparison_UV}. 
\cite{2022ApJ...938L..17Y} have preformed the surface brightness profile fittings 
for GL-z9-1 and GL-z12-1 
with Galight (\citealt{2020ApJ...888...37D}).
Our size and total magnitude measurement results are broadly consistent with their results 
although they fix the S\'ersic index at $n=1.0$, 
which is slightly different from our fixed value. 
For GL-z12-1, 
\cite{2022ApJ...940L..14N} have also presented size measurement results by using GALFIT.
Their obtained magnitude is roughly consistent, 
but the obtained size is much larger than our results.
A possible reason for this discrepancy is that they use a different filter image. 
They use F444W, while we use F200W. 
However, if we use the F444W image, we still obtain a smaller size than their result.
Another potential factor contributing to the difference is that 
they allow the S\'ersic index to vary as a free parameter in their fitting, 
resulting in a slightly different best-fit value ($n=1.0$) compared to our fixed value. 
However, when we perform the fitting with a fixed S\'ersic index value of $n=1.0$, 
the best-fit size is almost the same with only a marginal increase of about 
$10${\%}.\footnote{Although we also attempt the surface brightness fitting with the S\'ersic index as a free parameter, 
we encounter convergence issues with GALFIT in that case.}
Another possible reason is that 
they use theoretical PSF models generated with WebbPSF.  
This discrepancy may be due to their use of a PSF created with WebbPSF  
that deviates slightly from the actual PSF, 
resulting in their large $r_{\rm e}$ (See Section \ref{sec:data}).

\begin{figure}[h]
\begin{center}
   \includegraphics[width=0.5\textwidth]{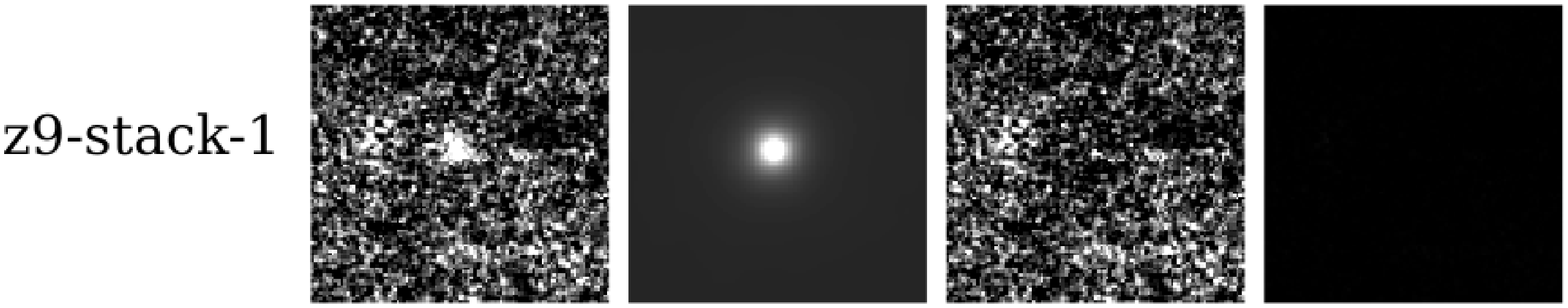}
   \includegraphics[width=0.5\textwidth]{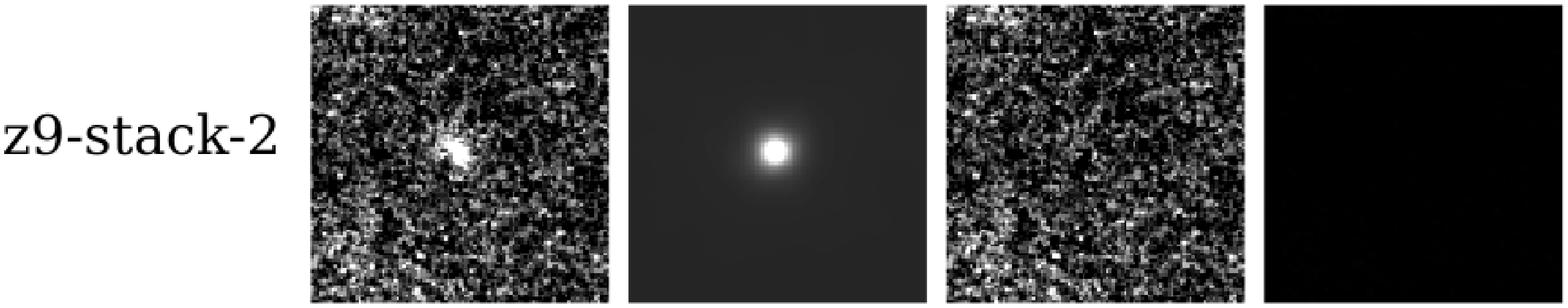}
   \includegraphics[width=0.5\textwidth]{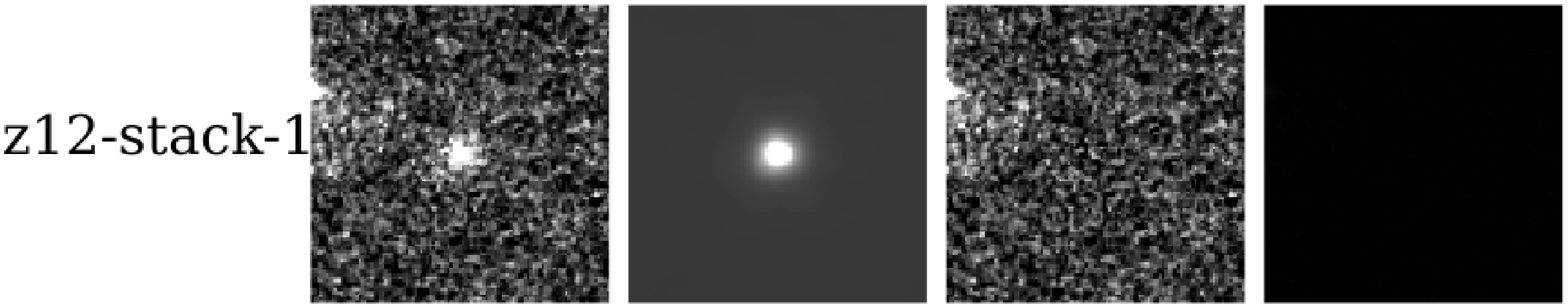}
\caption{
Same as Figure \ref{fig:fit_results_f277w_z17}, 
except that the objects are 
stacked F115W dropouts and F150W dropouts.  
}
\label{fig:fit_results_stack}
\end{center}
\end{figure}

In Figure \ref{fig:fit_results_stack}, 
we show the surface brightness profile fitting results 
for the stacked F115W-dropouts, 
whose UV luminosities are $L/L^\ast_{z=3} = 0.12$--$0.3$ and $0.048$--$0.12$, 
as well as the stacked F150W-dropout 
with UV luminosities $L/L^\ast_{z=3} = 0.12$--$0.3$. 
Because the axis ratios of the stacked objects should be close to unity, 
the axis ratio and position angle are fixed at $1$ and $0$, respectively.\footnote{Because the number of 
F150W-dropouts with $L/L^\ast_{z=3} = 0.12$--$0.3$ is small, 
we also perform the profile fitting with the axis ratio and position angle free, 
and confirm that the results are almost the same.}
The best-fit parameters of the surface brightness profile fitting 
for the stacked objects are also presented 
in Table \ref{tab:z9result} and Table \ref{tab:z12result}, 
where the systematic effects and statistical uncertainties 
in the size and total magnitude measurements are considered based on our MC simulation results.

In summary, our size measurement results indicate that 
most of the $z\sim9$--$16$ galaxy candidates have 
small sizes of around $200$--$500$ pc, 
which is comparable to previous results based on deep HST data 
for high-$z$ galaxies with similar UV luminosities 
(e.g., \citealt{2013ApJ...777..155O}; \citealt{2015ApJS..219...15S}; \citealt{2015ApJ...804..103K}). 
Interestingly, in our samples, 
GL-z12-1 is exceptionally compact with a half-light radius of only $39 \pm 11$ pc. 
We discuss its physical origin in Section \ref{sec:discussion}.

\begin{deluxetable*}{ccccccc} 
\tablecolumns{7} 
\tablewidth{0pt} 
\tablecaption{Surface Brightness Profile Fitting Results for $z \sim 9$ Galaxy Candidates (F115W-dropouts) 
\label{tab:z9result}}
\tablehead{
    \colhead{ID} 
    &  \colhead{$m_{\rm UV}$}
    &  \colhead{$M_{\rm UV}$}
    &  \colhead{$r_{\rm e}$}
    &  \colhead{$m_{\rm opt}$}
    &  \colhead{$M_{\rm opt}$}
    &  \colhead{$r_{\rm e,opt}$}
\\
    \colhead{ } 
    &  \colhead{(mag)}
    &  \colhead{(mag)}
    &  \colhead{(kpc)}
    &  \colhead{(mag)}
    &  \colhead{(mag)}
    &  \colhead{(kpc)}
\\
    \colhead{(1)} 
    &  \colhead{(2)}
    &  \colhead{(3)}
    &  \colhead{(4)}
    &  \colhead{(5)}
    &  \colhead{(6)}
    &  \colhead{(7)}
}
\startdata 
\multicolumn{7}{c}{$L/L^\ast_{z=3} = 0.3$--$1$} \\  
GL-z9-1 & $26.82^{+0.07}_{-0.07}$  &  $-20.72^{+0.07}_{-0.07}$  &  $0.30^{+0.03}_{-0.02}$  & 
$26.30^{+0.02}_{-0.02}$  &  $-21.25^{+0.02}_{-0.02}$  &  $0.28^{+0.02}_{-0.01}$ \\ \hline
\multicolumn{7}{c}{$L/L^\ast_{z=3} = 0.12$--$0.3$} \\  
z9-stack-1 & 
$28.21^{+0.20}_{-0.21}$  &  $-19.12^{+0.20}_{-0.21}$  &  $0.49^{+0.14}_{-0.09}$  & 
$27.89^{+0.09}_{-0.13}$  &  $-19.43^{+0.09}_{-0.13}$  &  $0.71^{+0.16}_{-0.12}$  \\
\hline
\multicolumn{7}{c}{$L/L^\ast_{z=3} = 0.048$--$0.12$} \\  
z9-stack-2 & 
$28.67^{+0.12}_{-0.14}$  &  $-18.65^{+0.12}_{-0.14}$  &  $0.29^{+0.09}_{-0.06}$ & 
$28.99^{+0.15}_{-0.18}$  &  $-18.34^{+0.15}_{-0.18}$  &  $0.24^{+0.15}_{-0.10}$ \\ 
\hline
\enddata 
\tablecomments{The systematic effects and statistical uncertainties 
in these obtained size and total magnitude measurements are considered 
based on our MC simulation results.  
(1) Object ID. 
(2) Total apparent UV magnitude measured by GALFIT. 
(3) Total absolute UV magnitude using $z_{\rm photo}$. 
For the stacked objects, we use $z_{\rm photo} = 9.0$. 
(4) UV circularized half-light radius $r_{\rm e} = a \sqrt{b/a}$, 
where $a$ is the radius along the semi-major axis and $b/a$ is the axis ratio.
(5) Total apparent optical magnitude measured by GALFIT. 
(6) Total absolute optical magnitude using $z_{\rm photo}$. 
For the stacked objects, we use $z_{\rm photo} = 9.0$. 
(7) Optical circularized half-light radius.
}
\end{deluxetable*} 

\begin{deluxetable}{cccc} 
\tablecolumns{4} 
\tablewidth{0pt} 
\tablecaption{Surface Brightness Profile Fitting Results for $z \sim 12$ Galaxy Candidates (F150W-dropouts) 
\label{tab:z12result}}
\tablehead{
    \colhead{ID} 
    &  \colhead{$m_{\rm UV}$}
    &  \colhead{$M_{\rm UV}$}
    &  \colhead{$r_{\rm e}$}
\\
    \colhead{ } 
    &  \colhead{(mag)}
    &  \colhead{(mag)}
    &  \colhead{(kpc)}
\\
    \colhead{(1)} 
    &  \colhead{(2)}
    &  \colhead{(3)}
    &  \colhead{(4)}
}
\startdata 
\multicolumn{4}{c}{$L/L^\ast_{z=3} = 0.3$--$1$} \\  
GL-z12-1 & $26.90^{+0.07}_{-0.06}$  &  $-20.87^{+0.07}_{-0.06}$  &  $0.04^{+0.01}_{-0.01}$  \\ 
CR2-z12-1 & $27.22^{+0.20}_{-0.21}$  &  $-20.45^{+0.20}_{-0.21}$  &  $0.36^{+0.08}_{-0.06}$  \\ \hline
\multicolumn{4}{c}{$L/L^\ast_{z=3} = 0.12$--$0.3$} \\  
CR2-z12-2 & $27.86^{+0.21}_{-0.21}$  &  $-19.88^{+0.21}_{-0.21}$  &  $0.18^{+0.05}_{-0.04}$  \\ 
CR2-z12-3 & $27.59^{+0.12}_{-0.14}$  &  $-20.11^{+0.12}_{-0.14}$  &  $0.28^{+0.08}_{-0.06}$  \\ 
z12-stack-1 & $28.13^{+0.12}_{-0.14}$  &  $-19.61^{+0.12}_{-0.14}$  &  $0.20^{+0.05}_{-0.04}$   \\  \hline
\enddata 
\tablecomments{The systematic effects and statistical uncertainties 
in these obtained size and total magnitude measurements are considered 
based on our MC simulation results.  
(1) Object ID. 
(2) Total apparent UV magnitude measured by GALFIT. 
(3) Total absolute UV magnitude using $z_{\rm photo}$. 
For the stacked objects, we use $z_{\rm photo} = 12.0$. 
(4) Circularized half-light radius $r_{\rm e} = a \sqrt{b/a}$. 
}
\end{deluxetable} 

\begin{deluxetable}{cccc} 
\tablecolumns{4} 
\tablewidth{0pt} 
\tablecaption{Surface Brightness Profile Fitting Results for $z \sim 16$ Galaxy Candidates (F200W-dropouts) 
\label{tab:z17result}}
\tablehead{
    \colhead{ID} 
    &  \colhead{$m_{\rm UV}$}
    &  \colhead{$M_{\rm UV}$}
    &  \colhead{$r_{\rm e}$}
\\
    \colhead{ } 
    &  \colhead{(mag)}
    &  \colhead{(mag)}
    &  \colhead{(kpc)}
\\
    \colhead{(1)} 
    &  \colhead{(2)}
    &  \colhead{(3)}
    &  \colhead{(4)}
}
\startdata 
S5-z16-1 & $26.56^{+0.18}_{-0.24}$  &  $-21.62^{+0.18}_{-0.24}$  &  $0.15^{+0.09}_{-0.06}$  
\enddata 
\tablecomments{The systematic effects and statistical uncertainties 
in these obtained size and total magnitude measurements are considered 
based on our MC simulation results.  
(1) Object ID. 
(2) Total apparent UV magnitude measured by GALFIT. 
(3) Total absolute UV magnitude using $z_{\rm photo}$. 
(4) Circularized half-light radius $r_{\rm e} = a \sqrt{b/a}$. 
}
\end{deluxetable} 

\begin{deluxetable*}{ccccccccc} 
\tablecolumns{9} 
\tablewidth{0pt} 
\tablecaption{Comparison of Our Profile Fitting Results with Previous Results 
\label{tab:result_comparison}}
\tablehead{
    \colhead{ID} 
    &  \colhead{Filter$_{\rm UV}$}
    &  \colhead{$M_{\rm UV}$}
    &  \colhead{$r_{\rm e}$}
    &  \colhead{Filter$_{\rm opt}$}
    &  \colhead{$M_{\rm opt}$}
    &  \colhead{$r_{\rm e,opt}$}
    &  \colhead{PSF}
    &  \colhead{Reference}
\\
    \colhead{ } 
    &  \colhead{ }
    &  \colhead{(mag)}
    &  \colhead{(kpc)}
    &  \colhead{ }
    &  \colhead{(mag)}
    &  \colhead{(kpc)}
    &  \colhead{ }
    &  \colhead{ }
\\
    \colhead{(1)} 
    &  \colhead{(2)}
    &  \colhead{(3)}
    &  \colhead{(4)}
    &  \colhead{(5)}
    &  \colhead{(6)}
    &  \colhead{(7)}
    &  \colhead{(8)}
    &  \colhead{(9)}
}
\startdata 
GL-z9-1 		& F150W 	& $-20.72^{+0.07}_{-0.07}$	& $0.30^{+0.03}_{-0.02}$ 	& F444W 	& $-21.25^{+0.02}_{-0.02}$  	& $0.28^{+0.02}_{-0.01}$ 	& empirical	& This Study \\ 
560 			& F150W 	& $-20.6 \pm 0.20$  			& $0.37 \pm 0.03$		& F444W 	& $-21.2 \pm 0.05$  			& $0.50 \pm 0.02$ 		& empirical	& \cite{2022ApJ...938L..17Y} \\ 
GL-z10		& --- 		& ---						& --- 					& F444W 	& $-20.7 \pm 0.2$ 			& $0.7$ 				& WebbPSF	& \cite{2022ApJ...940L..14N} \\ \hline
GL-z12-1		& F200W 	& $-20.87^{+0.07}_{-0.06}$  	& $0.04^{+0.01}_{-0.01}$  & --- 		& --- 						& --- 					& empirical	& This Study \\ 
5153			& F200W	& $-21.1 \pm 0.1$			& $0.12 \pm 0.01$		& --- 		& --- 						& --- 					& empirical	& \cite{2022ApJ...938L..17Y} \\ 
GL-z12		& F444W	& $-21.0 \pm 0.1$			& $0.5$				& --- 		& --- 						& --- 					& WebbPSF	& \cite{2022ApJ...940L..14N} \\ \hline
\enddata 
\tablecomments{(1) Object ID. 
(2) Filter used for the rest-frame UV continuum profile fitting. 
(3) Total absolute UV magnitude. 
(4) Half-light radius in the rest-frame UV.
(2) Filter used for the rest-frame optical continuum profile fitting. 
(6) Total absolute optical magnitude. 
(7) Half-light radius in the rest-frame optical.
(8) Method adopted in making PSFs. 
(9) Reference. 
}
\end{deluxetable*} 

\subsection{Surface Brightness Profile Fitting Results for the Rest-frame Optical Continuum} 

\begin{figure}[ht]
\begin{center}
   \includegraphics[width=0.5\textwidth]{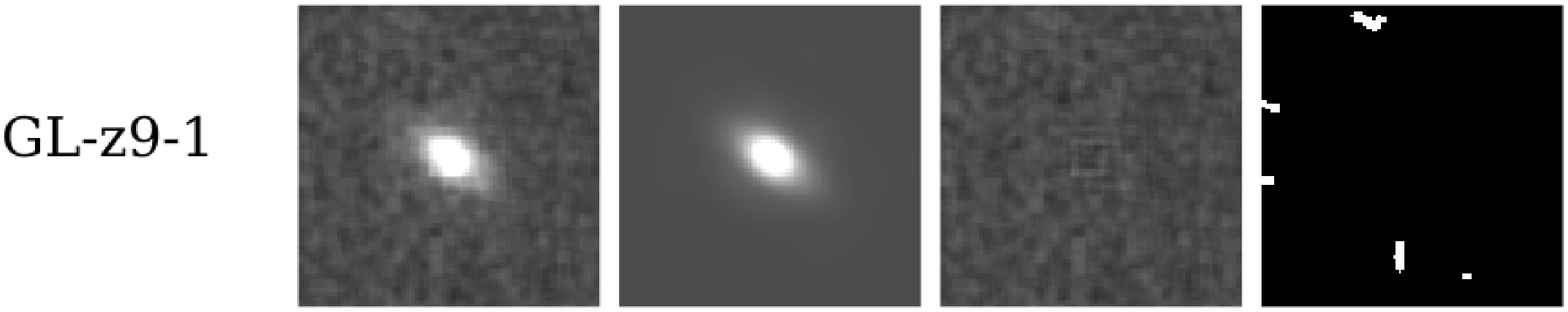}
   \includegraphics[width=0.5\textwidth]{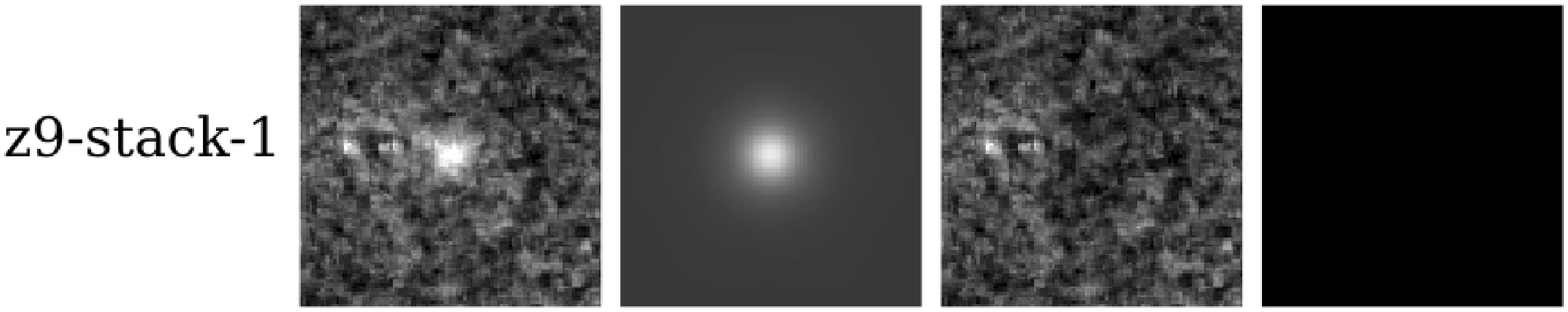}
   \includegraphics[width=0.5\textwidth]{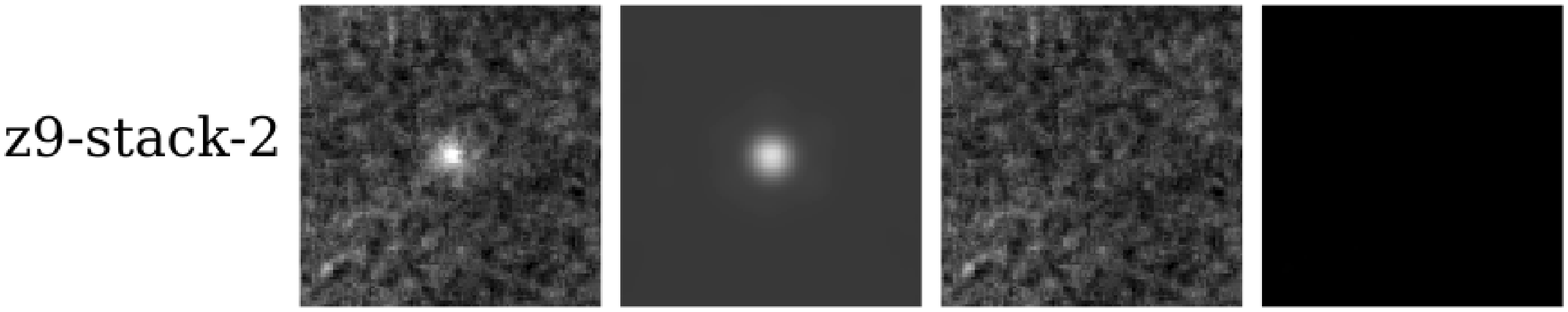}
\caption{
Same as Figure \ref{fig:fit_results_f150w_z9}, 
but in F444W for GL-z9-1 and the stacked F115W dropouts. 
}
\label{fig:fit_results_f444w_z9}
\end{center}
\end{figure}

For $z\sim9$ galaxy candidates, 
we perform the surface brightness profile fitting with the F444W images 
to obtain their sizes and total magnitudes in the rest-frame optical 
in the same manner as in the rest-frame UV. 
In Figure \ref{fig:fit_results_f444w_z9}, we present the results of the profile fitting 
for GL-z9-1, which is the only object in our sample showing an S/N $>10$ in F444W.  
In the same figure, we also show the profile fitting results for the stacked F115W-dropouts, 
whose UV luminosities are $L/L^\ast_{z=3} = 0.12$--$0.3$ and $0.048$--$0.12$. 
Their best-fit parameters of the profile fitting are also presented in Table \ref{tab:z9result}.

\begin{figure}[h]
\begin{center}
   \includegraphics[width=0.45\textwidth]{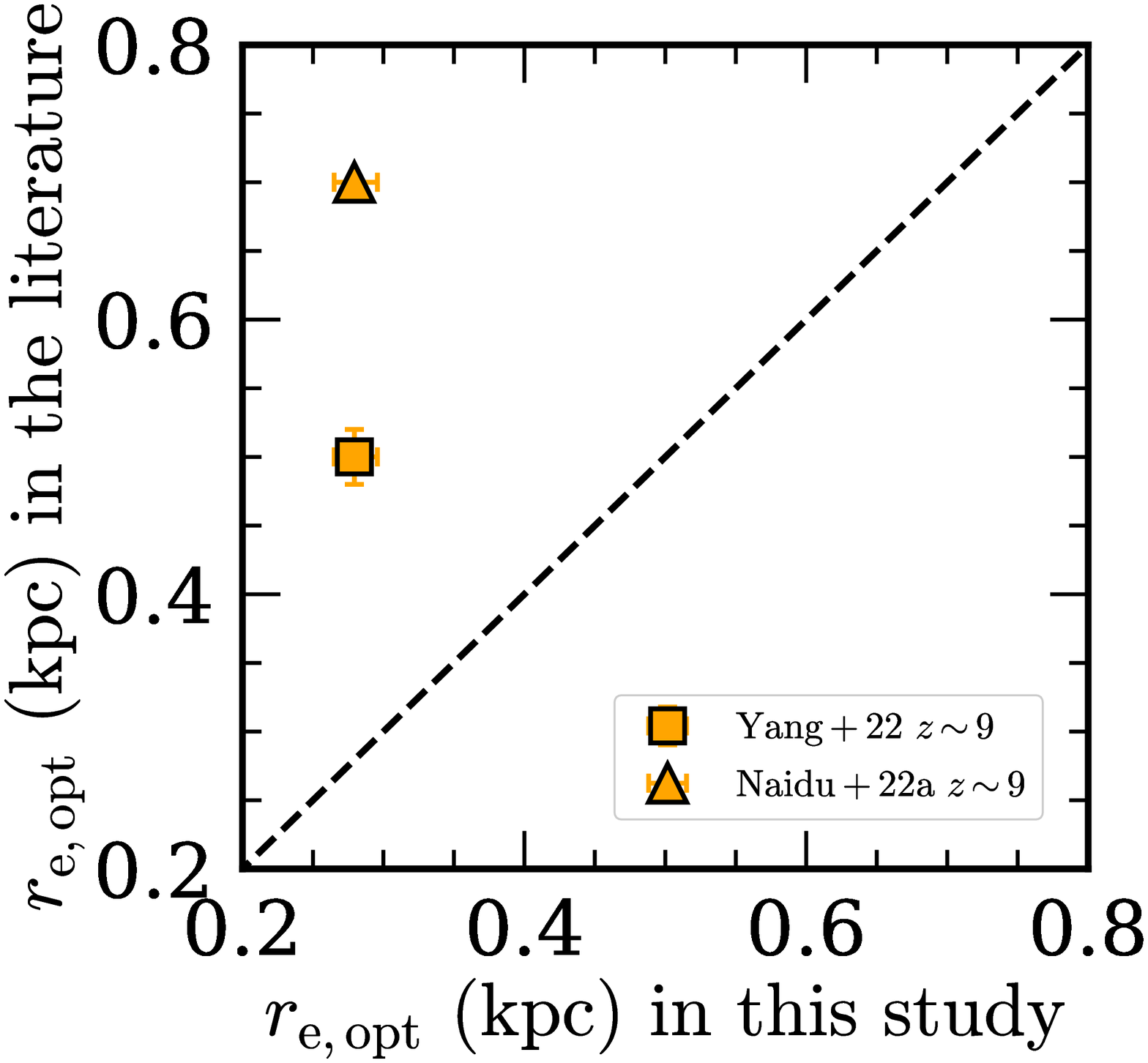}
   \includegraphics[width=0.45\textwidth]{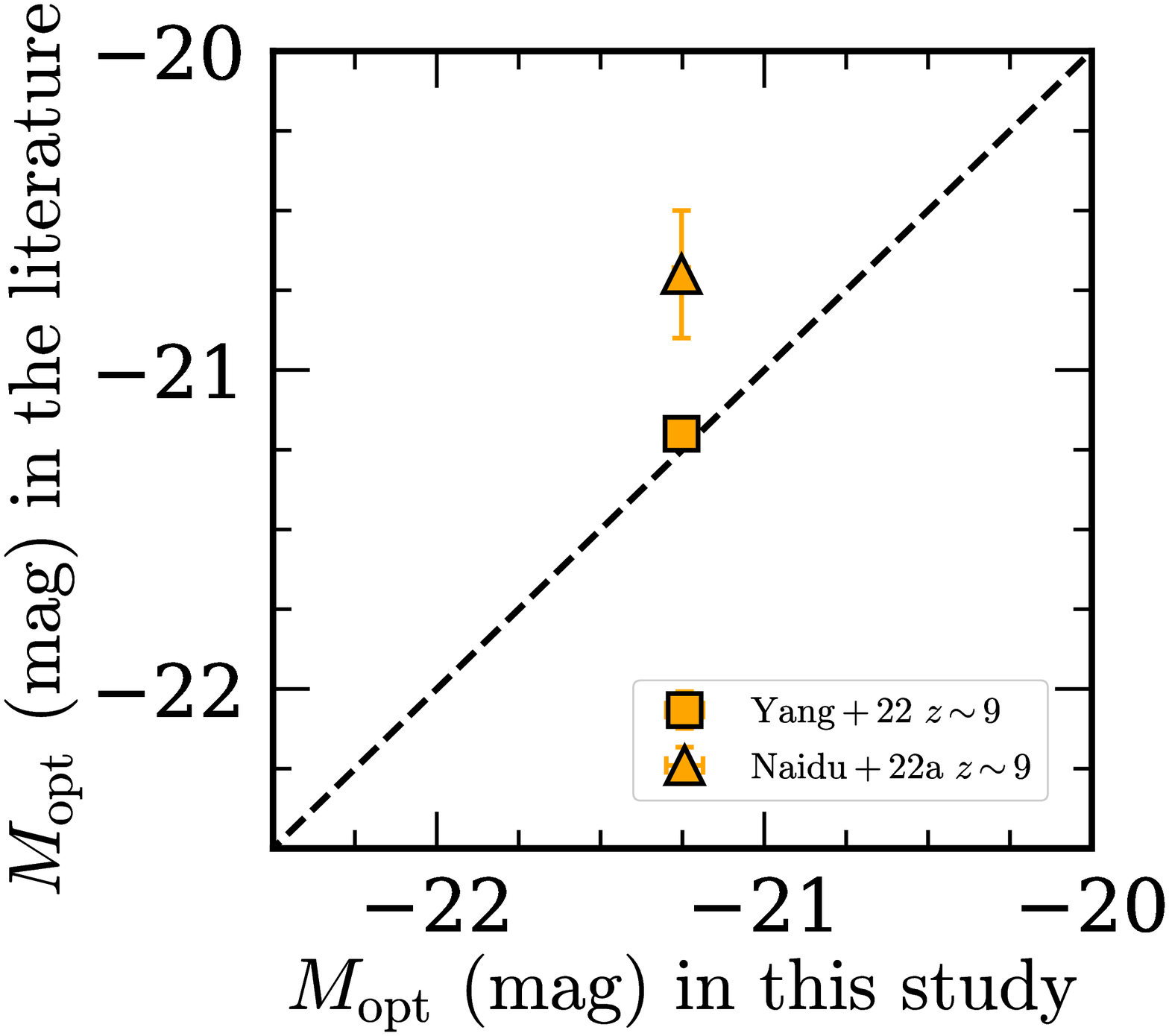}
\caption{
Same as Figure \ref{fig:comparison_UV}, 
but for the rest-frame optical. 
}
\label{fig:comparison_opt}
\end{center}
\end{figure}

The rest-frame optical continuum profile fitting for GL-z9-1 
has also been carried out in previous work 
as presented in Table \ref{tab:result_comparison} (See also Figure \ref{fig:comparison_opt}). 
Although our obtained size is smaller than that in \cite{2022ApJ...938L..17Y}, 
our obtained total magnitude is in good agreement with their result. 
\cite{2022ApJ...940L..14N} have also performed the profile fitting for GL-z9-1. 
Although their estimated size is much larger than our result, 
their obtained total magnitude is broadly consistent with our obtained value. 
A possible reason for the obtained size difference is the PSF difference as described above. 
Another possible reason is that 
the S\'ersic index is a free parameter in their profile fitting, 
although the best-fit S\'ersic index value ($n=0.8$ for GL-z9-1) 
is a disk-like value.

\begin{figure}[h]
\begin{center}
   \includegraphics[width=0.47\textwidth]{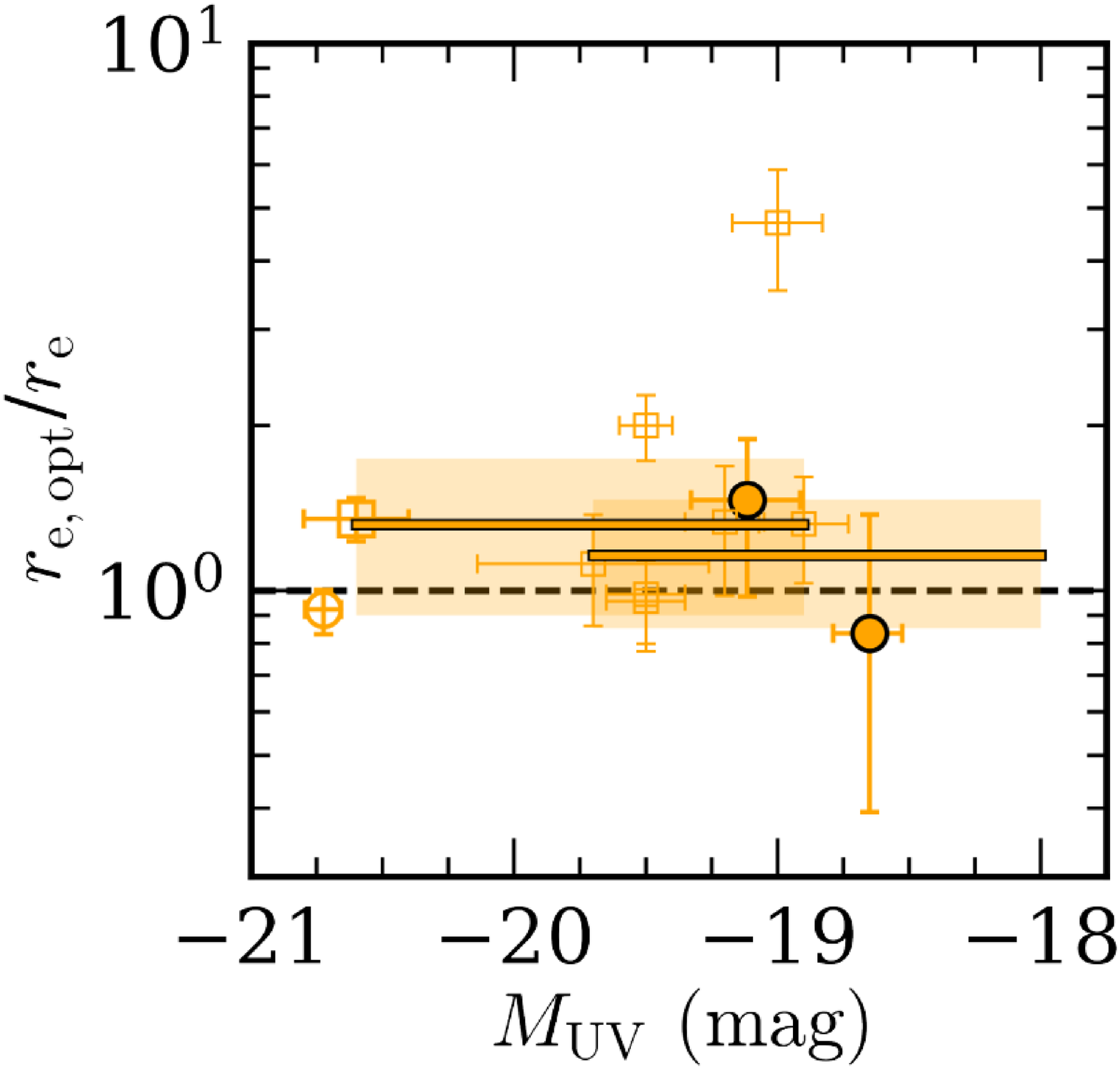}
\caption{
Ratio of size in the rest-frame optical to that in the rest-frame UV 
as a function of $M_{\rm UV}$. 
The large open circle denotes our results for GL-z9-1 
and the large filled circles corresponds to our results for stacked $z\sim 9$ galaxy candidates 
with $L/L^\ast_{z=3} = 0.12$--$0.3$ and $0.048$--$0.12$. 
The orange solid line and the orange shade in the faint magnitude range 
($-19.7 \lesssim M_{\rm UV} \lesssim -18.0$ mag)  
are the average size ratio of these two stacked objects and its uncertainty, respectively. 
The open squares are the previous results for $z\sim7$--$9$ galaxy candidates 
with size measurements in F150W and F444W (\citealt{2022ApJ...938L..17Y}). 
The large open square represents their results for GL-z9-1.
The orange solid line in the brighter magnitude range 
($-20.6 \lesssim M_{\rm UV} \lesssim -18.9$ mag) 
is the median size ratio presented in \cite{2022ApJ...938L..17Y}. 
The black dashed line denotes the case when the size ratio is unity. 
}
\label{fig:re_ratio}
\end{center}
\end{figure}

Figure \ref{fig:re_ratio} presents the size ratio of our $z\sim 9$ galaxy candidates 
between the rest-frame UV and optical as a function of $M_{\rm UV}$. 
We also plot the results of \cite{2022ApJ...938L..17Y}, 
because they have examined the size ratios for high-$z$ galaxies down to slightly lower redshifts with JWST. 
\cite{2022ApJ...938L..17Y} have reported that 
their $z\sim7$--$9$ galaxy candidates have $1.32 \pm 0.42$ 
times larger sizes 
in the rest optical than in the rest UV. 
Although their size ratio is consistent with unity within the $1\sigma$ uncertainties, 
if the size ratio is larger than unity, 
it would indicate that 
relatively old stars in $z\sim 7$--$9$ bright galaxies already have disk-like structures, 
and their on-going star formation is occurring in smaller regions around their center. 
Note that, in the case of these galaxies containing a considerable amount of dust, 
their size ratio may be explained by dust  
because the UV emission is relatively strongly absorbed.  
However, most of our candidates show blue UV slopes (i.e., $F150W-F277W < 0$, 
see Table 3 of \citealt{2023ApJS..265....5H}), 
suggesting that dust extinction is not very effective in these galaxies. 
Accurate measurements on their dust extinction can be obtained 
from the flux ratios of the Balmer emission lines by spectroscopy in the future.

Thanks to the stacking analysis, we investigate the size ratio 
down to a fainter magnitude range compared to the individual analyses. 
We find that the average size ratio is 
$r_{\rm e,opt}/r_{\rm e} = 1.16 \pm 0.31$ 
for our stacked $z\sim9$ galaxy candidates 
with the faint luminosity bins of $L/L^\ast_{z=3} = 0.12$--$0.3$ and $0.048$--$0.12$. 
Although our results for the faint stacked objects are consistent with 
those for brighter objects within the $1\sigma$ uncertainties, 
if the size ratio for the faint galaxies is smaller than that for the bright ones, 
it may indicate that the distributions of young massive and old less massive stars are not so different 
probably because of more recent star formation activities in faint galaxies.  
This point needs to be investigated for larger samples in future studies.

\begin{figure*}[ht]
\begin{center}
   \includegraphics[height=27em]{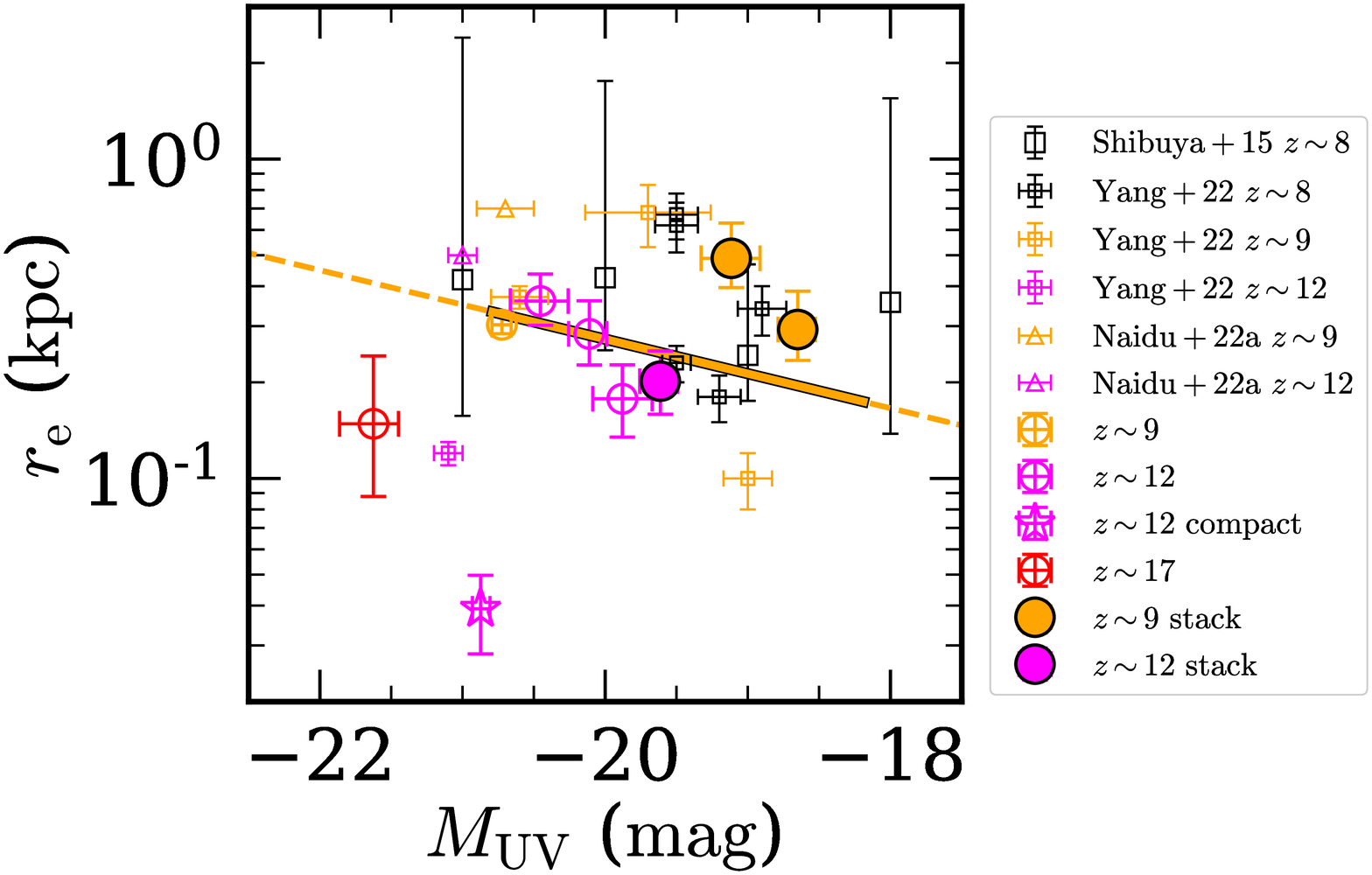}
   \includegraphics[height=27em]{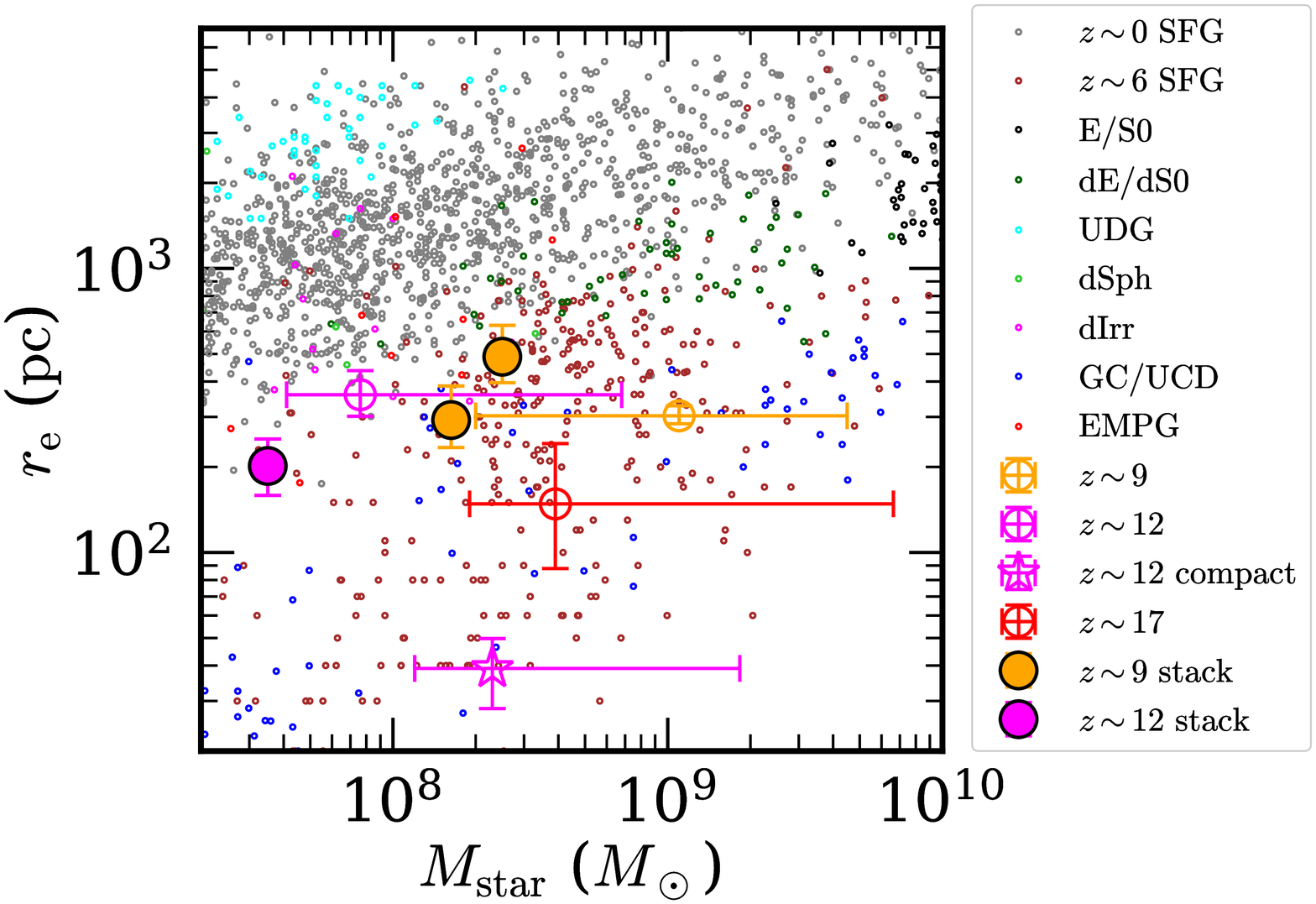}
\caption{
\textbf{Top}: 
Size--luminosity relation for $z\sim9$--$16$ galaxy candidates. 
The orange, magenta, and red large open circles 
are bright objects in our $z\sim9$, $z\sim12$, and $z\sim16$ galaxy candidate samples, respectively. 
The magenta open star denotes the very compact galaxy candidate, GL-z12-1.  
The orange (magenta) filled circles 
correspond to the stacked objects of our faint $z\sim9$ ($z\sim12$) galaxy candidates. 
The orange solid line is the best-fit size-luminosity relation for our $z\sim9$ galaxy candidates, 
and the orange dashed line is its extrapolation. 
The small orange and magenta symbols (open squares and triangles) 
denote previous JWST results for $z\sim9$ and $z\sim12$ galaxy candidates, respectively 
(\citealt{2022ApJ...938L..17Y}; \citealt{2022ApJ...940L..14N}).  
The black large open squares are previous results for $z\sim8$ galaxy candidates 
and their large error bars denote the $16$th and $84$th percentiles of the individual size distribution 
(\citealt{2015ApJS..219...15S}). 
The black small open squares are previous JWST results for galaxy candidates at $z<8.5$ 
in \cite{2022ApJ...938L..17Y}. 
\textbf{Bottom}: 
Size as a function of stellar mass for $z\sim9$--$16$ galaxy candidates. 
The large orange, magenta, and red symbols are the same as those in the top panel. 
The stellar masses of our stacked objects are calculated 
by using the stellar mass-to-luminosity ratios for the bright candidates. 
The other small symbols are compiled by \cite{2021ApJ...918...54I};  
the gray and brown dots are star-forming galaxies at $z\sim0$ and $z\sim6$, respectively 
(\citealt{2015ApJS..219...15S}; \citealt{2020ApJ...893...60K}); 
the black dots are local ellipticals (E/S0; \citealt{2014MNRAS.443.1151N}); 
the dark green dots are local dwarf ellipticals (dE/dS0; \citealt{2014MNRAS.443.1151N}); 
the cyan dots are local ultra diffuse galaxies (UDGs; \citealt{2015ApJ...798L..45V}; \citealt{2020JCAP...01..059H}); 
the light green dots are local dwarf spheroidals (dSph; \citealt{2012AJ....144....4M}); 
the magenta dots are local dwarf irregulars (dIrr; \citealt{2012AJ....144....4M}); 
the blue dots are globular clusters and local ultra compact dwarfs (GC/UCD; \citealt{2014MNRAS.443.1151N});  
and 
the red dots are local extremely metal-poor galaxies (EMPGs; \citealt{2021ApJ...918...54I}). 
}
\label{fig:re_Muv}
\end{center}
\end{figure*}

\subsection{Size -- UV Luminosity Relation } 

We investigate the size--luminosity ($r_{\rm e}$--$L_{\rm UV}$) relation 
for our high-$z$ galaxy candidates. 
The top panel of Figure \ref{fig:re_Muv} presents the size--luminosity relation 
for our $z \sim 9$ galaxy candidates. 
The $z\sim9$ data points are broadly consistent with the previous results 
for $z\sim8$ galaxies within the $1\sigma$ uncertainties. 
Our results for $z \sim 12$--$16$ are also presented, 
although the UV magnitude ranges that can be examined with the current samples are limited.

Following the previous work, 
we fit a power-law function to the data points, 
\begin{equation}
r_{\rm e}
	= r_0 \left( \dfrac{L_{\rm UV}}{L_0} \right)^\alpha, 
\end{equation}
where $r_0$ is the effective radius at the luminosity of $L_0$, 
corresponding to $- 21.0$ mag 
(\citealt{2013ApJ...765...68H}; \citealt{2015ApJS..219...15S}),
and $\alpha$ is the slope of the size--luminosity relation. 
Since the previous work has shown that 
the slope value is almost constant, 
$\alpha = 0.27 \pm 0.01$, over a wide redshift range of $z = 0$--$8$ 
(\citealt{2015ApJS..219...15S}), 
we fix it at this value and perform the fitting 
with $r_0$ as the only free parameter. 
As a result, we obtain $r_0 = 0.35 \pm 0.02$ kpc, 
which is broadly consistent with the best-fit function of 
$r_0 \propto (1+z)^{\beta_z}$ presented in Figure 10 of \cite{2015ApJS..219...15S}.

We also fit the size--luminosity relation in the same manner 
to the $z\sim12$ data points, 
although the observed range of UV magnitudes is limited. 
Note that the data point of GL-z12-1 with the exceptionally small size 
is not used for the fitting. 
The obtained $r_0$ value is $r_0 = 0.32 \pm 0.03$ kpc, 
which is also consistent with the extrapolation of 
the best-fit function of \cite{2015ApJS..219...15S}.

In addition to the UV luminosity, 
the relation between stellar mass and size is also interesting for comparisons with previous results. 
In the bottom panel of Figure \ref{fig:re_Muv}, 
we show our bright $z\sim9$--$16$ galaxy candidates whose stellar masses can be estimated. 
The stellar masses for our bright high-$z$ galaxy candidates are estimated 
with the flexible Bayesian inference spectral energy distribution (SED) fitting code PROSPECTOR 
(\citealt{2021ApJS..254...22J}; for details, see Section 3.3 of \citealt{2023ApJS..265....5H}). 
The stellar masses of our stacked objects are calculated 
with the stellar mass-to-luminosity ratios of the bright candidates in the same dropout samples.  
We also plot lower-$z$ star-forming galaxies including local objects investigated in the literature. 
We find that our $z\sim9$--$16$ galaxy candidates 
have similar sizes and stellar masses 
to those of $z\sim6$ star-forming galaxies 
(\citealt{2015ApJS..219...15S}; \citealt{2020ApJ...893...60K})
and local ultra compact dwarfs (UCDs; \citealt{2014MNRAS.443.1151N}).

\subsection{Size Evolution } \label{subsec:size_evolution}

\begin{figure*}[ht]
\begin{center}
   \includegraphics[height=29em]{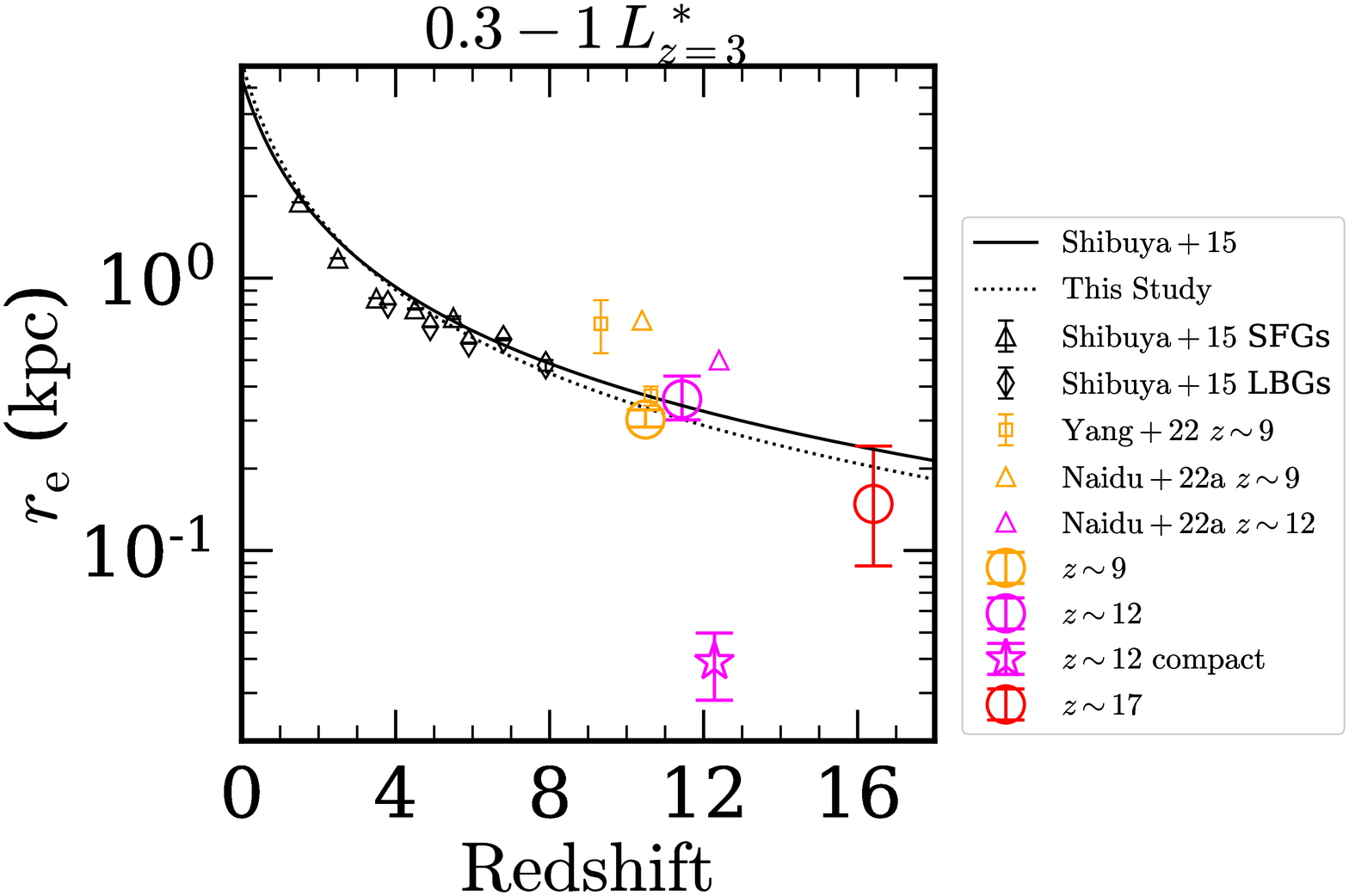}
   \includegraphics[height=29em]{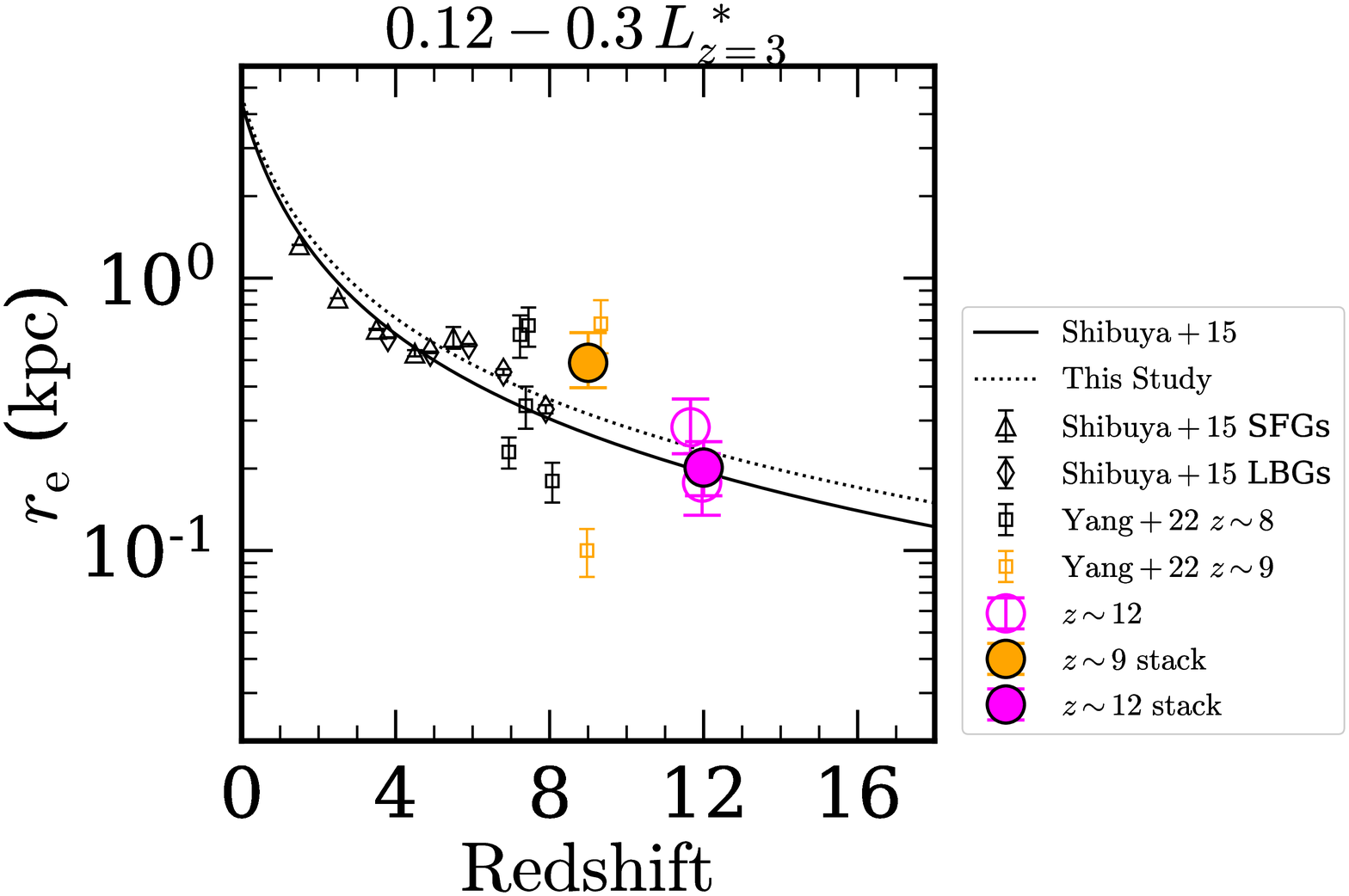}
\caption{
Evolution of the half-light radius of star-forming galaxies 
with UV luminosities of ($0.3$--$1$) $L^\ast_{z=3}$ (top) and ($0.12$--$0.3$) $L^\ast_{z=3}$ (bottom).
The orange, magenta, and red large open circles 
show bright objects in our $z\sim9$, $z\sim12$, and $z\sim16$ galaxy candidate samples, respectively. 
The magenta open star denotes the very compact galaxy candidate, GL-z12-1.  
The orange (magenta) filled circle 
denotes the stacked object of our faint $z\sim9$ ($z\sim12$) galaxy candidates. 
The small orange and magenta symbols (open squares and triangles) 
denote previous JWST results for $z\sim9$ and $z\sim12$ galaxy candidates, respectively 
(\citealt{2022ApJ...938L..17Y}; \citealt{2022ApJ...940L..14N}).  
The black small open squares are previous JWST results for galaxy candidates at $z<8.5$ 
in \cite{2022ApJ...938L..17Y}. 
The black open triangles and diamonds are previous results for lower-$z$ 
star-forming galaxies (SFGs) and Lyman break galaxies (LBGs), respectively 
(\citealt{2015ApJS..219...15S}). 
The solid and dotted curves correspond to the best-fit functions of 
$r_{\rm e} \propto (1+z)^s$ 
obtained in \cite{2015ApJS..219...15S} and in this study, respectively. 
}
\label{fig:re_redshift}
\end{center}
\end{figure*}

The previous studies have shown that 
the galaxy size evolution can be characterized with a functional form of $r_{\rm e} \propto (1+z)^s$ 
(e.g., \citealt{2004ApJ...600L.107F}; 
\citealt{2004ApJ...611L...1B}; 
\citealt{2006ApJ...653...53B}; 
\citealt{2008ApJ...673..686H}; 
\citealt{2010ApJ...709L..21O}; 
\citealt{2013ApJ...777..155O}; 
\citealt{2015ApJ...804..103K}; 
\citealt{2015ApJ...808....6H}; 
\citealt{2015ApJS..219...15S}; 
\citealt{2016MNRAS.457..440C}; 
\citealt{2017ApJ...834L..11A}; 
\citealt{2018ApJ...855....4K}; 
\citealt{2019ApJ...882...42B}; 
\citealt{2020AJ....160..154H}), 
where $s$ is the slope of the size evolution.  
Here we extend the previous work 
by adding the new JWST measurement results. 
Following the previous work, 
because of the size dependence on the luminosity, 
we compare the half-light radii of our high-$z$ galaxy candidates within fixed magnitude ranges. 
In Figure \ref{fig:re_redshift}, 
we present the half-light radii as a function of redshift 
for our high-$z$ galaxy candidates 
with $L/L^\ast_{z=3} = 0.3$--$1$ and $0.12$--$0.3$, 
as well as the HST results for $z = 0$--$8$ star-forming galaxies 
presented in \cite{2015ApJS..219...15S} 
and the JWST results for $z = 9$--$12$ galaxy candidates 
reported in \cite{2022ApJ...938L..17Y} and \cite{2022ApJ...940L..14N}.

Since the number of galaxies whose sizes are measured with JWST is limited compared to the previous HST studies, 
we first compare our results with the extrapolations of the size evolution based on the previous HST results. 
As can be seen from the top panel of Figure \ref{fig:re_redshift}, 
for $L/L^\ast_{z=3} = 0.3$--$1$, 
our results at $z \sim 10$--$16$ are broadly consistent with the extrapolation of the HST results 
within the $1\sigma$ uncertainties. 
Based on size measurement results up to higher redshifts with JWST,  
we find that the size evolution of galaxies with $L/L^\ast_{z=3} = 0.3$--$1$ 
shows almost the same trend beyond $z>10$. 
For the fainter luminosity bin of $L/L^\ast_{z=3} = 0.12$--$0.3$, 
as shown in the bottom panel of Figure \ref{fig:re_redshift}, 
the sizes of the stacked objects of $z\sim9$ and $z\sim12$ galaxy candidates  
are also roughly consistent with the extrapolations of the best-fit function obtained in the previous work (solid curve).
Note that GL-z12-1 at $z\sim12$ has an exceptionally compact size 
compared to the other candidates 
at similar redshifts and the extrapolation of the HST results. 
We discuss possible physical interpretations of its very compact size in Section \ref{sec:discussion}.

Next, although the number of our high-$z$ galaxy candidates with size measurements available is small, 
we fit the functional form of $r_{\rm e} \propto (1+z)^s$ to the data points 
by taking into account our results 
for GL-z9-1 and CR2-z12-1 for the brighter bin ($L/L^\ast_{z=3} = 0.3$--$1$), 
and CR2-z12-2, CR2-z12-3, z9-stack-1, and z12-stack-1 for the fainter bin ($L/L^\ast_{z=3} = 0.12$--$0.3$) 
as well as the previous HST results.  
We do not use the result for GL-z12-1, because it is exceptionally compact 
and may be affected by a possible faint active galactic nucleus (AGN) contribution 
as discussed in Section \ref{sec:discussion}. 
The $z\sim16$ galaxy candidate is also not used, 
because it is relatively luminous ($L/L^\ast_{z=3}  \simeq 2$). 
As in \cite{2015ApJS..219...15S}, for the uncertainties of the data points of \cite{2015ApJS..219...15S}, 
we use the $68$th percentiles of the individual size distributions.  
For the error bars of our data points, 
we use the individual size measurement uncertainties 
because of the limited number of our high-$z$ galaxy candidates whose sizes are measured. 
The obtained slope value for $L/L^\ast_{z=3} = 0.3$--$1$ is 
$s = -1.22^{+0.17}_{-0.16}$ 
and that for the fainter luminosity bin of $L/L^\ast_{z=3} = 0.12$--$0.3$ is 
$s = -1.17^{+0.16}_{-0.16}$,  
which are consistent with the previous results for $z=0$--$8$ galaxies, 
as expected from the good agreement of the data points with the extrapolation of the previously obtained size evolution. 
These fittings should be improved by obtaining size measurements 
for a larger sample of high-$z$ galaxies through upcoming deep JWST observations.

\begin{figure*}[ht]
\begin{center}
   \includegraphics[width=0.7\textwidth]{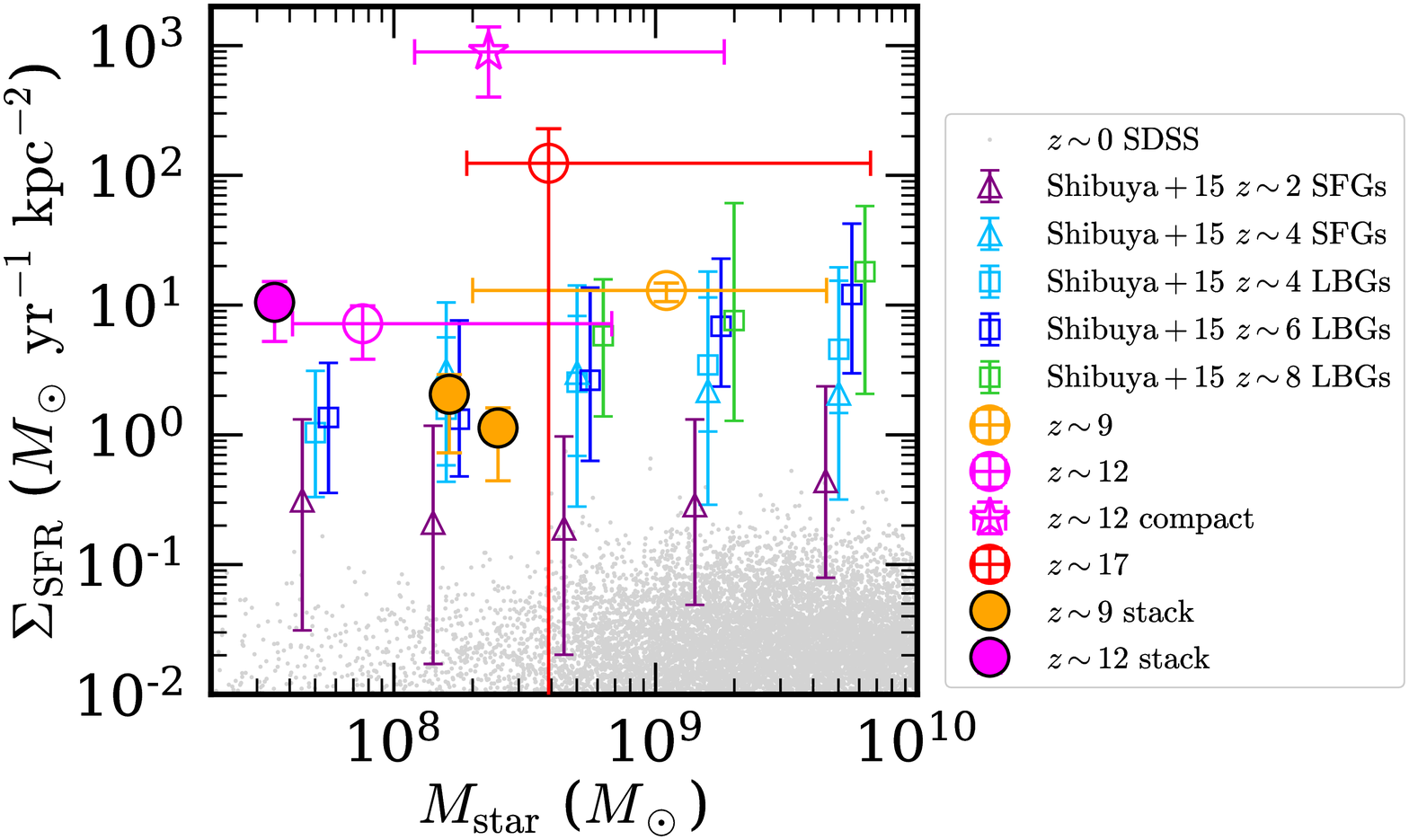}
\caption{
SFR surface density $\Sigma_{\rm SFR}$ vs. stellar mass $M_{\rm star}$. 
The orange, magenta, and red open circles 
show bright objects in our $z\sim9$, $z\sim12$, and $z\sim16$ galaxy candidate samples, respectively. 
The orange (magenta) filled circle 
denotes the stacked object of our faint $z\sim9$ ($z\sim12$) galaxy candidates. 
The open triangles indicate  
SFGs at $z\sim2$ (purple) and $z\sim4$ (cyan). 
The open squares denote 
LBGs at $z\sim4$ (cyan), $z\sim6$ (blue), and $z\sim8$ (green). 
The gray dots are SDSS galaxies compiled by \cite{2015ApJS..219...15S}; 
the $\Sigma_{\rm SFR}$ values are calculated based on the catalog of \cite{2012MNRAS.421.2277L}, 
and the stellar masses for the SDSS galaxies are taken from 
\cite{2003MNRAS.341...33K}, \cite{2004MNRAS.351.1151B}, and \cite{2007ApJS..173..267S}. 
}
\label{fig:SigmaSFR_SFR}
\end{center}
\end{figure*}

\subsection{SFR Surface Density} 

We compare the SFR surface densities, $\Sigma_{\rm SFR}$, 
of our high-$z$ galaxy candidates with previous results for lower-$z$ galaxies. 
The SFRs are calculated by using Equation (1) of \cite{1998ApJ...498..541K}, 
\begin{equation}
{\rm SFR} 
	= 1.4 \times 10^{-28} \alpha_{\rm SC} L_\nu, 
\end{equation}
where $L_\nu$ is the rest UV luminosity density in units of erg s$^{-1}$ Hz$^{-1}$. 
We multiply by $\alpha_{\rm SC} = 0.63$ (\citealt{2014ARA&A..52..415M})
to convert from the Salpeter initial mass function (IMF; \citealt{1955ApJ...121..161S}) 
to the Chabrier IMF (\citealt{2003PASP..115..763C}). 
We then calculate the SFR surface density $\Sigma_{\rm SFR}$ 
in units of $M_\odot$ yr$^{-1}$ kpc$^{-2}$ 
as the average SFR in a circular region whose half-light radius is $r_{\rm e}$, 
\begin{equation}
\Sigma_{\rm SFR} 
	= \dfrac{\rm SFR}{2 \pi r_{\rm e}^2}. 
\end{equation}
The multiplicative factor $1/2$ is applied 
because the SFR is estimated from the total luminosity 
while the area is calculated with the half-light radius 
(e.g., \citealt{2008ApJ...673..686H}; \citealt{2013ApJ...768...74T}; \citealt{2016ApJ...833...70D}).

In Figure \ref{fig:SigmaSFR_SFR}, 
we plot the $\Sigma_{\rm SFR}$ values as a function of stellar mass 
for  our high-$z$ galaxy candidates and star-forming galaxies at lower redshifts. 
At lower redshifts of $z=0$--$8$, 
the average $\Sigma_{\rm SFR}$ values increase with redshift 
at a fixed stellar mass, as reported in Section 5.3 of \cite{2015ApJS..219...15S}. 
Our study extends it toward higher redshifts.  
At fixed stellar masses of $\sim 10^{8-9} M_\odot$, 
we find that our high-$z$ galaxy candidates have consistent or higher $\Sigma_{\rm SFR}$ values 
compared to $z\sim6$--$8$ star-forming galaxies. 
In particular, the very compact GL-z12-1 has an exceptionally high $\Sigma_{\rm SFR}$, 
about $900$ $M_\odot$ yr$^{-1}$ kpc$^{-2}$.

\section{Discussion} \label{sec:discussion}

\subsection{Very Compact Star Formation at $z\sim12$} \label{subsec:discuss_compact}

As presented in Section \ref{sec:results}, 
the size of GL-z12-1 is very compact 
compared to the other candidates 
and those expected from the size evolution based on the previous HST studies. 
There are two possible physical origins of such galaxies with very compact morphologies.

One possible physical origin is very compact star formation.
Qualitatively, very compact star formation can happen at high redshifts 
because collapsed haloes are more compact and dense at higher redshifts.
Recent 3D radiation hydrodynamic simulations of star cluster formation 
have suggested that, 
when an initial gas surface density is sufficiently high, 
most of the gas in molecular gas clouds 
can be converted into stars with very high star formation efficiencies 
thanks to the inefficiency of star formation suppression by ionizing radiation in high density regions 
(\citealt{2021MNRAS.506.5512F}; see also, \citealt{2018ApJ...859...68K}; \citealt{2020MNRAS.497.3830F}), 
and such objects may be observed as very compact star forming galaxies. 
In fact, the $\Sigma_{\rm SFR}$ value of GL-z12-1 is very high, about $900$ $M_\odot$ yr$^{-1}$ kpc$^{-2}$. 
If it hosts very compact star formation, 
it may correspond to actively forming young massive star clusters in a compact region 
some of which would evolve into present-day globular clusters (e.g., \citealt{2014CQGra..31x4006K}).
For lower mass galaxies, recent theoretical studies on globular cluster formation 
based on cosmological simulations have been reported in \cite{2022arXiv220400638S}. 
It would be interesting to compare with these theoretical results 
if high-$z$ galaxies with similar low masses are found in future JWST observations 
such as for galaxy cluster regions where gravitational lensing effects are strong.

\begin{figure}[h]
\begin{center}
   \includegraphics[width=0.5\textwidth]{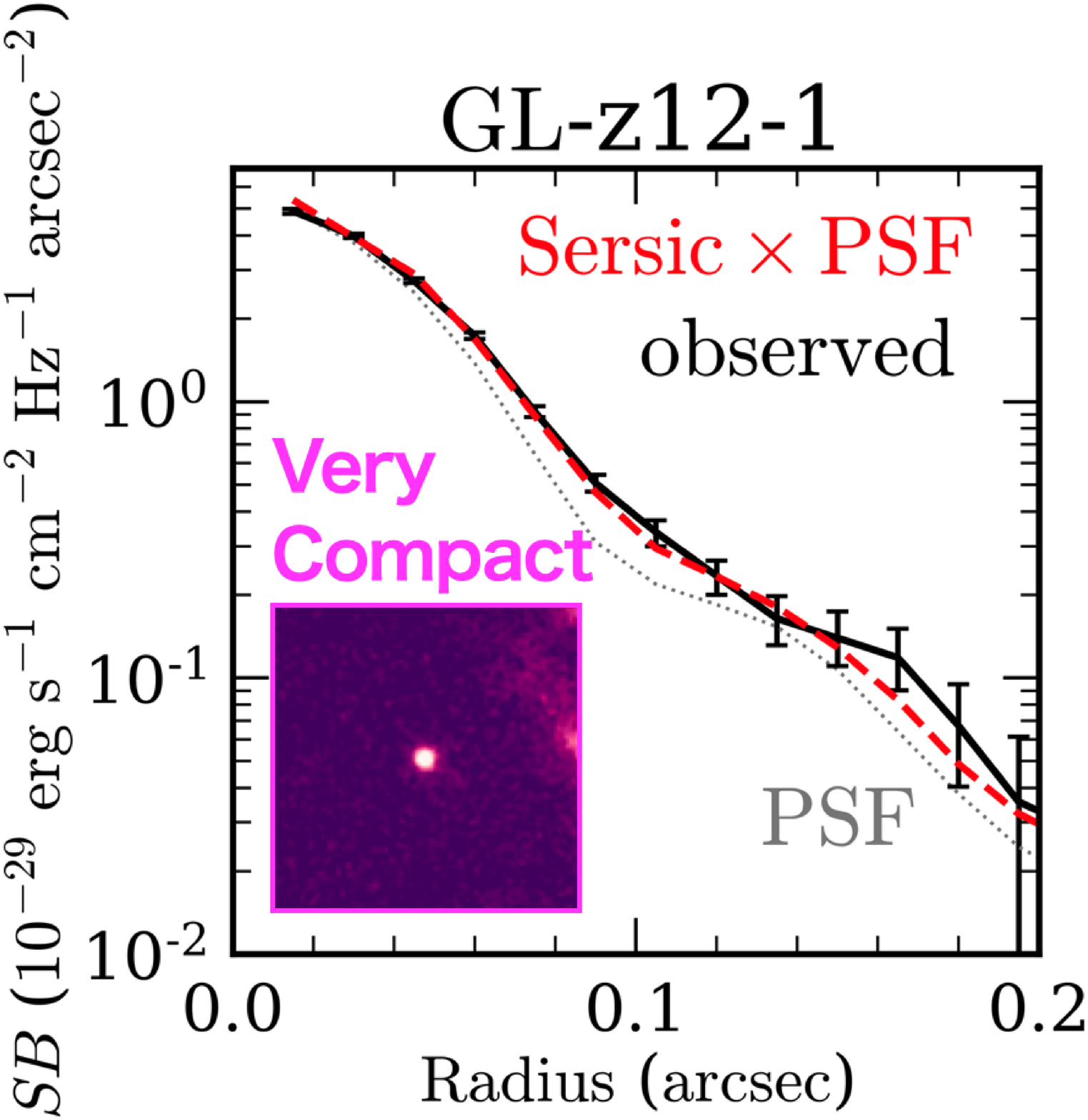}
\caption{
UV continuum surface brightness radial profile of GL-z12-1. 
The black solid curve 
corresponds to the observed surface brightness profile in F200W. 
The $1\sigma$ uncertainties are calculated based on the $68$ percentiles 
of radial profiles obtained at randomly selected positions in the F200W image. 
The red dashed curve denotes the best-fit S\'ersic profile convolved with the PSF. 
The gray dotted curve represents the PSF profile 
whose peak is normalized by the peak of the observed profile 
to clarify the profile difference. 
The inset panel is the $1\farcs5 \times 1\farcs5$ cutout of the original image. 
}
\label{fig:radial_SB_GL_z12_1}
\end{center}
\end{figure}

\begin{figure}[h]
\begin{center}
   \includegraphics[width=0.5\textwidth]{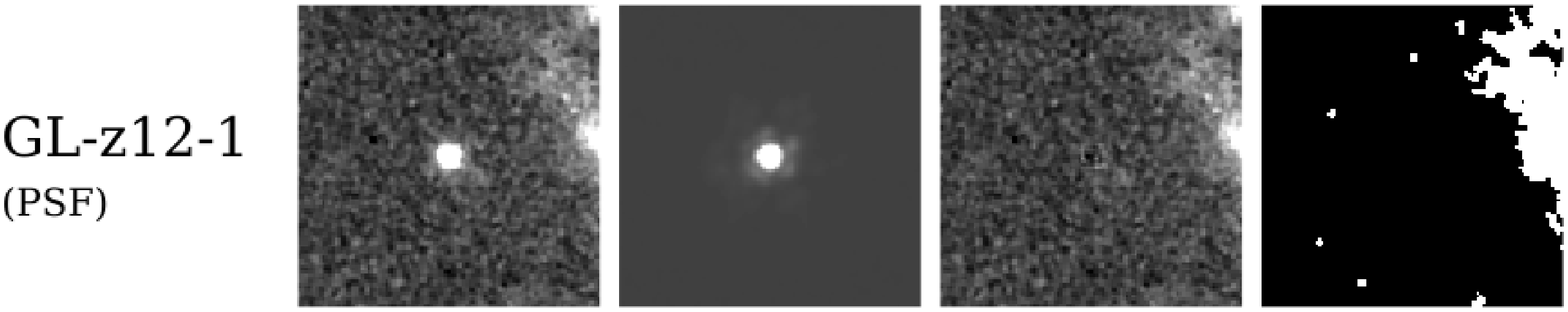}
   \includegraphics[width=0.5\textwidth]{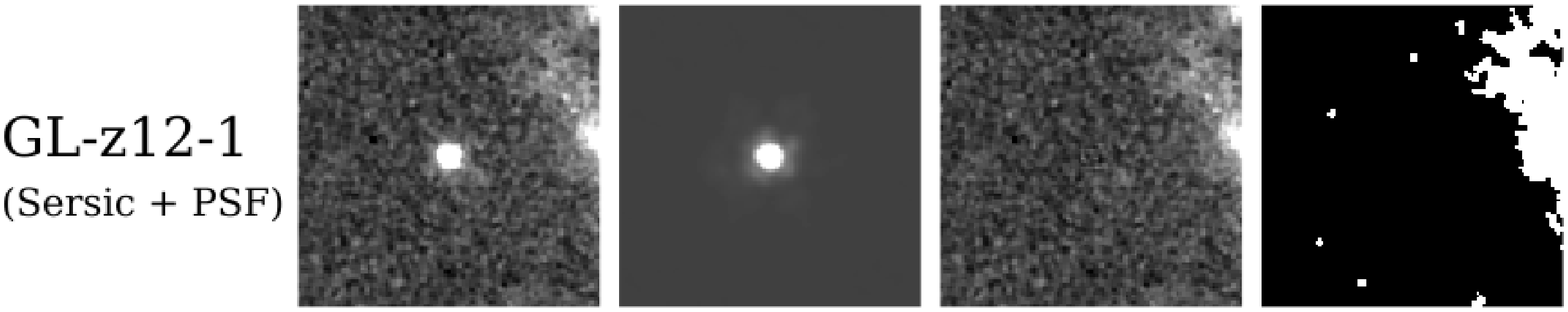}
\caption{
Further surface brightness profile fitting results for GL-z12-1. 
The top panels present the results with a zero-size PSF for GL-z12-1. 
The bottom panels show the results for GL-z12-1 
with a superposition of a zero-size PSF for the possible AGN contribution 
and a S\'ersic profile for the host galaxy. 
}
\label{fig:fit_results_f200w_z12_2comp}
\end{center}
\end{figure}

The other possible physical origin is AGNs. 
If the AGN contribution is dominant, 
their surface brightness profiles are expected to be mostly explained with the PSF profiles. 
However, as shown in Figure \ref{fig:radial_SB_GL_z12_1}, 
we find that its radial profile is clearly more extended than the PSF profile. 
This can also be confirmed by the surface brightness profile fitting. 
In the top panels of Figure \ref{fig:fit_results_f200w_z12_2comp}, 
we present the results of the profile fitting with a PSF for GL-z12-1. 
In this case, the free parameters are the centroid coordinates and the total magnitude. 
The residual image clearly shows a systematically negative region around the center, 
indicating that the PSF fitting result is not as good as the S\'ersic profile fitting result. 
These results suggest that the AGN contribution is not dominant to its surface brightness profile.

The possibility that GL-z12-1 hosts a faint AGN cannot be ruled out 
based on the comparison only with the PSF.  
To examine it, 
we perform the surface brightness profile fitting 
with a composite of two components: a PSF profile for the possible faint AGN component 
and a S\'ersic profile for the host galaxy component. 
The bottom panel of Figure \ref{fig:fit_results_f200w_z12_2comp} 
presents the results of the two component profile fitting. 
Because there is almost no systematically negative or positive regions around center of the residual image, 
the surface brightness profile of GL-z12-1 is characterized well with the two components. 
The obtained total magnitudes in the two component fitting are $M_{\rm UV} = -20.0^{+0.4}_{-0.4}$ mag 
for the PSF component 
and $M_{\rm UV} = -20.2^{+0.3}_{-0.3}$ mag 
for the S\'ersic component. 
Importantly, even in this two component fitting, 
the size of the S\'ersic component is still small, $r_{\rm e} = 48^{+38}_{-15}$ pc, 
which means that even with a faint AGN, 
a very compact star-forming component is needed 
to explain the surface brightness profile of GL-z12-1.

In the discussion above, 
we consider compact star formation and AGNs 
as possible physical origins for the very compact size of GL-z12-1. 
Based on the detailed analyses with its surface brightness profile, 
we find that, even if GL-z12-1 hosts a faint AGN, it still needs 
a very compact S\'ersic component, 
suggesting that we detect extremely compact star formation at $z\sim12$ for the first time  
thanks to the great sensitivity and resolution of JWST.

\begin{figure}[h]
\begin{center}
   \includegraphics[width=0.5\textwidth]{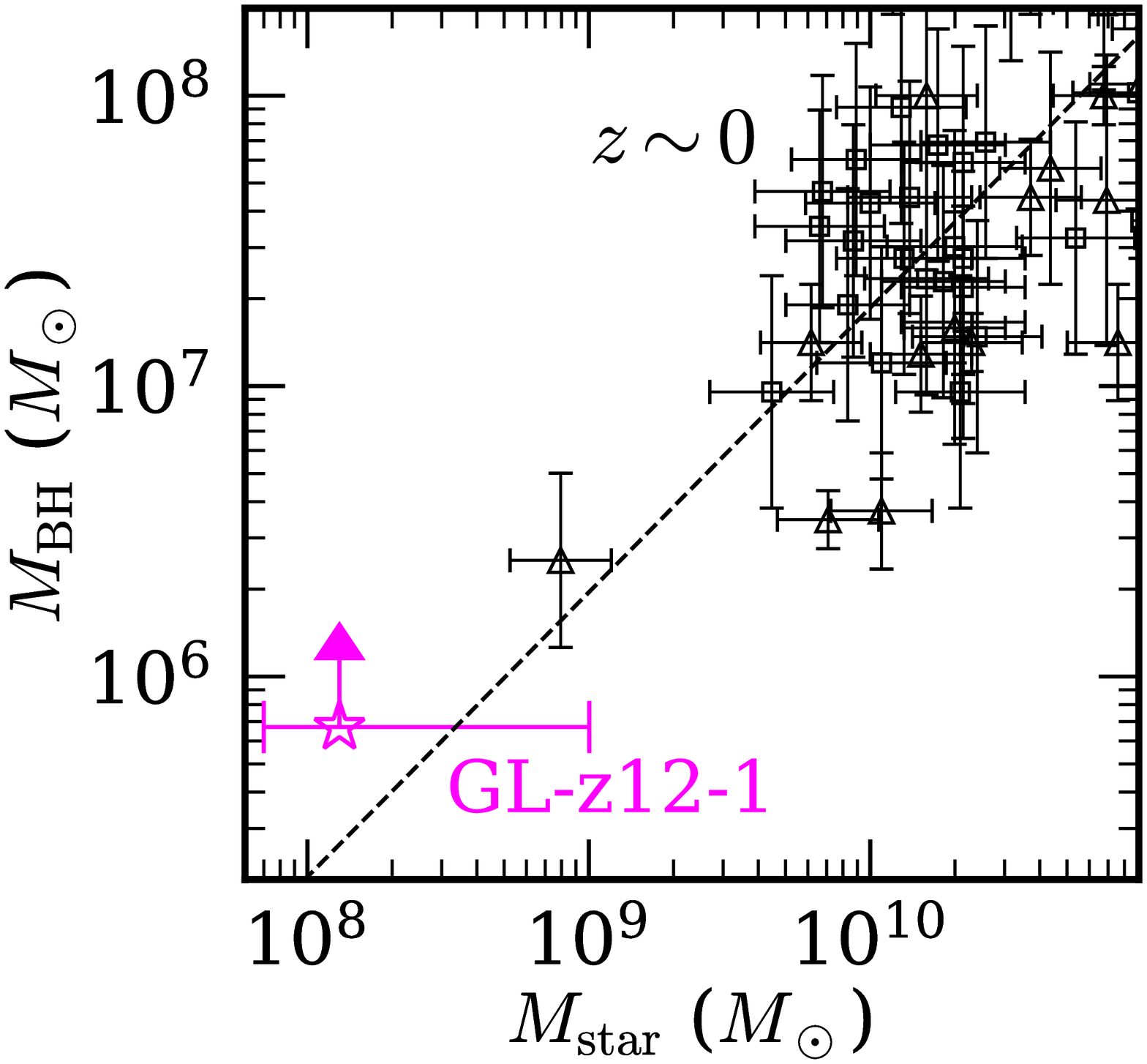}
\caption{Central black hole mass vs. stellar mass. 
The magenta open star denotes GL-z12-1, 
whose black hole mass lower limit is calculated from the UV luminosity 
in the case of the Eddington limit. 
The black open triangles and squares are local galaxies 
in \cite{2004ApJ...604L..89H} and \cite{2011ApJ...726...59B}, respectively, 
compiled by \cite{2020ApJ...888...37D}. 
The dashed line denotes the best-fit local relation, 
$\log (M_{\rm BH} / 10^7 M_\odot) = 0.27 + 0.98 \log (M_{\rm star} / 10^{10} M_\odot)$,
presented in Figure 7 of \cite{2020ApJ...888...37D}. 
}
\label{fig:MBH_Mstar}
\end{center}
\end{figure}

\begin{figure}[h]
\begin{center}
   \includegraphics[width=0.5\textwidth]{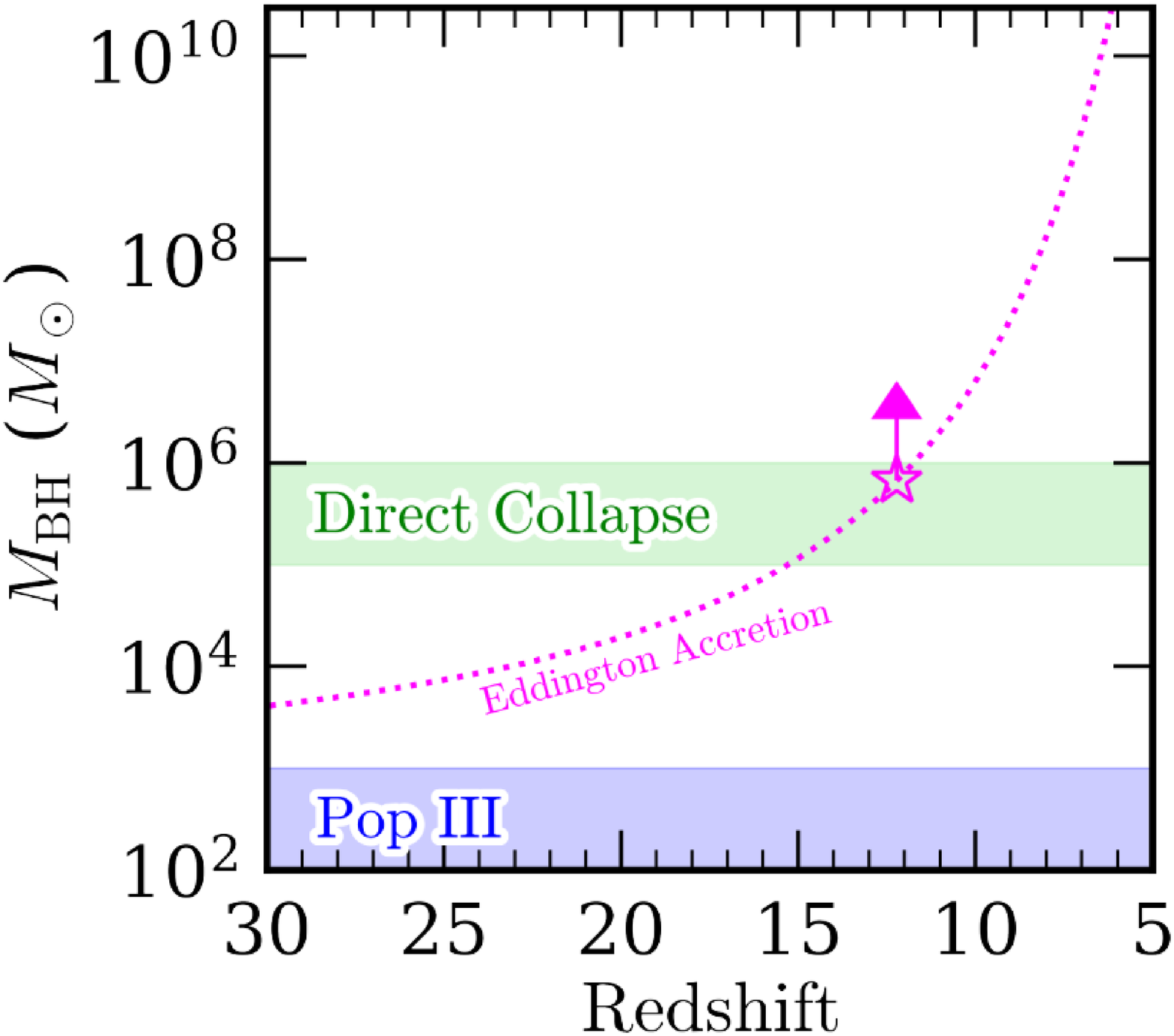}
\caption{Central black hole mass as a function of redshift. 
The magenta open star denotes GL-z12-1, 
whose black hole mass lower limit is calculated from the UV luminosity 
in the case of the Eddington limit. 
The magenta dotted curve corresponds to 
the expected growth history of the central black hole in GL-z12-1 
in the case of the Eddington limit. 
The green and blue shaded regions denote 
the seed mass range of direct collapse black holes 
($M_{\rm seed} \sim 10^{5-6} M_\odot$; \citealt{2016PASA...33...51L}) 
and that of Pop III remnant black holes 
($M_{\rm seed} \lesssim 10^{3} M_\odot$; \citealt{2014ApJ...781...60H}), 
respectively. 
}
\label{fig:MBH_redshift}
\end{center}
\end{figure}

\subsection{A Possible Faint AGN at $z\sim12$?} \label{subsec:discuss_faintAGN}

Although the two component fitting result for GL-z12-1
does not strongly suggest the existence of a faint AGN in it, 
we compare this possible faint AGN 
with the Magorrian relation (e.g., \citealt{2013ARA&A..51..511K}). 
The bolometric luminosity of the PSF component of GL-z12-1 
converted from the UV luminosity is $L_{\rm bol} \simeq 8.7 \times 10^{43}$ erg s$^{-1}$. 
In the case of the Eddington limit,  
$L_{\rm bol} = L_{\rm Edd} = 1.3 \times 10^{38} M_{\rm BH}$ 
(e.g., \citealt{1979rpa..book.....R}),  
where $L_{\rm Edd}$ is the Eddington luminosity 
and $M_{\rm BH}$ is the central black hole mass, 
we obtain $M_{\rm BH} \simeq 6.7 \times 10^5 M_\odot$, 
corresponding to the lower limit of the central black hole mass in GL-z12-1.   
%
%
%
In \cite{2023ApJS..265....5H}, 
the stellar mass of GL-z12-1 is estimated to be 
$\simeq 2.3 \times 10^8 M_\odot$ 
based on the SED fitting (see their Section 3.3 and Table 10). 
Because this stellar mass is derived from the SED with the total luminosities, 
by scaling the stellar mass by the UV luminosity based on the two component fitting result, 
we obtain the stellar mass of the host galaxy, 
$M_{\rm star} \simeq 1.3 \times 10^8 M_\odot$. 
Figure \ref{fig:MBH_Mstar} compares the stellar and central black hole masses of GL-z12-1 
with those of local objects in the literature compiled by \cite{2020ApJ...888...37D} 
(see also, \citealt{2004ApJ...604L..89H}; \citealt{2011ApJ...726...59B}). 
The results of GL-z12-1 are consistent with the extrapolation of the relation 
between $M_{\rm BH}$ and $M_{\rm star}$ in the nearby Universe to the low mass regime.

\begin{figure*}[ht]
\begin{center}
   \includegraphics[width=0.8\textwidth]{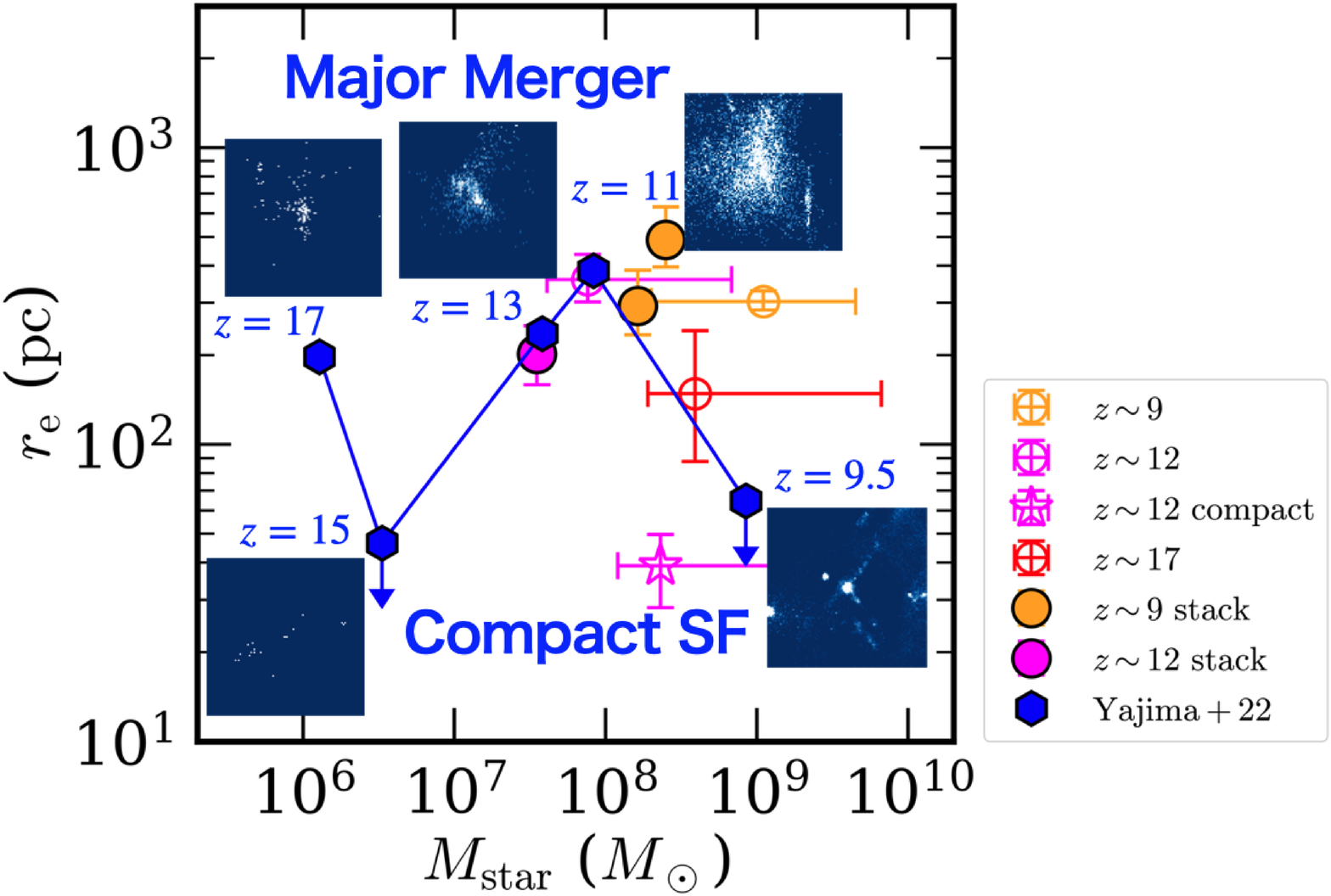}
   \includegraphics[width=0.8\textwidth]{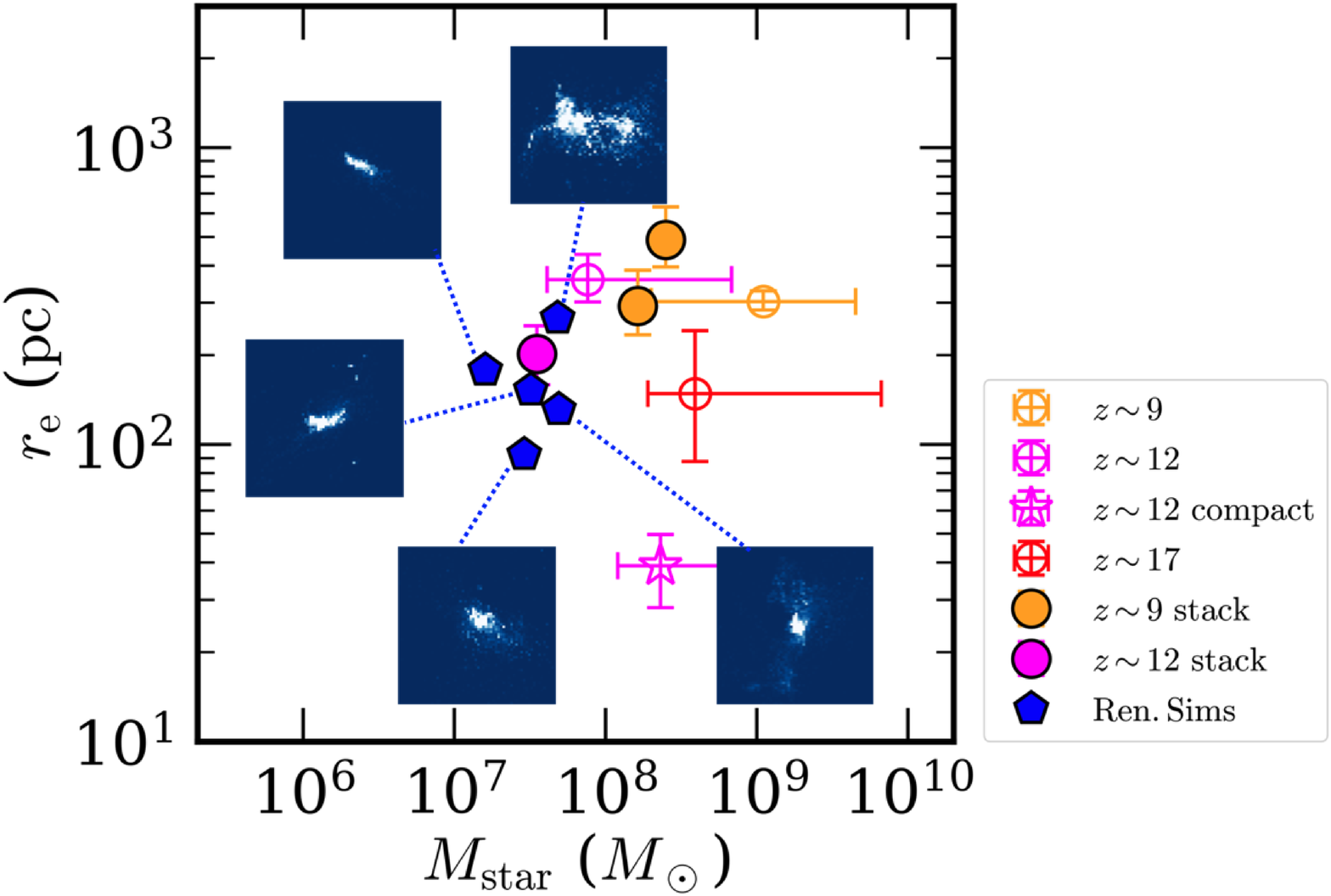}
\caption{
Comparison with the results of theoretical studies on the half-light radius versus stellar mass plane. 
\textbf{Top}: 
The blue hexagons represent the evolution of a star-forming galaxy 
from $z=17$ to $z=9.5$ 
with similar stellar masses and redshifts to our candidates 
in a cosmological hydrodynamics zoom-in simulation (\citealt{2022MNRAS.509.4037Y}).
Each inner image is 
a $1\farcs5 \times 1\farcs5 $ snapshot 
of the 2D projected stellar mass distribution at each redshift. 
The other symbols are the same as the bottom panel of Figure \ref{fig:re_Muv}. 
\textbf{Bottom}: 
The blue pentagons denote star-forming galaxies at $z=10$--$15$ 
with similar stellar masses to our candidates 
extracted from the Renaissance Simulations 
(\citealt{2016ApJ...833...84X}; \citealt{2017MNRAS.469.4863B}). 
Each inner image corresponds to 
a $\simeq 1\farcs \times 1\farcs $ snapshot 
of the surface brightness of each simulated galaxy.
The other symbols are the same as the bottom panel of Figure \ref{fig:re_Muv}.
}
\label{fig:comparison_theory}
\end{center}
\end{figure*}

It is also interesting to check 
whether such a massive black hole can be formed by $z\sim12$. 
The black hole mass $M_{\rm BH}$ increases exponentially with time $t$ 
from the seed black hole with mass of $M_{\rm seed}$ at a fixed Eddington ratio 
(e.g., \citealt{2018Natur.553..473B}; \citealt{2019ApJ...880...77O}), 
\begin{equation}
M_{\rm BH}
	= M_{\rm seed} \exp \left( \dfrac{t}{\tau} \right), 
\end{equation}
where $\tau$ is the $e$-folding timescale, 
\begin{equation}
\tau
	= 0.45 \left( \dfrac{\eta}{1-\eta} \right) \left( \dfrac{L_{\rm bol}}{L_{\rm Edd}} \right)^{-1} \, {\rm Gyr}.
\end{equation}
We adopt the radiation efficiency of $\eta = 0.1$, 
which corresponds to a standard thin accretion disk (\citealt{1976MNRAS.175..613S}). 
In this case, $\tau = 50$ Myr. 
The estimated growth history of the black hole with the estimated mass at $z\sim12$
at the Eddington limit is plotted in Figure \ref{fig:MBH_redshift}. 
If this black hole grows at the Eddington limit toward lower redshifts, 
the expected black hole mass at $z\sim6$--$7$ is $\sim 10^{9-10} M_\odot$, 
which is comparable to those of central super massive black holes in $z\sim 6$--$7$ QSOs 
found in the previous work 
(e.g., \citealt{2015Natur.518..512W}; \citealt{2018Natur.553..473B}; \citealt{2019ApJ...880...77O}; 
\citealt{2019ApJ...873...35S}; \citealt{2020ApJ...897L..14Y}; \citealt{2021ApJ...907L...1W}; \citealt{2021ApJ...923..262Y}). 
For higher redshifts, even if the mass of this black hole increases at the Eddington limit, 
it does not fall within the mass range of seed black holes formed from Pop III remnants, 
$M_{\rm seed} \lesssim 10^3 M_\odot$ (e.g., \citealt{2014ApJ...781...60H}), 
at $z<30$, when the first stars and galaxies are thought to have formed (e.g., \citealt{2011ARA&A..49..373B}). 
If the black hole mass of GL-z12-1 at $z\sim12$ is even larger 
and/or the black hole growth rate is smaller than the Eddington limit, 
it is more difficult to explain its seed black hole with Pop III remnants. 
This may suggest the possibility that the central black hole of GL-z12-1 is formed 
by direct collapse of primordial gas clouds, which can result in more massive seed black holes 
with $\sim 10^{5-6} M_\odot$ (e.g., \citealt{2016PASA...33...51L}).

At this moment, we cannot distinguish whether GL-z12-1 hosts 
very compact star formation with or without a faint AGN based on the currently available data. 
Although the number density of faint AGNs at $M_{\rm UV} \simeq -20.0$ mag 
at $z\sim12$ is expected to be very low compared to that of galaxies 
(e.g., Figure 4 of \citealt{2022ApJ...938...25F}), 
implying that the probability of finding a faint AGN at this high redshift is not high, 
deep follow-up observations for exceptionally compact high-$z$ candidates  
may be helpful to examine the existence of a very high-$z$ intermediate-mass black hole (IMBH) 
at an early phase of black hole formation. 


\subsection{Comparison with Cosmological Zoom-In Simulation Results} \label{subsec:discuss_cw_theory}

As described above, we have found that 
one of our $z\sim12$ galaxy candidate, GL-z12-1, has a very compact size of $40$ pc 
in the rest-frame UV continuum.
We have also found that the average size of $z\sim9$--$16$ galaxy candidates is around $200$--$500$ pc. 
Here we compare these observational results 
with two theoretical study results for galaxies with comparable stellar masses at similar redshifts.

First, we compare our results with those in \cite{2022MNRAS.509.4037Y}, 
who have performed cosmological hydrodynamics zoom-in simulations to study galaxy formation and evolution 
in a large comoving volume of $(714 \,\,{\rm Mpc})^3$ 
for the project named FOREVER22 
(See also, \citealt{2022arXiv221112970Y}).  
Based on their results of the most high-resolution run 
with a mass resolution of $\sim 8 \times 10^3 M_\odot$ (First run), 
we focus on a single galaxy with stellar masses of about $10^{8-9} M_\odot$ at $z\sim9$--$12$ 
that are comparable to those of our galaxy candidates. 
We extract its 2D projected stellar mass density distributions from $z=17$ to $z=9.5$ for comparison.

The evolution of its half-light radius and stellar mass is compared with 
our observational results in the top panel of Figure \ref{fig:comparison_theory}. 
The stellar mass of the simulated galaxy increases with decreasing redshift, while the size becomes smaller and larger. 
Interestingly, at $z=9.5$, the simulated galaxy has a comparable stellar mass 
and a similar or smaller size compared to GL-z12-1.
In the top panel of Figure \ref{fig:comparison_theory}, 
we also present the stellar mass density distribution for the simulated galaxy 
at each redshift as an inset panel near each data point. 
At $z=9.5$, the simulated galaxy is relatively isolated 
with intense matter inflow from the surroundings along the filaments 
involving minor mergers. 
Our finding of the very compact nature of GL-z12-1 would be 
the first observational evidence of very compact star forming galaxies at high redshifts  
that naturally form in theoretical studies.

These physical processes may be similar to those taking place as the compaction, 
which is discussed for the formation of $z\sim2$--$4$ compact galaxies, so-called blue nuggets 
(\citealt{2016MNRAS.457.2790T}; \citealt{2015MNRAS.450.2327Z}). 
Note that the size of GL-z12-1 is much smaller than those of the simulated galaxies with comparable stellar masses at $z=2$ 
in \cite{2016MNRAS.457.2790T} (See their Table 1). 
Such very compact star formation can happen at high redshifts, 
because collapsed haloes at higher redshifts are more compact and dense.

At $z=11$--$13$, 
the simulated galaxy by \cite{2022MNRAS.509.4037Y} 
has similar stellar masses and sizes to those of our faint stacked objects. 
As shown in the inset panels in the top panel of Figure \ref{fig:comparison_theory}, 
at these redshifts, 
major mergers and/or tidal interactions with surrounding objects are taking place, 
which would imply that most of our high-$z$ galaxy candidates 
are experiencing major merger and/or interaction events. 
The major merger rate increases with increasing redshift, 
and galaxies experience $\sim 1$--$2$ major mergers on average by $z=10$ 
based on the Illustris simulation results  
(\citealt{2015MNRAS.449...49R}). 
Because the age of the universe is only $\sim500$ Myr at $z=10$, 
which corresponds to several times the dynamical time at $z=10$ ($\sim 100$ Myr), 
morphologies of galaxies at $z\gtrsim10$ are expected to be strongly affected by major mergers 
for most of the time since their formation.

Next, we also compare our observational results with theoretical results of 
another zoom-in simulation suite that is focused on high-redshift galaxy formation 
called the Renaissance Simulations 
(\citealt{2016ApJ...833...84X}; \citealt{2017MNRAS.469.4863B}). 
We extract five simulated galaxies at $z=10$--$15$ 
with similar stellar masses to our galaxy candidates, $M_{\rm star} \sim 10^{7-8} M_\odot$, 
and calculate their half-light radii for comparison 
as presented in the bottom panel of Figure \ref{fig:comparison_theory}. 
We calculate their surface brightnesses and half-light radii 
using the stellar spectra in the rest-frame UV at $1500${\AA}.
These simulated galaxies have similar stellar masses and sizes 
to those of our $z\sim12$ stacked object.
As shown in the inset panels in the bottom panel of Figure \ref{fig:comparison_theory}, 
all of these simulated galaxies either show an elongated morphology 
or are accompanied by diffuse structures, 
indicating evidence of mergers and/or interactions with nearby objects.
This may suggest that most of our high-$z$ galaxy candidates are 
undergoing mergers and/or interactions, 
in a similar way to the results of \cite{2022MNRAS.509.4037Y}.

The high sensitivity and resolution of JWST 
have made it possible to investigate morphological properties of galaxies at such high redshifts 
that can be compared with cosmological hydrodynamics zoom-in simulation results.  
However, 
at this early stage of JWST explorations, the number of high-$z$ galaxy candidates is limited.  
Deeper JWST images and/or larger high-$z$ galaxy samples 
would be needed for a more robust discussion on their morphological properties.

\section{Summary} \label{sec:summary}

In this study, we have presented 
the surface brightness profile fitting results 
for the $z\sim9$--$16$ galaxy candidates, 
which are securely selected with the conservative photo-$z$ determination criteria $\Delta \chi^2 > 9$ 
as well as the conventional color criteria  
with the deep JWST NIRCam images 
taken by the four ERS and ERO programs, 
i.e., ERS GLASS, ERS CEERS, ERO SMACS J0723, and ERO Stephan's Quintet. 
One of these candidates has been spectroscopically identified at $z=11.44$ 
by recent JWST/NIRSpec spectroscopy.
In the same manner as the previous work, 
we correct for the systematic effects that galaxy sizes and luminosities are systematically underestimated for faint objects 
by performing Monte Carlo simulations. 
Our main results are as follows. 

\vspace{1em}

\begin{enumerate}

\item Our surface brightness profile fitting results indicate that 
most of our $z\sim9$--$16$ galaxy candidates have sizes (half-light radii) of 
$r_{\rm e} \sim 200$--$500$ pc, 
which is comparable to 
previous HST results for high-$z$ galaxies with similar UV luminosities. 
We have also found that 
one of our $z\sim12$ galaxy candidate, GL-z12-1, 
shows an exceptionally compact size of $39 \pm 11$ pc.

\item The sizes of our $z\sim9$ galaxy candidates in the rest-frame optical are 
comparable to those in the rest-frame UV. 
The average size ratio between the rest UV and optical 
for our faint stacked objects is consistent with unity, 
suggesting that 
the distributions of massive stars and less massive ones in faint galaxies are similar, 
probably because of recent star formation activities in low-mass galaxies.

\item The UV size--luminosity relation at $z\sim9$--$12$ 
does not change significantly from the previous results at $z\sim8$. 
The stellar masses and sizes of 
our bright galaxy candidates at $z \sim 9$--$16$ are comparable 
to previously reported values for $z \sim 6$ star-forming galaxies and local UCDs.

\item The sizes of our $z\sim10$--$16$ galaxy candidates with 
$L/L^\ast_{z=3} = 0.12$--$0.3$ and $L/L^\ast_{z=3} = 0.3$--$1$ 
are broadly consistent with the extrapolation of the galaxy size evolution 
based on the previous HST results at $z\sim0$--$10$.

\item The SFR surface densities of our bright high-$z$ galaxy candidates 
are consistent with or higher than those of $z\sim6$--$8$ star-forming galaxies 
at fixed stellar masses. 
In particular, GL-z12-1 shows an exceptionally high $\Sigma_{\rm SFR}$ value 
because of the very compact nature.

\item The surface brightness profile of the very compact galaxy candidate GL-z12-1 
requires a compact S\'ersic component 
even when we consider the possibility of having a faint AGN. 
Our results indicate that GL-z12-1 hosts very compact star formation with or without a faint AGN.

\item If GL-z12-1 has a faint AGN, 
its central black hole mass and stellar mass expected from its luminosity and SED 
are consistent with the extrapolation of the local $M_{\rm BH}$--$M_{\rm star}$ relation. 
It is difficult to explain the seed black hole mass of GL-z12-1 with Pop III remnants, 
which may suggest the possibility of the direct collapse scenario.

\item Recent cosmological simulations for a galaxy with comparable stellar masses at $z\sim9$--$12$ 
indicate that it can host very compact star formation that is consistent with GL-z12-1, 
and can also have star formation 
whose sizes are comparable to those of our stacked objects, 
depending on its evolution phase. 
The comparisons with the simulation results suggest that 
very compact star formation as observed in GL-z12-1 
corresponds to a relatively isolated phase with intense accretion of material, 
while star formation with $200$--$500$ pc sizes 
corresponds to a phase of major merger and/or interaction with the surrounding objects. 

\end{enumerate}

\vspace{1em}

\section*{Acknowledgements}

We thank Marcio B. Mel\'endez 
for giving us helpful advice on how to use WebbPSF. 
This work is based on observations
made with the NASA/ESA/CSA James Webb Space Telescope. 
The data were obtained from the Mikulski Archive for Space Telescopes 
at the Space Telescope Science Institute, 
which is operated by the Association of Universities for Research in Astronomy, Inc., 
under NASA contract NAS 5-03127 for JWST.
These observations are associated with programs 2732, 2736, 1324, and 1345. 
We acknowledge the ERO, GLASS, and CEERS teams 
led by Klaus M. Pontoppidan, Tommaso Treu, and Steven L. Finkelstein, respectively, 
for developing their observing programs with a zero-exclusive-access period.
This work was partially performed using the computer facilities of
the Institute for Cosmic Ray Research, The University of Tokyo. 
This work was supported 
by the World Premier International
Research Center Initiative (WPI Initiative), MEXT, Japan, 
as well as 
KAKENHI Grant Numbers 
15K17602, 
15H02064, 
17H01110, 
17H01114, 
19K14752, 
20H00180, 
21H04467, 
and 22K03670 
through the Japan Society for the Promotion of Science (JSPS). 
JHW is supported by NSF grants OAC-1835213 and AST-2108020 and NASA grants 80NSSC20K0520 and 80NSSC21K1053. 
This work was partially supported by the joint research program of 
the Institute for Cosmic Ray Research (ICRR), University of Tokyo. 

\software{GALFIT (\citealt{2002AJ....124..266P}; \citealt{2010AJ....139.2097P}), 
SExtractor (\citealt{1996A&AS..117..393B}), 
IRAF (\citealt{1986SPIE..627..733T,1993ASPC...52..173T}),\footnote{IRAF is distributed by the National Optical Astronomy Observatory, 
which is operated by the Association of Universities for Research in Astronomy (AURA) 
under a cooperative agreement with the National Science Foundation.} 
SAOImage DS9 \citep{2003ASPC..295..489J},
Numpy \citep{2020Natur.585..357H}, 
Matplotlib \citep{2007CSE.....9...90H}, 
Scipy \citep{2020NatMe..17..261V}, 
Astropy \citep{2013A&A...558A..33A,2018AJ....156..123A},\footnote{\url{http://www.astropy.org}}, 
Ned Wright's Javascript Cosmology Calculator \citep{2006PASP..118.1711W}.\footnote{\url{http://www.astro.ucla.edu/~wright/CosmoCalc.html}}
}

\bibliographystyle{aasjournal}
\bibliography{ref}

\appendix


\section{Monte Carlo Simulation Results for Varying Input S\'ersic Index Values} \label{sec:MCsimulation_for_varying_Sersic_Index}

In Section \ref{sec:measurements}, 
we perform the 2D surface brightness profile fittings for our high-$z$ galaxy candidates at $z\sim9$--$16$ 
with a fixed S\'ersic index of $n=1.5$, 
which is the median value obtained in previous work with HST 
for star-forming galaxies with similar UV luminosities to those of our high-$z$ galaxy candidates 
(\citealt{2015ApJS..219...15S}). 
However, the constraints on the S\'ersic index of 
star-forming galaxies at high redshifts comparable to our candidates are still limited. 
There is a possibility that the S\'ersic index values of such high-$z$ galaxies have a large scatter, 
which would result in a non-negligible systematic uncertainty in our measurements.  
To demonstrate the effect of such uncertainty,  
we perform Monte Carlo simulations 
with the F150W data for the GLASS field, 
following the same procedure as in Section \ref{sec:measurements} 
but with the input S\'ersic index randomly chosen between $0.5$ and $5.0$ 
as an extreme example.

\begin{figure}[ht]
\begin{center}
   \includegraphics[width=0.24\textwidth]{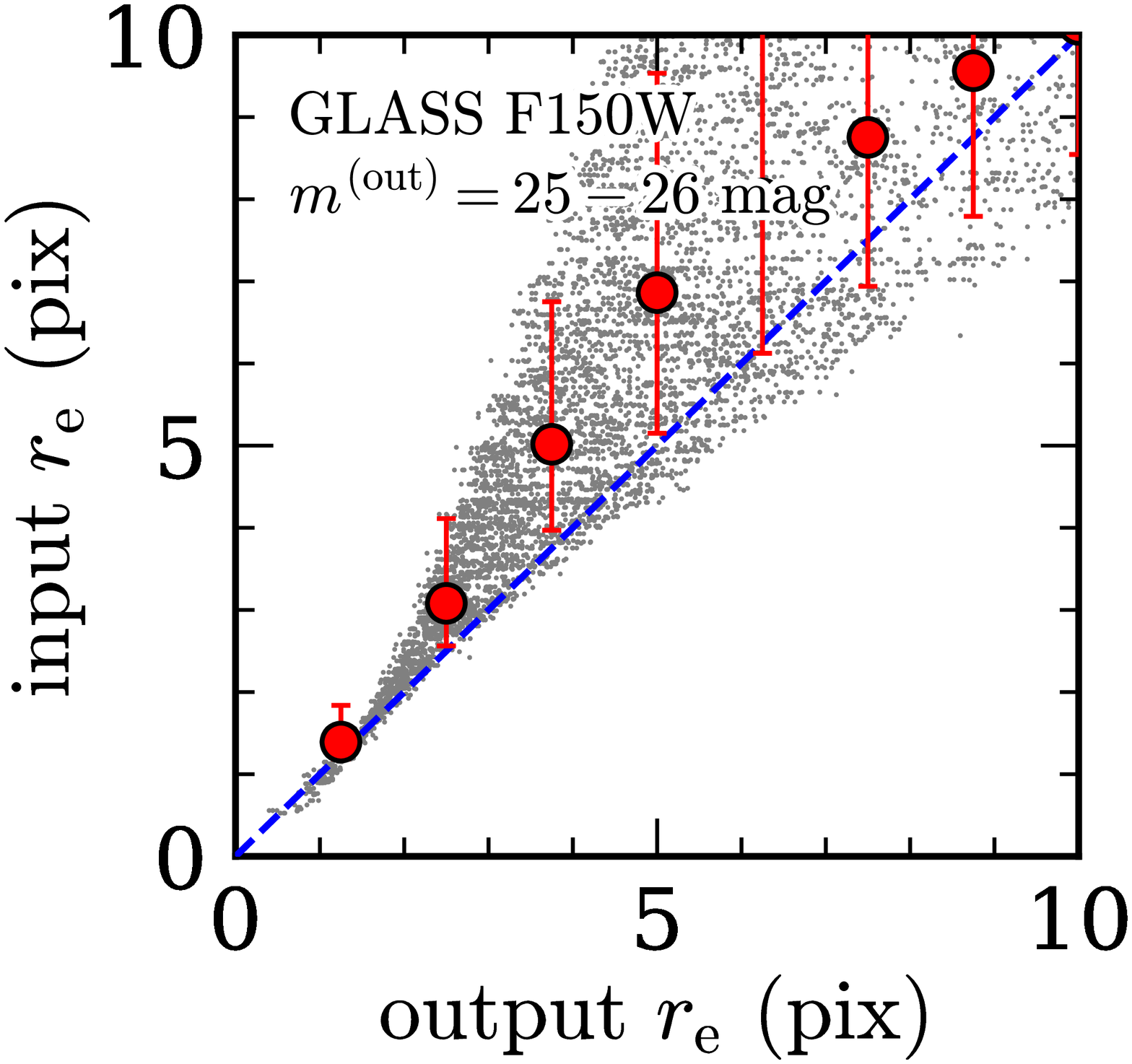}
   \includegraphics[width=0.24\textwidth]{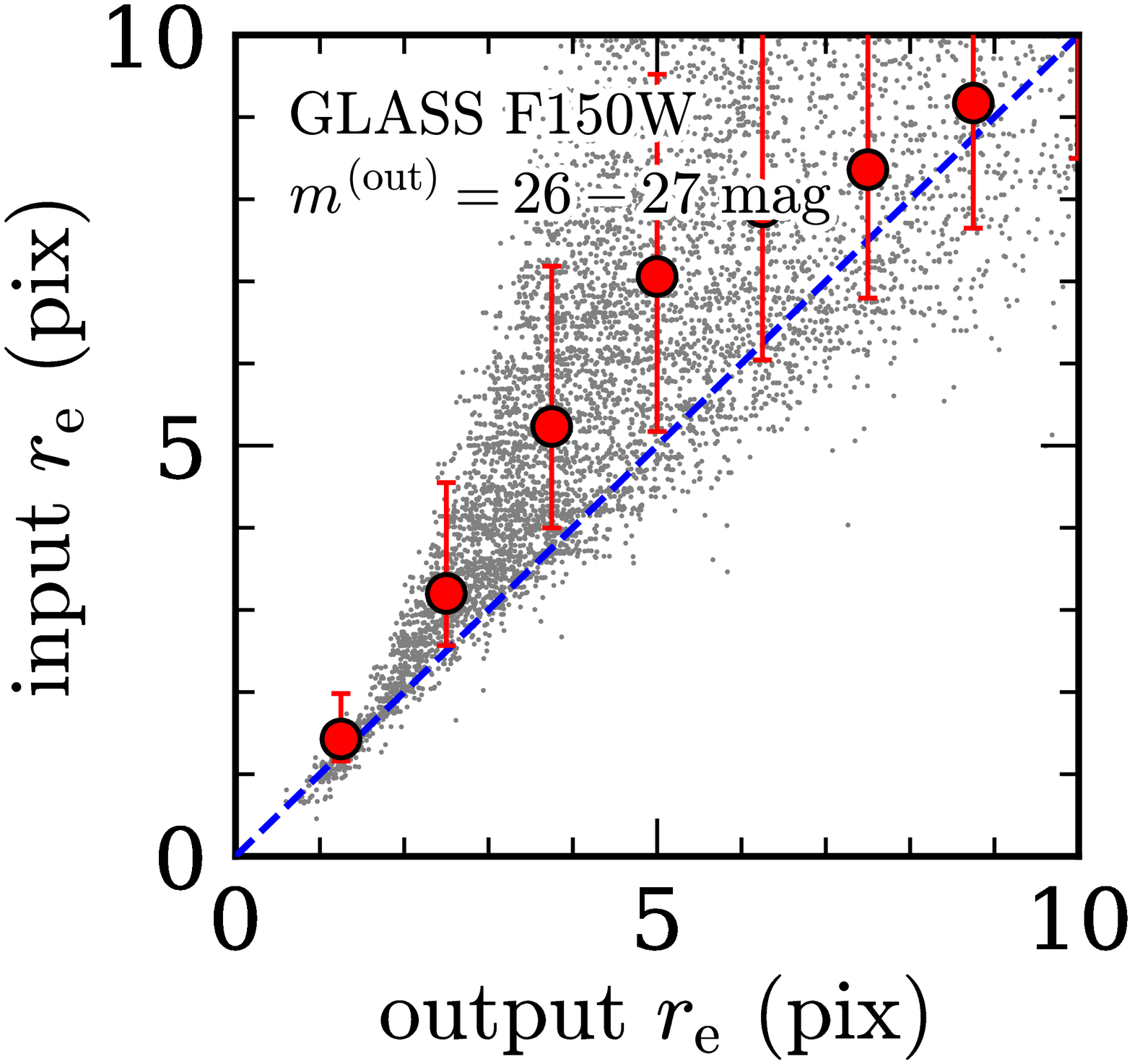}
   \includegraphics[width=0.24\textwidth]{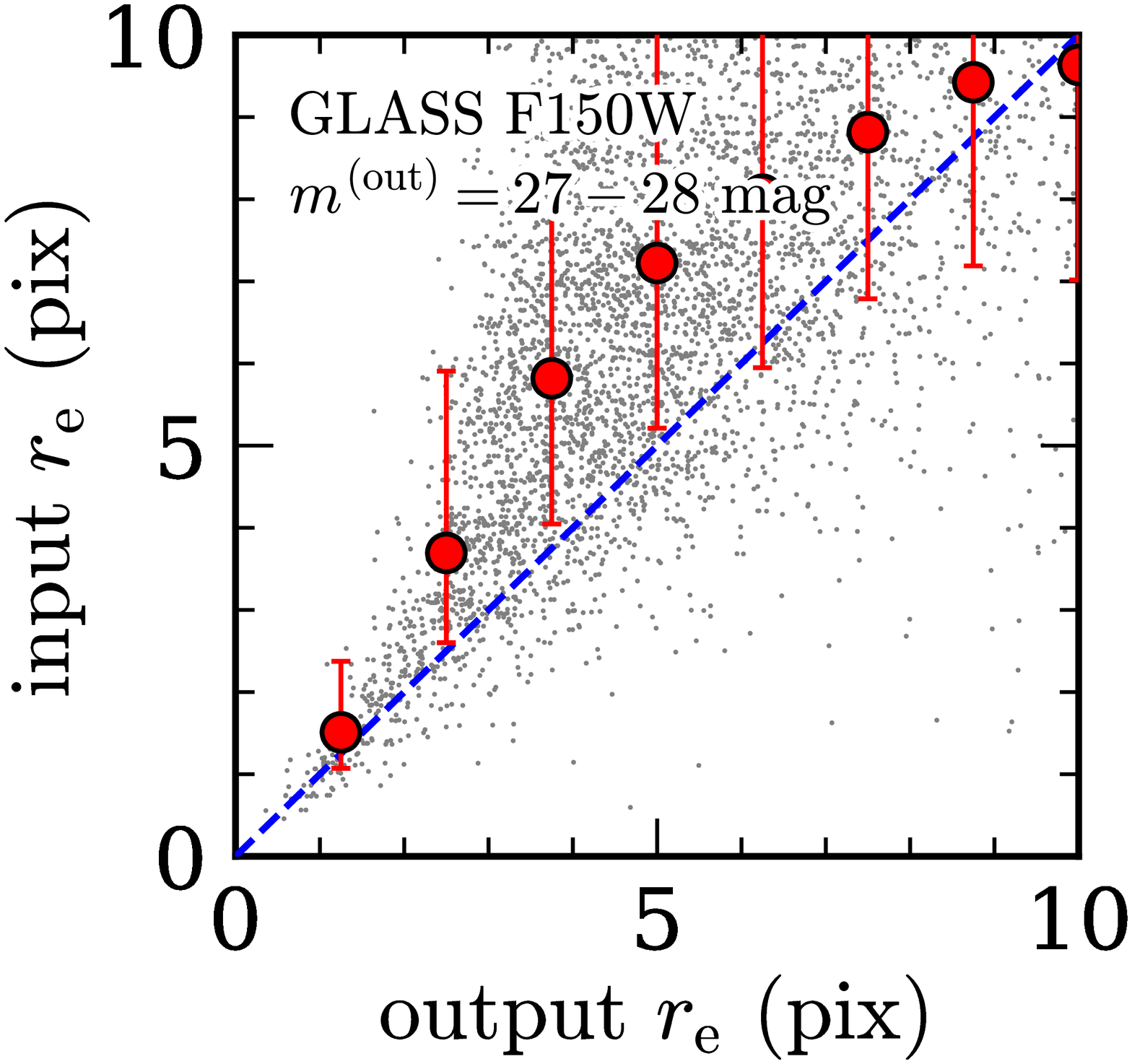}
   \includegraphics[width=0.24\textwidth]{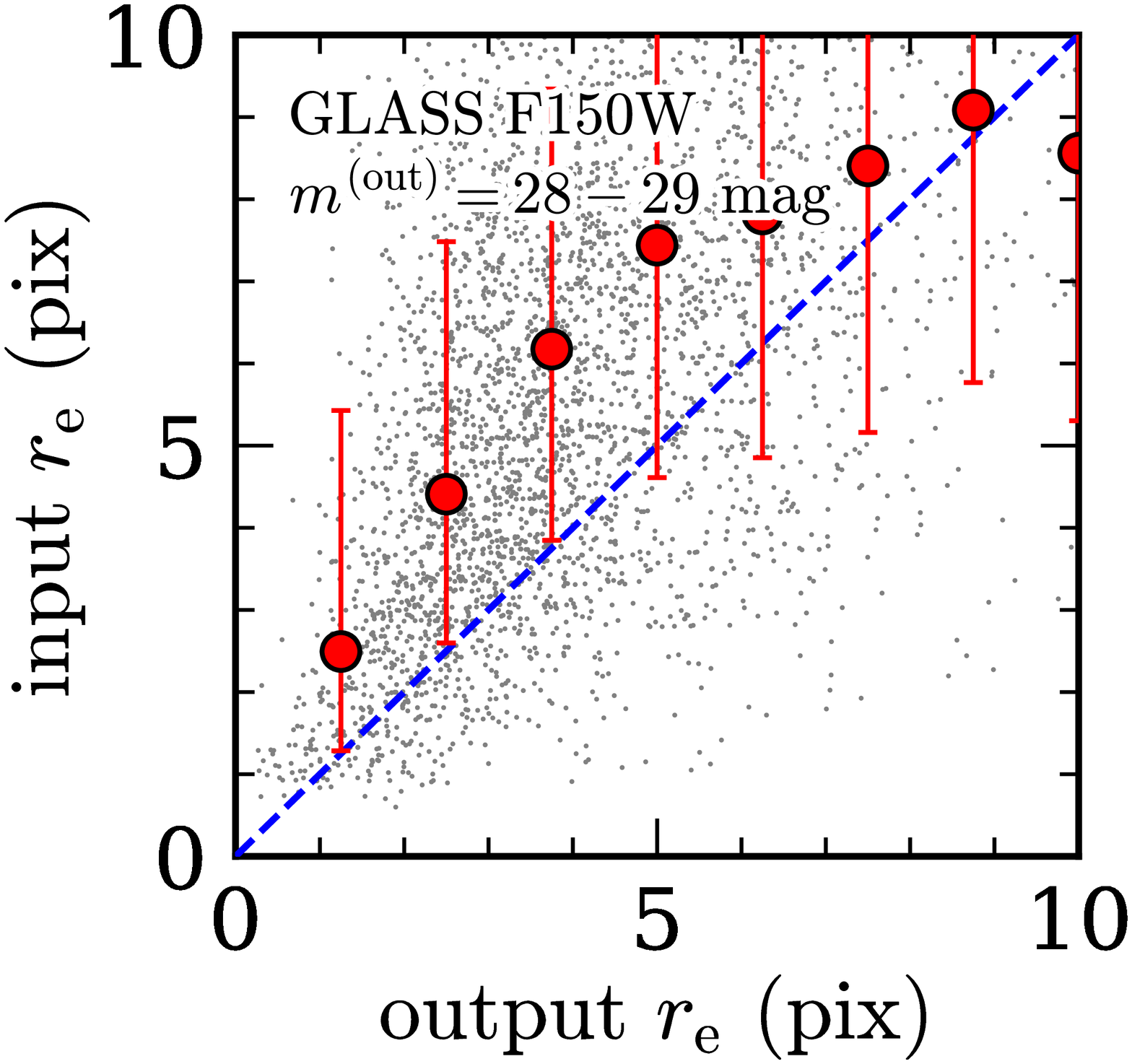}
\caption{
Same as Figure \ref{fig:input_output_re}, 
except that the input S\'ersic index values range from $n=0.5$ to $5.0$. 
}
\label{fig:input_output_re_Sersic0p5_5p0}
\end{center}
\end{figure}

\begin{figure}[h]
\begin{center}
   \includegraphics[width=0.23\textwidth]{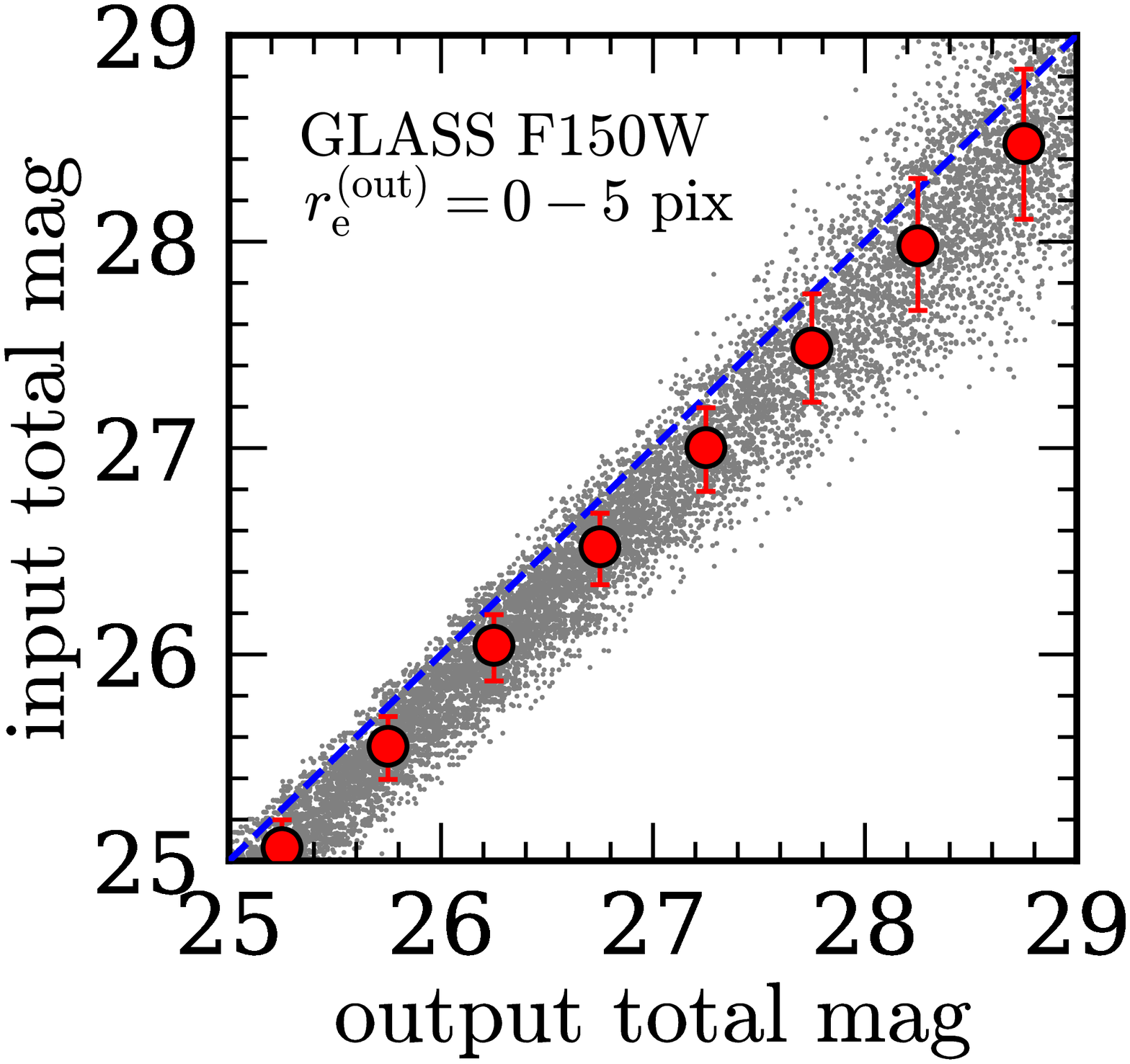}
   \includegraphics[width=0.23\textwidth]{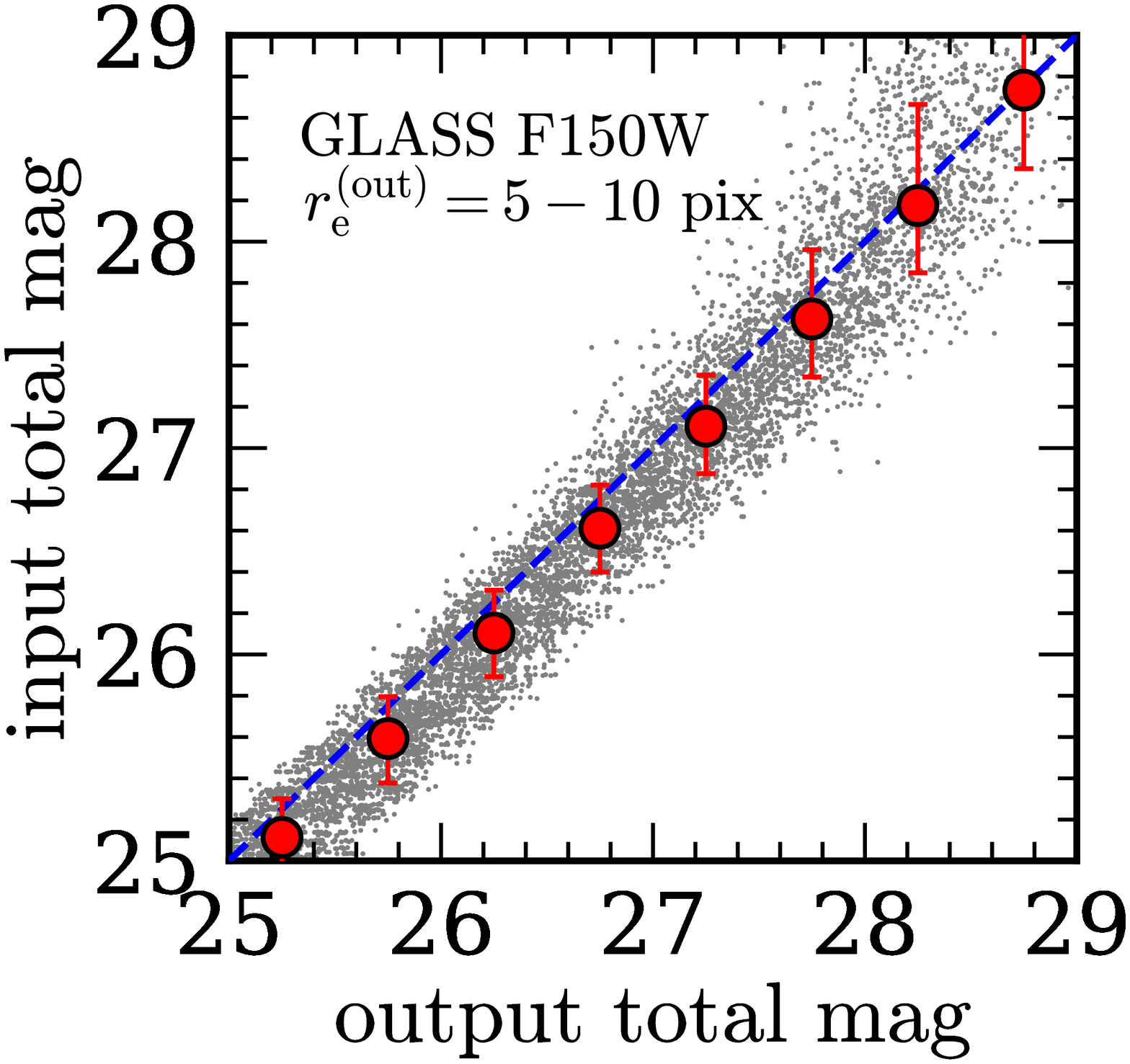}
\caption{
Same as Figure \ref{fig:input_output_mag}, 
but with the input S\'ersic index $n$ randomly chosen between $0.5$ and $5.0$. 
}
\label{fig:input_output_mag_Sersic0p5_5p0}
\end{center}
\end{figure}

Figure \ref{fig:input_output_re_Sersic0p5_5p0} shows 
the results of size measurements for our MC simulated galaxies. 
The dispersions of the data points are significantly larger  
compared to Figure \ref{fig:input_output_re}, 
as expected from the large scatter of the input S\'ersic index values. 
Similarly, Figure \ref{fig:input_output_mag_Sersic0p5_5p0} 
presents the results of total magnitude measurements for our MC simulated galaxies, 
indicating that the dispersions of the data points are larger than 
those in Figure \ref{fig:input_output_mag} as expected again.  
These MC simulation results imply that 
if the S\'ersic index of high-$z$ galaxies has a uniform distribution from $0.5$ to $5.0$, 
size and total magnitude measurements 
obtained with a fixed S\'ersic index of $n=1.5$ have large dispersions. 
However, if the distribution of the S\'ersic index is not so broad, 
the dispersions of the measurements are not as large as those obtained in these MC simulations. 
To address this point, we need to obtain better constraints on the S\'ersic index of high-$z$ galaxies 
based on sufficiently high S/N data in the future.

\end{document}